\documentclass[%
reprint,
longbibliography,
bibnotes,
 amsmath,amssymb,
 aps,
rmp,
showkeys]{revtex4-2}

\usepackage{siunitx}
\usepackage{graphicx}
\usepackage{dcolumn}
\usepackage{array}
\usepackage{amsmath,amssymb,amsfonts}
\usepackage{bm}
\usepackage{mathtools}
\usepackage{tabularx}
\usepackage{upgreek}
\usepackage{array}
\newcolumntype{Y}{>{\centering\arraybackslash}X}
\usepackage[compact]{titlesec}
\usepackage[percent]{overpic}
\titlespacing{\section}{0pt}{*3}{*2}
\titlespacing{\subsection}{0pt}{*1}{*1}
\titlespacing{\subsubsection}{0pt}{*1}{*1}

\usepackage{xcolor}
\newcommand{\vect}[1]{\boldsymbol{#1}}
\newcommand{\eqnreft}[2]{{Eq.~(\ref{#1})#2}}
\newcommand{\figreft}[2]{{Fig.~\ref{#1}(#2)}}
\newcommand{\figo}[1]{Fig.~\ref{#1}}
\newcommand{\figurereft}[2]{{Figure~\ref{#1}(#2)}}
\newcommand{\figureo}[1]{{Figure~\ref{#1}}}

\DeclareUnicodeCharacter{2212}{-}

\begin{document}
\title{Dynamic high-resolution optical trapping of ultracold atoms}

\author{Guillaume Gauthier}
\affiliation{Australian Research Council Centre of Excellence for Engineered Quantum Systems, School of Mathematics and Physics, University of Queensland, St. Lucia, QLD 4072, Australia.}
\author{Thomas A. Bell}
\affiliation{Australian Research Council Centre of Excellence for Engineered Quantum Systems, School of Mathematics and Physics, University of Queensland, St. Lucia, QLD 4072, Australia.}
\author{Alexander B. Stilgoe}
\affiliation{Australian Research Council Centre of Excellence for Engineered Quantum Systems, School of Mathematics and Physics, University of Queensland, St. Lucia, QLD 4072, Australia.}
\author{Mark Baker}
\affiliation{School of Mathematics and Physics, University of Queensland, St. Lucia, QLD 4072, Australia.}
\author{Halina Rubinsztein-Dunlop}
\affiliation{Australian Research Council Centre of Excellence for Engineered Quantum Systems, School of Mathematics and Physics, University of Queensland, St. Lucia, QLD 4072, Australia.}
\author{Tyler W. Neely}
\affiliation{Australian Research Council Centre of Excellence for Engineered Quantum Systems, School of Mathematics and Physics, University of Queensland, St. Lucia, QLD 4072, Australia.}

\date{\today}

\begin{abstract}
All light has structure, but only recently it has become possible to construct highly controllable and precise potentials so that most laboratories can harness light for their specific applications. In this chapter, we review the emerging techniques for high-resolution and configurable optical trapping of ultracold atoms. We focus on optical deflectors and spatial light modulators in the Fourier and direct imaging configurations. These optical techniques have enabled significant progress in studies of superfluid dynamics, single-atom trapping, and underlie the emerging field of atomtronics. The chapter is intended as a complete guide to the experimentalist for understanding, selecting, and implementing the most appropriate optical trapping technology for a given application. After introducing the basic theory of optical trapping and image formation, we describe each of the above technologies in detail, providing a guide to the fundamental operation of optical deflectors, digital micromirror devices, and liquid crystal spatial light modulators. We also describe the capabilities of these technologies for manipulation of trapped ultracold atoms, where the potential is dynamically modified to enable experiments, and where time-averaged potentials can realise more complex traps. The key considerations when implementing time-averaged traps are described.
\end{abstract}

\keywords{Ultracold atom trapping, Time-averaged trapping potentials, Optically trapped Bose-Einstein condensates, Atomtronics, Digital micromirror devices, Spatial light modulators, Acousto-optic deflectors, Beam deflectors}
\maketitle

\tableofcontents


\section{General Considerations}
\label{sec:Introduction}

\subsection{Introduction}
Ultracold dilute gases of atoms and Bose-Einstein condensates (BECs) provide nearly ideal experimental systems for a wide variety of condensed matter physics studies~\cite{cooper2019topological,stamper2013spinor,chien2015quantum}, while also providing a useful medium for developing precision quantum sensors~\cite{degen2017quantum,amico2017focus,seaman2007atomtronics}. Much of the utility of ultracold atoms is provided through their isolation from the external environment, leading to long coherence times, enhanced by the dilute nature of the system. In addition, the large number of control techniques have resulted in an extensive and adaptable toolbox for experimentalists.

Among experimental techniques, trapping and manipulation of ultracold atoms through far-detuned light fields is a mainstay technique of increasing prominence due to the refinement of spatial-light modulator technology in recent decades. These technologies have enabled new capabilities for the near-arbitrary configuration of trapped ultracold atoms. The aim of this Chapter is to highlight the development of  these techniques for trapping cold atoms and to provide the experimenter with a clear resource in understanding and implementing these devices, along with an overview of the experimental progress achieved thus far. The first section summarizes optical dipole trapping and imaging theory, before moving on to detailed descriptions of the operational theory and practical techniques for beam deflection devices and spatial light modulators (SLMs). We describe several different technologies that have risen to prominence, focusing on:
\begin{itemize}
    \item acousto-optic deflectors (AODs) and electro-optic deflectors (EODs).
    \item SLMs in the direct imaging plane, such as digital micromirror devices (DMD-SLMs).
    \item SLMs in the Fourier plane, such as DMD-SLMs or phase-modifying liquid crystal spatial light modulators (LCD-SLMs).
    \item Throughout the review, we highlight a range of scientific experiments in ultracold atoms in which the associated technologies have been used.
\end{itemize}


\subsection{Overview of Optical Trapping}
There are primarily two ways in which light can interact with atoms. The first is through the radiative force resulting from the scattering of photons by the atoms through absorption and re-emission, forming the basis for laser cooling and absorption imaging of the atoms. The second is through the dipole force which complements the radiative force, arising from the interaction between an induced dipole in the electric field and the gradient of the light field. While the dipole force is relatively weak close to resonance, at large detunings of the optical field the dipole force is the dominant component of the light-atom interaction. This dipole force enables most optical trapping in cold atoms with temperatures below the mK range~\cite{grimm2000optical}.

The optical dipole force is directly proportional to the intensity gradient of the projected light pattern. This can provide a high level of control, and there are multiple methods that can be used to shape the intensity gradients of light fields for application in optical dipole trapping of atoms. The most common is the use of simple Hermite-Gaussian or Laguerre-Gaussian beams to create different trapping geometries~\cite{gaunt_bose_2013,tempone2017high,meyrath2005high,carrat2014long}.

While dipole traps formed using single optical beams commonly use the transverse electromagnetic TEM$_{00}$ Gaussian spatial mode, realizing an approximately harmonic trapping potential, higher order modes can be created using static phase masks. Standard optical lenses may resize the resulting fields to form more complex potentials. Further flexibility is achieved using beam deflection devices, which may control the trapping beam center about a transverse plane. This technique realizes dynamical potentials with complex spatial structure, effectively through the combination of multiple Gaussian beams, either in a static or time-averaged implementation~\cite{henderson2009experimental,Schnelle:08,bell2016bose}.

More recently, dynamically reconfigurable devices have supplemented these techniques as their technical state of the art improves. We include here  devices that operate in both the Fourier and image planes of the optical system, and modulate either the phase or amplitude of the light field, or both. With this control, complex optical potentials can be formed at the focus of a final projection lens or optical system. These devices are usually flat plane devices with small control elements, known as pixels, which can be programmed electronically to discretely control the phase or the amplitude of the light within the region of space. The modified wavefront is then projected onto another plane to provide a light intensity gradient to interact with the atoms. Generally, the axial confinement along the projection direction is provided by an independent beam, resulting in a 2D trapping geometry. Structured trapping beams have also been created using diffractive optical elements, often termed `holograms'~\cite{boiron1998cold,newell2003dense,bakr2009quantum,tempone2017high}, and static masks that can be directly imaged onto the atom plane~\cite{eckel2016contact,scherer2007vortex}. 

We now provide a brief outline of the Chapter. Section~\ref{sec:Introduction} provides a general overview of optical potentials for trapping neutral atoms (Sec.~\ref{sec:ODT}), and then describes the imaging equations used to determine the results of placing an SLM in either the Fourier or direct image planes (Sec.~\ref{sec:ImageEquation}, \ref{sec:AddtImage}). The techniques and considerations for producing time-averaged optical potentials are discussed in Sec.~\ref{sec:TimeAvg}. The introductory section of the Chapter concludes with a listing of key experimental requirements and considerations (Sec.~\ref{sec:ExpRequirements}).

Subsequent sections of the Chapter focus on describing the specific technologies of interest: Sec.~\ref{sec:AOD} describes the theory of beam deflection devices and their experimental implementations and Sec.~\ref{sec:DMDs} introduces the theory of DMD-SLMs and their implementation in the direct-imaging and Fourier planes of an optical system. Section~\ref{sec:SLMs_details} describes the operation and implementation of LCD-SLMs in detail, focusing on the operation the commonly-used nematic-type. Section~\ref{sec:conclusion} provides summary remarks and future directions, and includes Table~\ref{table:Comparisons} comparing the devices, providing further guidance to the experimenter in choosing the appropriate technology for their application of interest.

\subsection{Optical Dipole Trapping}
\label{sec:ODT}

We begin by reviewing the atom-light interactions leading to the optical dipole trapping potential. More extensive reviews of the theory of optical dipole traps for trapping neutral atoms can be found elsewhere~\cite{grimm2000optical}. Here we highlight the key parameters by employing  the semiclassical treatment using atomic polarizabilities~\cite{grimm2000optical}. Other approaches include the optical Bloch approach~\cite{PhysRevA.20.224}, dressed states~\cite{Dalibard:85} and Stark energy shifts~\cite{StarkAtom68}.

\subsubsection{Atomic Polarizability}

The interactions between the optical trapping fields and the confined atoms are mediated by the bound electron states. These states have total magnetic moments formed via coupling between the electron spin $S$, orbital $L$ and total nuclear angular momenta  $I$. The electron spin and orbital angular momenta couple most strongly to form the fine-structure states associated with the total electron moments $J = S+L$. Hyperfine-structure states result from the weaker magnetic interactions between the electron and nucleus, and are associated with compound atomic moments $F = J + I$. For atom trapping applications, the energy differences between adjacent hyperfine states are proportionately small when compared with the typically far-detuned trapping fields, as required for low scattering rates. The theory of optical dipole traps can thus be developed using the fine-structure~\cite{grimm2000optical}.

Our treatment commences by considering a generalized atomic energy level structure that features degenerate ground states. The relevant energy level structures for specific atomic species need not incorporate all these features as the ensemble might be polarized in the presence of a magnetic field. The optical trapping field couples the degenerate ground states to the excited states. The composite interaction between the trapping field and atoms is modeled by calculating the total polarizability $\alpha$, using the ground to excited state transition polarizabilities $\alpha_{ge}$ and strengths $k_{ge}$. The ground states are assumed to have equal occupations~\cite{grimm2000optical,PhysRevA.70.023414}.

The generalized two-level transition frequency is denoted by $\omega_{ge}$. The optical field with angular frequency  $\omega$  is detuned by $\Delta_{ge}=\omega-\omega_{ge}$. Atoms in excited states can also undergo spontaneous decay with rate $\Gamma_{\hspace{-0.5ex}ge}$. This spontaneous decay rate  satisfies the condition $\Gamma_{\hspace{-0.5ex}ge}\ll\left|\Delta_{ge}\right|\ll\omega_{ge}$. In this case, the semi-classical atomic polarizability is given by 
\begin{align}
\label{eqn:AtomPolariseMany}
\alpha &= \left\langle \sum_e k_{ge} \hspace{0.2ex} \alpha_{ge} \right\rangle_{\hspace{-1ex}g} \,,\\[1ex]
\alpha_{ge} &= \frac{6\pi\epsilon_0 \hspace{0.1ex} c^3\hspace{0.2ex}\Gamma_{\hspace{-0.5ex}ge}}{\omega_{ge}^2 \left(\omega_{ge}^2-\omega^2\right)} \left(1+ \frac{i\hspace{0.2ex}\omega^3\Gamma_{\hspace{-0.5ex}ge}}{\omega^2_{ge} \left( \omega_{ge}^2-\omega^2\right)}  \right), 
\end{align}
where $\epsilon_0$ is the permeability of free space, and $c$ the speed of light. 

\subsubsection{Transition Strengths}
Transition strengths $k_{ge}$ are calculated using Clebsch-Gordan coefficients, which represent the overlap between  basis representations.  In order to calculate optical transition strength between the ground $|{J_g,m_g}\rangle$ and excited $|{J_e,m_e}\rangle$ states, we consider the overlap $\langle{J_g,m_g}|{J_e,m_e}\rangle$, where $m_a$ represents the projection quantum number for angular momentum state $J_a$. The Clebsch-Gordan coefficients can be calculated in a number of ways, including tables and recursive functions \cite{ModernQM}. Using the Wigner 3-j representation \cite{EdmundsAM1957,Rotenberg:1959zi}, the transition strengths are
\begin{align}
\label{eqn:LineStrengthODT}
k_{ge} &= \left(2J_e+1\right)
\left(
\begin{array}{ccc}
  J_g		&\Delta L   	&J_e   \\
  m_g	&\Delta m   	&-m_e   
\end{array}
\right)^{\hspace{-0.6ex}2}~.
\end{align}

\subsubsection{Dipole Potential for Alkali Atoms}

We now consider an ensemble of alkali atoms in their  $S_{1/2}$ fine-structure ground state that are illuminated with a linearly polarized optical field that results in $\Delta m =0$ transitions. In a zero magnetic background field environment, the $|{m_J=\pm1/2}\rangle$ projection states are degenerate. For the alkali atoms, there are two strong electric dipole transitions which contribute to the optical dipole confinement, termed the D1 and D2 lines. The optical potential $V(\vect{r})$ and dissipative scattering rate $S(\vect{r})$ associated with these atomic transitions and the optical intensity profile $I(\vect{r})$ are given by
\begin{eqnarray}
\label{eqn:OdtTrap}
V(\vect{r}) &=& \left[\frac{-\Re(\alpha)}{2\epsilon_0c}\right] I(\vect{r}) \label{eqn:Potential}\\[0ex]
&=& \left[\frac{\pi c^2 \hspace{0.2ex} \Gamma_{\hspace{-0.4ex}D1}}{\omega_{\hspace{-0.3ex}D1}^2\hspace{0.2ex}(\omega^2\hspace{-0.5ex}-\hspace{-0.5ex}\omega^2_{\hspace{-0.3ex}D1})} + \frac{2\pi c^2 \hspace{0.2ex} \Gamma_{\hspace{-0.4ex}D2}}{\omega_{\hspace{-0.3ex}D2}^2\hspace{0.2ex}(\omega^2\hspace{-0.5ex}-\hspace{-0.5ex}\omega_{\hspace{-0.3ex}D2}^2)}\right] I(\vect{r}) \,,\nonumber\\[2.5ex]
\label{eqn:OdtScat}
S(\vect{r}) &=& \left[\frac{+\Im(\alpha)}{\hbar\hspace{0.1ex}\epsilon_0c}\right] I(\vect{r}) \label{eqn:Scattering} \\[0ex]
&=& \left[\frac{2\pi c^2 \omega^3\hspace{0.2ex} \Gamma_{\hspace{-0.4ex}D1}^2}{\hbar \hspace{0.2ex} \omega_{\hspace{-0.3ex}D1}^4\hspace{0.2ex}(\omega^2\hspace{-0.5ex}-\hspace{-0.5ex}\omega^2_{\hspace{-0.3ex}D1})^2} + \frac{4\pi c^2 \omega^3\hspace{0.2ex} \Gamma_{\hspace{-0.4ex}D2}^2}{\hbar \hspace{0.2ex} \omega_{\hspace{-0.3ex}D2}^4\hspace{0.2ex}(\omega^2\hspace{-0.5ex}-\hspace{-0.5ex}\omega^2_{\hspace{-0.3ex}D2})^2}\right] I(\vect{r})\,,\nonumber
\end{eqnarray}
\noindent where $\Gamma_{Di}$ are the excited state spontaneous decay rates, and $\omega_{Di}$ the angular frequencies of the D-lines. 

Attractive or repulsive optical potentials are formed by large negative or positive optical detunings $\Delta_{Di}$, respectively. This is understood by considering the denominators in the equations~(\ref{eqn:OdtTrap}-\ref{eqn:OdtScat}), which approximately satisfy $\omega^2-\omega_{\hspace{-0.2ex}Di}^2 = (\omega+\omega_{\hspace{-0.2ex}Di})(\omega-\omega_{\hspace{-0.2ex}Di})\approx2\omega\Delta_{\hspace{-0.2ex}Di}$. Since the scattering rate reduces as $\Delta_{\hspace{-0.2ex}Di}^{-2}$, while the potential strength decreases as $\Delta_{\hspace{-0.2ex}Di}^{-1}$, far detuned optical fields can be used to create trapping potentials with long trap lifetimes. Typical detunings range on the order of hundreds of THz, and watt-level powers can achieve \si{\milli\kelvin} trap depths for focused beams. These trap depths are sufficient to confine both ultracold atoms and BECs.

\subsubsection{Gaussian Beam Traps}
\label{sec:GaussianTraps}
From \eqnreft{eqn:Potential}{}, one can create an optical trap by engineering the intensity distribution $I(\vect{r})$ of far-detuned light. Given the single spatial mode output of lasers and optical fibers, a single lens can be used to bring the light to a focus and achieve sufficient trapping potential depth. Repulsive potentials can be achieved for $\Delta_{ge}>0$ while attractive potentials result for $\Delta_{ge}<0$.

For a TEM$_{00} $ Gaussian beam, the trapping potential corresponds to the spatial intensity distribution
\begin{align}
\label{eqn:AtomPolariseMany_Gauss}
I(x,y,z) &= \left( \frac{2 P}{\pi w_x w_y}\right) \exp\left(-\frac{2 x^2}{w_x^2} - \frac{2y^2}{w_y^2}\right),\\[1.5ex]
\label{eqn:AtomPolariseMany_GaussWaist}
w_{x,y}(z) &= \sigma_{x,y} \sqrt{1+\left(\frac{z\lambda}{\pi\sigma_{x,y}^2}\right)^2} ,
\end{align}

where $\sigma_i$ represent the minimum focused waists along the transverse axes, $P$ the optical power and $\lambda$ the optical wavelength. The potential is then given by \eqnreft{eqn:Potential}{}. Expanding the Gaussian potential to 2$^\text{nd}$ order about its center describes a harmonic potential form which simplifies several theoretical considerations. The trap frequencies can be found through a Taylor expansion about the potential minimum, defining $\omega_i = [\frac{\partial^2 V}{\partial i^2}/m]^{1/2}$, where $m$ is the mass of the trapped atom. For the Gaussian potential of \eqnreft{eqn:AtomPolariseMany_Gauss}{}, the effective transverse trap frequencies are $\omega_{x,y} = [4\hspace{0.05ex}V_0/m\hspace{0.1ex}\sigma_{x,y}^2]^{1/2}$, where $V_0 = V(0,0)$ is the trap depth.

Common optical dipole traps (ODTs) use single or crossed Gaussian beams, where $\Delta_{ge} <0$, forming an attractive trap for the atoms. When trapping using a single dipole beam with $\sigma_x = \sigma_y = \sigma$, the confinement along the propagation direction of the beam is determined by the Rayleigh length $z_R = \pi \sigma^2 / \lambda$~\cite{grimm2000optical}, and the trapping frequency can be  determined through a Taylor expansion as above. With the exception of high-numerical aperture projection objectives, often located inside the vacuum system~\cite{Bergamini2004Nov,barredo_synthetic_2018}, limited optical access to the atoms mean that $z_R$ is much larger than $\sigma$ and thus trapping along the propagation direction is weak. SLM and deflector-based traps thus do not usually achieve sufficient trapping along their projection direction. For this reason, confining potentials are usually formed by adding a `light sheet' dipole trap~\cite{davidson1995long} that has weak trapping in the plane of the SLM projection, but has a tight waist and trapping perpendicular to the projection plane, confining the atoms at the focus of the SLM pattern [c.f. \figreft{fig:AOD_Atoms}{a}]. The tight confinement of the light sheet is typically oriented with gravity.

TEM$_{00}$ Gaussian beams form the basis of configurable deflector controlled optical potentials. Deflector devices may displace the trapping center \cite{henderson2009experimental,Schnelle:08,bell2016bose,yavuz2006fast,nogrette_single-atom_2014}, or simultaneously form multiple trapping sites \cite{boyer_dynamic_2006,shin2004atom,meyrath2005bose}.  The flexibility of these methods depends on the spatial resolution of the optical projection system, and temporal resolution of the deflector, as discussed in detail in Sec.~\ref{sec:AOD}.

 \subsubsection{Higher Order Gaussian Traps}
While the majority of trapping uses TEM$_{00}$ Gaussian modes, higher order Hermite Gaussian and Laguerre-Gaussian modes offer more control over the spatial configuration of the trapped atoms. Innovative optical traps have been created from these beams, for example, homogeneous BECs have been created from the combination of repulsive [$\Delta_{ge} >0$] TEM$_{01}$ and LG$_{01}$ modes~\cite{gaunt_bose_2013}. Laguerre-Gaussian modes can be also used to transfer orbital angular momentum to trapped BECs through Raman transitions utilizing multiple electronic states of the atoms~\cite{ryu2007observation,beattie2013persistent,moulder2012quantized}.

The interference between two Gaussian beams, creating an optical lattice, has also been extensively used for trapping cold atoms; such approaches are beyond the scope of the current discussion and we refer the reader to the existing reviews of these techniques~\cite{bloch2005ultracold,morsch2006dynamics,windpassinger2013engineering}. One-dimensional optical lattices have been used to confine BECs into planes to be compatible with SLM-based optical trapping along the orthogonal direction to the lattice~\cite{ville_loading_2017}, similar to the trap achieved with a TEM$_{01}$ mode~\cite{meyrath2005high,plisson2011coherence}. This technique has the notable advantage of permitting the waist of the atom trap to be precisely and dynamically controlled -- closely matching the depth of focus of projection optics, and furthermore permitting the achievement of effective 2D dynamics~\cite{ville_loading_2017,ville_sound_2018}.

\subsection{Imaging Equations}
\label{sec:ImageEquation}

The Fourier-transforming property of a lens is of key importance to image formation and optical trapping when using SLMs to shape the optical dipole field. In the next three sections, we describe the general theory and two devices for image formation, where we use  \textit{direct imaging} or  \textit{Fourier plane imaging}. We begin with a brief review  of the theory of image formation from a Fourier optics perspective~\cite{goodman2005introduction}.

\subsubsection{Diffraction Integrals}
\label{sec:Diffraction_ints}
Although more exact approaches to solving the scalar diffraction from an aperture can be used~\cite{goodman2005introduction}, we first describe the Fresnel diffraction approximation as it can be applied when the distance from the aperture $z \gg \lambda$. This is almost always the case, as the wavelength of light used for optical trapping is extremely small in comparison to propagation distances through optical systems. Let $U(\xi, \eta, z=0 )$ define a scalar field at the input aperture of the optical imaging system. For distances $z \gg \lambda$ from the aperture, scalar fields may be approximated by the Fresnel diffraction integral to describe the optical field $U(x,y,z)$,
\begin{align}
\label{eqn:Fresnel}
\small
    U(x,y,z) =& 
    \frac{e^{ikz \left[1 + \frac{(x^2 + y^2)}{2z^2}\right]}}{i \lambda z} \times \nonumber\\[0.5ex]
    &\int_{-\infty}^{\infty}\hspace{-0.6ex} U(\xi, \eta,0)e^{i\frac{k}{2z}(\xi^2 + \eta^2)-i\frac{2 \pi}{ \lambda z}(x\xi + y\eta)}\, \text{d}\xi \text{d}\eta~,
\end{align}
where $k = 2 \pi/\lambda$. Provided the secondary plane is far from the aperture such that $z \gg k(\xi^2 + \eta^2)/2$, the Fresnel integral may be approximated by the Fraunhofer diffraction integral,
\begin{align}
\label{eqn:Fraunhofer}
\small
    U(x,y,z) =&
    \frac{e^{ikz \left[1 + \frac{(x^2 + y^2)}{2z^2}\right]}}{i \lambda z} \times \nonumber\\[0.5ex] 
    &\int_{-\infty}^{\infty} U(\xi, \eta,0) e^{-i\frac{2 \pi}{ \lambda z}(x\xi + y\eta)}\, \text{d}\xi \text{d}\eta~.
\end{align}
This equation, with the exception of the leading term, can be recognized as the Fourier transform of the field at the aperture: $U(x,y,z) \propto {\cal F}[U(\xi,\eta, z=0)]$, where the spatial frequencies are given by $x/\lambda z, y/\lambda z$.

While the Fraunhofer approximation holds sufficiently far from the aperture, placing a thin lens just after the aperture, and setting the observation plane to $z=f$, the focal length of the lens, applies a phase factor of $e^{-i\frac{k}{2f}(\xi^2 + \eta^2)}$, exactly canceling the phase term within the Fresnel diffraction integral \eqnreft{eqn:Fresnel}{}. The resulting field then has exactly the same form as \eqnreft{eqn:Fraunhofer}{}, where the propagation distance is $z=f$.

\begin{figure}[t!]
\includegraphics[width=\columnwidth]{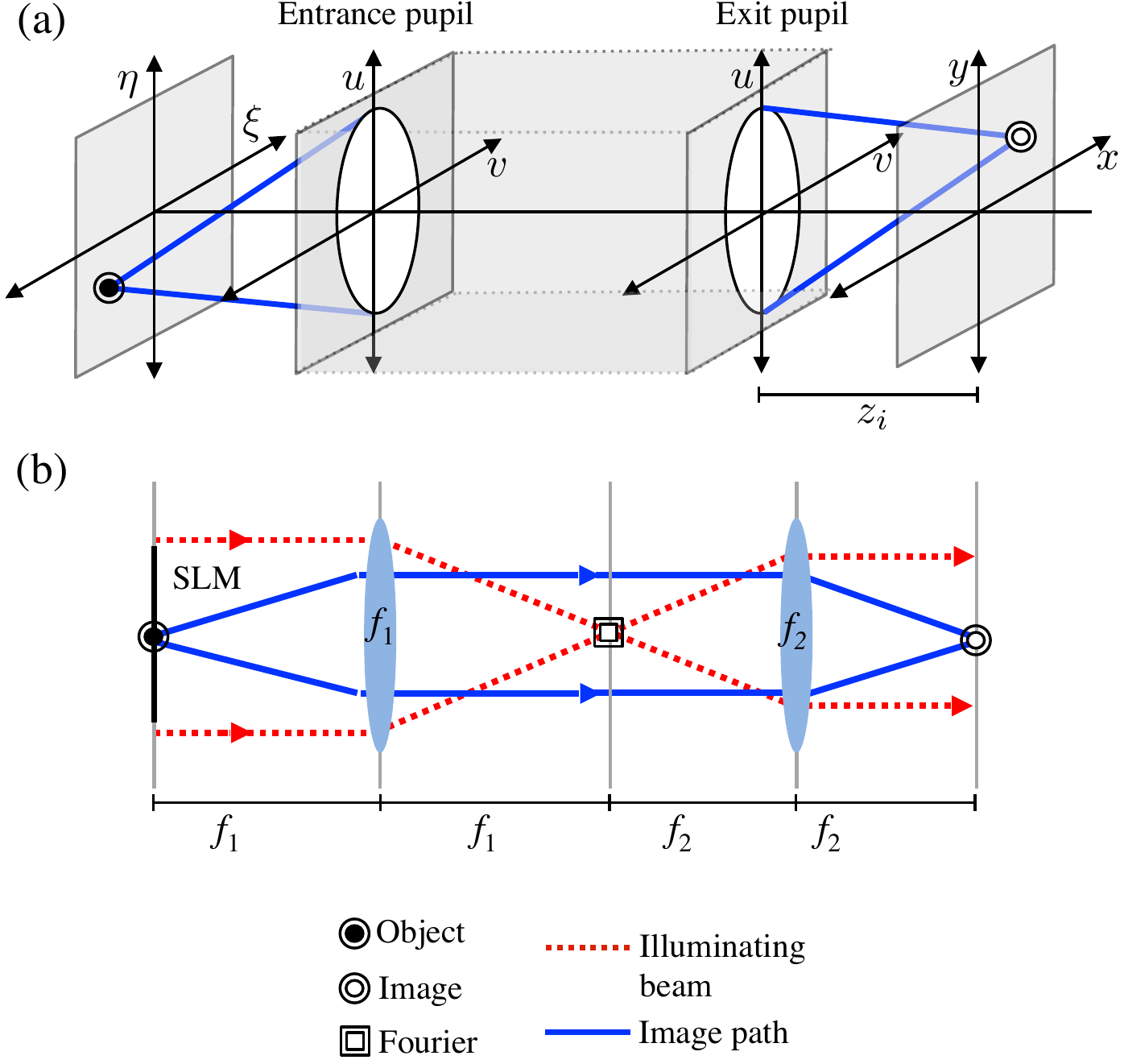}
\caption{(a) Model of an imaging system. The image resolution is determined by the Fraunhofer diffraction pattern of the exit pupil. The intermediate optical system can consist of a number of elements. The diffraction properties of the system can be determined by considering the entrance and exit pupils, which are determined by imaging the aperture  through the system. (b) Simple two-lens infinite-conjugate system for direct imaging of a SLM with spatially coherent illumination. The first lens results in the Fourier transform of the SLM at the intermediate plane.  This plane and the SLM-shaped input field can be considered Fourier transform pairs. When using coherent illumination, care must be taken in both implementations of the optical design to consider the path of the illuminating beam (dashed red) and the image path (blue).}
\label{fig:SLMTwopaths}
\end{figure}

\begin{figure*}[t]
\includegraphics[width=\textwidth]{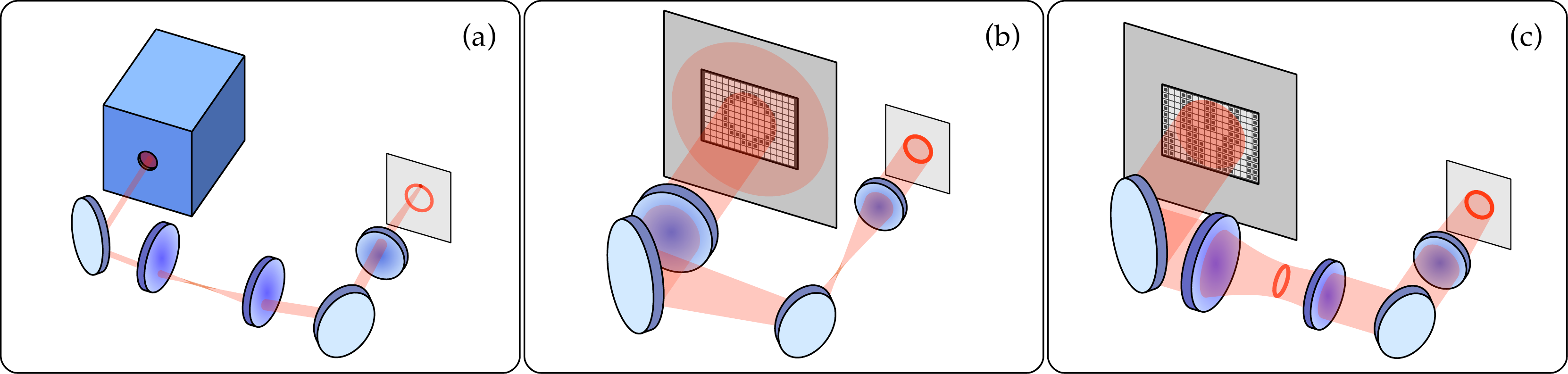}
\caption{Configurable optical trapping technologies which enable the formation of ring shaped potentials.
(a)~A bi-directional beam deflector (AOD or EOD) where, the rapid scanning of the focused beam `paints' the time-averaged potential.
(b)~Direct imaging of a digital micromirror device (DMD-SLM) where the mirrors are configured to produce a ring-shaped mask that is directly imaged to the atom plane. 
(c)~A liquid-crystal spatial light modulator (LCD-SLM)  produces a ring trap by changing the spatial properties of light in the Fourier plane which is then projected by the optical system into a field in the image plane.}
\label{fig:SLMTypes}
\end{figure*}

\subsubsection{Direct Imaging}
The diffraction integral approach introduced in Sec.~\ref{sec:Diffraction_ints} can be applied to the generalized optical system that is shown in \figreft{fig:SLMTwopaths}{a}. The optical system can consist of any number of optical elements, but can be considered in terms of its entrance and exit pupils, these are found by imaging the system's aperture to the object and image planes. The transmittance of the pupil is defined by the following function:
\begin{equation}
    P (u,v) =  
    \begin{cases}
      1 & \text{inside the pupil aperture}\\
      0 & \text{otherwise}.
    \end{cases}       
\end{equation}
\noindent Assuming an ideal plane-wave input, the resulting amplitude at the image plane is defined through
\begin{equation}
\small
h(x,y;\xi,\eta) = \frac{A}{\lambda z_i}\int_{-\infty}^{\infty} \hspace{-2ex}P(u,v)e^{-i\frac{2 \pi}{\lambda z_i}\left[(x-M\xi)v + (y-M\eta)u\right]}\, \text{d}u\,\text{d}v   
\end{equation}

\noindent where $A$ is the input amplitude, $z_i$ defines the distance from the exit pupil to the image plane, and $M$ defines the magnification that is obtained from a geometrical optics treatment of the image formation of the system. By introducing the reduced coordinates, $\tilde{\xi} = M\xi$ and $\tilde{\eta} = M\eta$, the ideal geometrical image $U_g(\tilde{\xi},\tilde{\eta})$ is defined as
\begin{equation}
    U_g(\tilde{\xi},\tilde{\eta}) = \frac{1}{|M|}U\left(\frac{\tilde{\xi}}{M},\frac{\tilde{\eta}}{M}\right).
\end{equation}

\noindent The function $h(x,y;\xi,\eta)$, obtained through the Fourier transform of the pupil function, is also known as the point spread function (PSF) of the optical system and describes the spatial frequencies transmitted by an ideal, aberration free system. The resulting image is then a convolution of the the ideal geometric image with the PSF,
\begin{equation}
\label{equ:PSFconvolution}
    U_i(x,y) = \int_{-\infty}^{\infty} h(x,y;\tilde{\xi},\tilde{\eta})U_g(\tilde{\xi},\tilde{\eta}) \, \text{d}\tilde{\xi}\text{d}\tilde{\eta}.
\end{equation}

We can thus determine the effect of the direct imaging system shown in \figreft{fig:SLMTwopaths}{b} by simply calculating the PSF for the system  along with the magnification determined by the ideal image output. In the typical applications relevant to trapping cold atoms with SLMs, large demagnification (high minification) factors are desired to realize a focused and high-intensity pattern, which can result in sufficiently deep potentials at the far-detuned wavelengths. In the configuration shown in \figreft{fig:SLMTwopaths}{b}, the magnification is given by the ratio of the focal lengths of the objectives, $M = f_2/f_1$. A typical application would utilize an infinity-corrected microscope objective to realize the high minification factor. 

For binary devices such as DMD-SLMs, the smoothness of the projected pattern depends on the number of mirrors contained within each resolution element of the image. This can be controlled through careful selection of the magnification relative to the resolution, given the size of the individual elements of the SLM device and the desired resolution at the image plane of the optical system. A complete description of these techniques is found in Sec.~\ref{sec:DMDs}.

\subsubsection{SLM in the Fourier Plane}
\label{Sec:Fourier_Transform_General}

We now consider the SLM and first lens in \figreft{fig:SLMTwopaths}{b}, where the SLM is placed $f_1$ away from the lens. In this case, one considers the entrance pupil in \figreft{fig:SLMTwopaths}{a}, and propagates the field from the SLM to the entrance pupil in \figreft{fig:SLMTwopaths}{a} which acts as a low-pass filter to the input field from the SLM. Moving to the focal point of the lens, the Fraunhofer approximation shown in \eqnreft{eqn:Fraunhofer}{} is valid, realizing the Fourier transform of the SLM-defined input field. In typical implementations, this intermediate image is often then reimaged on the trapping plane, as shown in \figreft{fig:SLMTypes}{c}.

As described in Sections~\ref{sec:DMDs},~\ref{sec:SLMs_details}, configurable optical patterns can be realized by implementing a phase or amplitude modulation of the input field with the SLM that is the Fourier transform of the desired pattern. This method has the advantage of directly modulating a complex optical field, rather than just its amplitude as described above for direct imaging.

\begin{figure}
\includegraphics[width=0.99\columnwidth]{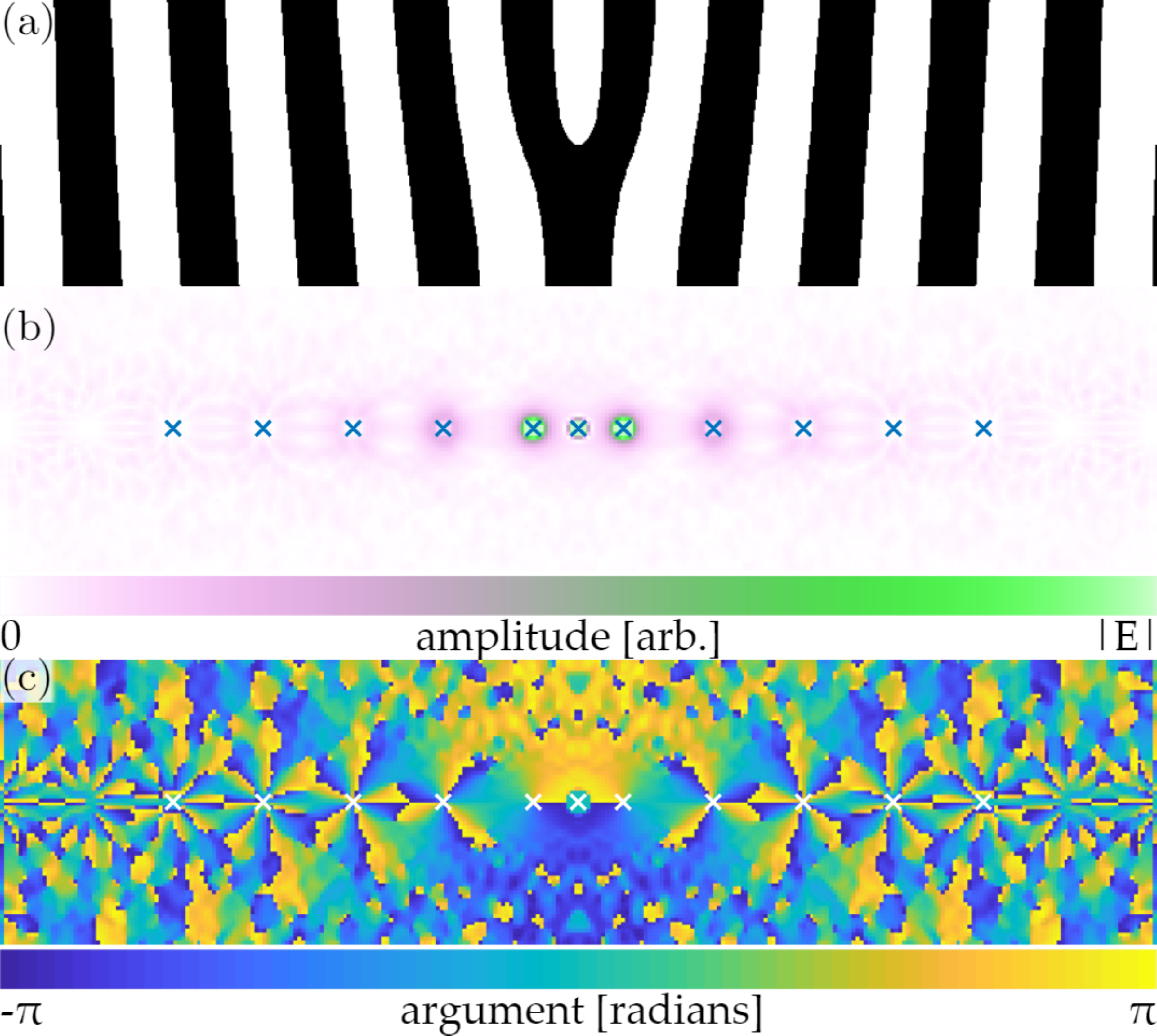}
\caption{Binary forked grating and scattering of a Gaussian beam. (a) Transmissive forked grating. (b) Amplitude of light scattered off the grating into the far-field (color bar spans 1/4 of the peak amplitude). Crosses denote the center of the diffraction orders, $0,\,\pm(2n-1)$. (c) Phase at the same plane. Around each cross are azimuthal phase ramps with local angular momentum corresponding to the spot locations, $0,\,\pm(2n-1)\hbar$.}
\label{fig:fourierexplained}
\end{figure}

The general complex modulation property can be understood with the example of a binary amplitude forked diffraction grating, pictured in \figreft{fig:fourierexplained}{a}. Between the plane of the grating and some far-field collection plane [see \figreft{fig:SLMTwopaths}{a}] we have
\begin{equation}
U_i(x,y) = \int_{-\infty}^{\infty} \mathcal{K}(x,y; \xi, \eta) H(\xi, \eta)U(\xi, \eta)\mathrm{d}\xi\mathrm{d}\eta,
\end{equation}

\noindent where $H(\xi, \eta)=\frac{1}{2}-\frac{1}{2}\mathrm{sgn} \left[\sin \left(k_{\xi} \xi+ k_{\eta} \eta + \arctan{\eta/\xi}\right)\right]$ is the transmission function of the forked diffraction grating, $\mathcal{K}(x,y; \xi, \eta)$ is the integrating kernel [e.g. Fraunhofer, \eqnreft{eqn:Fraunhofer}], $U(\xi, \eta)$ is the incident complex field amplitudes, e.g. a Gaussian smaller than the pattern, propagating normal to the surface and $U_i(x,y)$ is the complex field amplitudes after a long distance of diffraction. Light scattered off a forked pattern splits diffraction into multiple orders placed at $x=0,\,\pm(2n-1)k_{\xi}$, from the example. Each of the diffracted orders have well normalized linear momentum shifts, $\pm(2n-1)$, and orbital angular momentum, $\pm(2n-1)\hbar$ per photon, with an associated complex field structure, seen in \figo{fig:fourierexplained}(b)--(c). The result is a spectrum of light of different linear and angular momenta that can be used to manipulate matter. Instead of a long translation distance, a lens at one focal length from the diffraction grating will translate the spectrum of momentum produced into a real spatial coordinate at the focus.

With the simple example of the binary diffraction grating in mind, we now consider the example a nematic liquid crystal SLM, which can change the phase of a field with near zero losses. Many other SLM approaches that modify an input field are described in later sections. In each case, an output field can be `sculpted' out of the incident light by changing the field such that it is no-longer an eigenstate of the Fourier transform operator. This becomes clearer if we consider the difference between the solid and dashed paths in \figreft{fig:SLMTwopaths}{b}. Each point or small pixel on the SLM can be represented with the solid line path and uniform illumination on the dashed line path. The solid line path has a delocalized effect on the Fourier plane and the dashed line path a localized one. Local approximations of any field satisfying the optical transfer properties of the optical system are possible by controlling each element of the SLM at the expense of generating other non-zero amplitude components in the Fourier plane.

Imaging in the Fourier plane is also known as Fourier holography~\cite{wyrowski_diffraction_1990} in which the a Fourier transform relationship exists between the shaped wavefront and the diffraction pattern,
\begin{equation}
\mathcal{F}_{f_{X}}\left\{g(x)\right\} = G\left(f_{X}\right),
\label{eq:Fourier_Plane_Imaging}
\end{equation}
where $\mathcal{F}_{f_X}$ is the Fourier transform from coordinates $x$ to coordinates $f_X$, $g(x)$ is the wavefront, and $G(f_X)$ is the diffraction pattern.

Both LCD-SLMs and DMD-SLMs can be used to spatially modulate the waveform in the Fourier plane. DMD-SLMs are limited to binary modulation of the incident wavefront, LCD-SLMs are more varied with several amplitude and/or phase-level modulations possible. DMD-SLMs come with the advantage of being able to display truly static images.

The biggest advantage of using the Fourier imaging over direct imaging is the inverse scaling of the features when performing beam shaping:
\begin{equation}
\mathcal{F}_{f_{X}}\left\{g(ax)\right\}= \frac{1}{\left|a\right|} G\left(\frac{f_{X}}{a}\right),
\label{eq:Fourier_Transform_Scaling}
\end{equation}
\noindent shows that large features in the beam shaping plane contribute to small features in the projected plane and vice versa. This is in contrast to direct imaging, \eqnreft{equ:PSFconvolution}{}, where length scales map linearly between the beam shaping and projection plane. This means that using Fourier plane imaging improves the power efficiency of the system in the creation of small features, and allows for the contribution of many degrees of freedom to the final shape of these small features. These aspects are ideal for applications where the goal is to create one or multiple small traps, such as creating a single lattice site or addressing trapping sites in an atomic lattice.

Performing imaging in the Fourier plane with a finite-size device suffers from all of the problems familiar from fast Fourier transforms performed with a finite length vector. Large features or small features distributed over a large spatial region require high frequency resolution; on the other hand, sharp features require access to high frequency components. 

An SLM can be scaled up in size, but unless the device gained more elements there is an inverse relationship between the two requirements, whereby increasing the frequency resolution decreases the highest frequency accessible and vice versa. Another drawback is that for large sharp features most of the high frequency components tend to be zero, which means that the Fourier approach will be inefficient. A better approach in that case is the direct imaging approach.

The goal of all holography techniques is to shape a field, $U\left(\xi, \eta\right)$, at a given location in the system such that when propagated through the optical system it gives the target image plane field $U_i\left(x,y\right)$. In the case of cold atoms trapping, the intensity distribution of the optical field is the parameter of interest $\left|U_i\left(x,y\right)\right|^2$ which means we are free to choose the phase of the electrical field. There are two main challenges to this problem~\cite{goodman2005introduction}, the first being the computational problem which calculates the required optical field and the second is the physical device limitation problem which will limit the ways in which the complex-valued field can be implemented. Sec.~\ref{sec:BinaryFourierImaging} discusses the specific challenge associated with Fourier plane holography using binary DMD-SLM devices and their associated optimization algorithms, while Sec.~\ref{sec:LCSLM_BeamShaping} focuses on phase based SLMs such as LCD-SLMs.
 
\subsection{Additional Imaging and Illumination Considerations}
\label{sec:AddtImage}
\subsubsection{Wavefront Aberrations}

The above description assumes ideal, aberration-free imaging. In practice, this is not the case, even for the best-designed optical systems. However, the typical approach has been to design the optical system to minimize aberrations, and this has led to nearly-diffraction limited optical trapping/imaging~\cite{gauthier_direct_2016,sherson2010single}. 

For SLMs in the Fourier plane, controlling  the complex optical field can also compensate for subsequent aberrations of the optical system, provided the trapping potential can be probed with high accuracy. This has been accomplished through either examining the atom density~\cite{bell2016bose,zupancic_ultra_2016,tajik_designing_2019}, or by measuring the wavefront of the trapping light, typically after the optical system ~\cite{nogrette_single-atom_2014,bruce2015feedback, zupancic_ultra_2016}. A full treatment of optical aberrations is beyond the scope of this review, and the reader is referred to relevant texts~\cite{sasian2013introduction,boreman2001modulation,goodman2005introduction}.

\subsubsection{Imaging With Spatially Coherent Light}

The imaging equations discussed above explicitly assume monochromatic illumination, but not illumination that is spatially coherent. In practice, monochromatic illumination is typically achieved for optical trapping through laser illumination of the SLM or deflection device, which is typically nearly monochromatic and also consists of a single spatial mode, usually TEM$_{00}$. More recently, diffractive optical elements and wavefront shaping have provided avenues for enhancing the uniformity of the illumination for SLMs beyond TEM$_{00}$ optical beams~\cite{turunen1988holographic}. While high spatial coherence of the laser is advantageous for achieving high intensity illumination of the SLM, and the corresponding optical trapping plane, additional care must be taken in considering the laser beam propagation through the imaging system in addition to the SLM imaging path. For example, \figreft{fig:SLMTwopaths}{b} shows the propagation of both the SLM image and the collimated illuminating laser through the optical system.  Similarly, consideration of the beam propagation through the system requires careful thought in the design of the optical system to ensure high optical throughput to the image plane, which can be accomplished through the ABCD matrix equations or equivalent methods.

Coherent illumination also leads to speckle and the effects of edge diffraction in the images. In subsequent sections, we describe how SLMs in both the Fourier plane and in a direct imaging configuration can be used to mitigate these deleterious effects through feed-forward approaches.

\subsubsection{High numerical aperture optics}
Under focusing with high-NA optics, the paraxial approximation no longer applies and thus the map between the image plane and elsewhere in the optical system cannot be accurately performed using paraxial Fourier transforms. A simple example of the difference is that Gaussian light profiles (TEM) focus to symmetric Gaussian spots under small angle diffraction, but not for large angle diffraction. When the angle of convergence is no longer small the difference between the TE and TM modes becomes apparent. Different interference effects and mode shapes are observed ~\cite{richards1959electromagnetic}. When modeling projected light under tight focusing the diffraction integrals/transforms must thus be modified to account for the actual angular spectrum from the optics and the vector nature of light by calculating the appropriate angle weighted transforms for each field component.

\subsection{Time-Averaging}
\label{sec:TimeAvg}
The underlying principle of time-averaged trapping is straightforward -- by modulating the potential sufficiently rapidly atoms are effectively confined in a pseudo-static potential equivalent to the instantaneous potential time-averaged over the modulation period. A familiar example of time-averaged potentials applied to cold atom systems is the time-orbiting potential (TOP), which enabled some of the first experimental demonstrations of Bose-Einstein condensation~\cite{anderson1995observation}. This technique was developed to counter the atom loss through non-adiabatic spin-flips in the vicinity of the magnetic field zero present in a linear magnetic quadrupole trap~\cite{petrich1995stable}. The addition of a rapidly rotating magnetic bias field and quadrupole field transformed the potential from linear to harmonic, with a non-zero constant magnetic bias.

More recently, the principles of time-averaging have been applied to optical dipole traps. By rapidly modulating the position and intensity of the optical beam, a wide variety of time-averaged trapping geometries have been demonstrated, including line traps, connected reservoirs, lattices, and ring traps~\cite{henderson2009experimental,Schnelle:08,bell2016bose}.

Time-averaged optical traps have been derived from the fast beam deflection of acousto-optical devices, but as described below, these time-average techniques can be extended to more complex optical potentials derived from spatial light modulators.

\subsubsection{Equations of Motion}
\label{sec:TimeAvgEqnMotion}

Similar to the magnetic case described above, in the context of optical traps, time-averaging requires that the modulation of the beam be sufficiently rapid such that the resulting trap is pseudo-static with respect to relevant time scales for the trapped atoms. Since the creation of a time-averaged optical potential often involves rapid spatial scanning of a beam, as is the case for deflector based approaches, rather than simply modulating an optical intensity in time, the resulting modulation has both a spatial and temporal nature.

The pseudo-static requirement can be rephrased as requiring that no net kinetic energy is added over the course of the cyclic modulation of the potential. For cold atoms with \si{\micro\kelvin} or less temperatures, the velocities can be as low as a few \si{\milli\meter/\second}, which sets the minimum modulation rate for containment, as the scanned beam must `paint' the overall potential before the atoms can drift outside the range of the beam. The center of mass of the atoms will then follow the average potential.

For simplicity, we will confine our treatment of time-averaged potentials to the confinement of Bose-condensed gases, described by a well defined wave function. In the mean-field approximation, the condensate wave function $\Psi(\mathbf{r},t)$ obeys the Gross-Pitaevski Equation (GPE)~\cite{dalfovo1999theory},
\begin{equation}
    \small
    i\hbar \frac{d\Psi(\mathbf{r},t)}{dt} = \left[\frac{-\hbar^2}{2m}\nabla^2 + V(\mathbf{r},t)+g |\Psi(\mathbf{r},t)|^2 \right]\Psi(\mathbf{r},t).
    \label{eqn:GPE}
\end{equation}

\noindent The Laplacian term describes the kinetic energy, $V(\mathbf{r},t)$ the time-varying potential, and the two-body interaction energy is governed by the constant $g = 4 \pi \hbar^2 a/m$, with $a$ the s-wave scattering length, $m$ the atomic mass and $\hbar$ Planck's constant. $\Psi(\mathbf{r})$ obeys the normalization condition $\int|\Psi(\mathbf{r})|^2 d\mathbf{r} = \int n(\mathbf{r})\, d\mathbf{r} = N$, where $n(\mathbf{r})$ and $N$ are the density and total atom number respectively.

In equilibrium, condensates with chemical potential $\mu$, held in a physically time-independent potential $V(\mathbf{r})$, are described by the wave-function $\Psi(\mathbf{r},t) = \Psi(\mathbf{r})e^{-i\mu t/\hbar}$ and the stationary states $\Psi(\mathbf{r})$ are solutions of the time-independent GPE,
\begin{equation}
    \small
    \mu\Psi(\mathbf{r}) = \left[\frac{-\hbar^2}{2m}\nabla^2 + V(\mathbf{r})+g |\Psi(\mathbf{r})|^2 \right]\Psi(\mathbf{r})\,.
    \label{eqn:TI-GPE}
\end{equation}

Time-averaged potentials conceptually aim to form an effectively time-independent potential using a rapidly modulated time-dependent potential. Consider the case where the modulation frequency is sufficient that the $V(\mathbf{r},t)$ evolves faster than the kinetic and interaction terms in \eqnreft{eqn:GPE}{}. Time-averaged trapping requires that appreciable energy cannot be continuously added to the system. Thus, during the cyclic modulation of the potential, the kinetic and interaction terms in the GPE are approximately constant, and can therefore be neglected. Condensates in time-averaged traps then approximately obey $i \hbar\hspace{0.2ex} \partial \Psi(\mathbf{r},t)/ \partial t = V(\mathbf{r},t)\Psi(\mathbf{r},t)$.

The condensate wave function can also be written in an intuitive form through the Madelung representation, $\Psi(\mathbf{r},t) = \sqrt{n(\mathbf{r},t)}e^{i\phi(\mathbf{r},t)}$, where $n(\mathbf{r},t)$ is the condensate density that may vary spatially and with time, and $\phi(\mathbf{r},t)$ the corresponding spatial and time-dependent phase. This representation must preserves the normalization condition $\int \textrm{d}\mathbf{r}|\Psi(\mathbf{r},t)|^2 = N$, at each time $t$. Combining the Madelung representation and time-averaged trapping equation of motion, and then collecting real and imaginary components, leads to the phase and density equations of motion,
\begin{equation}
    \hbar \frac{ \partial \phi(\mathbf{r},t)}{\partial t} = - V(\mathbf{r},t), ~~~ \frac{\partial \hspace{0.1ex} n(\mathbf{r},t)}{\partial t} = 0.
    \label{eqn:phaseDensEqns}
\end{equation}

\subsubsection{Scanning Frequency Requirements}
\label{sec:ScanningFrequencyRequirements}

In the approximation described by \eqnreft{eqn:phaseDensEqns}, the density of the condensate is constant, and effectively confined using the instantaneous potential time-averaged over one complete scanning cycle,
\begin{equation}
\overline{V}(\vect{r}) = \frac{1}{T} \int_{0}^{T} V(\vect{r},t) \hspace{0.4ex} dt~.
\label{eq:Time_Averaged_Potential}
\end{equation}
However, the time evolution of the potential results in an imprint of the condensate phase. This effect has been utilized to imprint excitations into a condensate such as solitons~\cite{burger1999dark,denschlag2000generating,aycock2017brownian} and persistent currents~\cite{kumar2018producing,PhDGauthier2019,bell2020PhD,zheng2003classical}. 

For a time-averaged trap, the resulting continuous phase imprint can be determined by specifying the time dependent potential $V(\vect{r},t)$ and then solving \eqnreft{eqn:phaseDensEqns}{} for the phase $\phi(\vect{r},t)$. It should be noted that, although the magnitude of the phase imprint is reduced through more rapid modulation of the potential, it remains a non-vanishing contribution to the local phase of the condensate. The additional phase evolution of the condensate imposed by the time-averaged trap may be an important consideration for particular applications that are phase-sensitive, such as interferometry. 

As the modulation frequency of the trap is decreased, the kinetic and interaction terms in the GPE become non-negligible and the condition of constant density no longer holds. As a result, the density begins to respond to the instantaneous modulated potential. This excitation of the condensate can lead to atom loss as the kinetic energy per particle,
\begin{equation}
\label{eqn:KenSingle}
T(\mathbf{r},t) =  \left(\frac{-\hbar^2}{2\hspace{0.1ex}m}\right)\frac{\Psi^{*} \nabla^2 \Psi}{\Psi^{*} \Psi}~,
\end{equation}   
\noindent exceeds the time-averaged trap depth $\overline{V}(\infty)$. 

While \eqnreft{eqn:KenSingle}~holds generally, this formalism derives the well known `parametric heating' limit $f_m \gg 2 f_i$ for harmonic potentials. Here $f_m$ represents the modulation frequency and $f_i$ the time-averaged trapping frequencies along the modulated confinement directions, where it is assumed that trapping along the axial direction is provided by the light sheet described above~\cite{bell2018phase,bell2020PhD}. Close to the boundary between trapped and untrapped conditions, parametric heating of the condensate periodically ejects atoms from the potential~\cite{bell2018phase}. 

\subsubsection{Phase Imprinted Micromotion}
\label{sec:PhaseMicromotion}

While the average position of the BEC is constant in a rapidly varying potential, instantaneous forces result in micromotion within the condensate. These effects were first encountered for neutral atoms trapped in TOP~\cite{petrich1995stable}. While for many applications the time-independent zeroth-order approximation adequately describes the TOP, more exact treatments reveal the atoms undergo micromotion. Micromotion has been observed using time-of-flight center-of-mass measurements~\cite{muller2000atomic}. 
The phase imprinting resulting from the time-varying optical field, introduced in the previous section, similarly results in a continuously evolving velocity field through 
\begin{equation}
\vect{v}(\mathbf{r},t) = \frac{\hbar}{m} \hspace{0.5ex}\nabla \phi(\mathbf{r},t)~.
\label{eq:CondensateVelocityField}
\end{equation}
While closed trajectories must be formed by the velocity field for the atoms to remain trapped, the displacement trajectories need not be closed over the time-averaging cycle. This allows the atoms to migrate around the trap, for example through azimuthal precession around a ring geometry~\cite{bell2018phase,bell2020PhD}.

\subsection{Consideration of Experimental Requirements}
\label{sec:ExpRequirements}

The several technologies introduced above provide for a variety of approaches for production of configurable optical potentials for trapping cold atoms. The typical experimental implementations of AODs, SLMs in the Fourier plane, and SLMs in the direct imaging configuration are shown in \figo{fig:SLMTypes}. Depending on the particular application of interest, various factors may determine which approach is best suited. Here, we outline the key technical considerations and how those are addressed by each of the different technologies. This section aims to guide the reader in making a preliminary assessment as to which of the available technologies most likely satisfies their specific experimental requirements. 

\subsubsection{Condensate density in optical potentials}
\label{SpatialTrapUniform}

We consider here the distribution of ultra-cold trapped atoms, such as a BEC, trapped within an arbitrary optical trapping potential $V(\mathbf{r})$. For sufficiently high atom number, the interaction energy between atoms is much larger than the kinetic energy term in \eqnreft{eqn:GPE}{}. Under these conditions, known as the Thomas-Fermi approximation, the kinetic term is set to zero, resulting in the following time-independent form of the equation:
\begin{equation}
    \small
    \left[V(\mathbf{r})+g |\Psi(\mathbf{r})|^2 \right]\Psi(\mathbf{r}) = \mu \Psi(\mathbf{r}).
    \label{eqn:TF-Approx}
\end{equation}
\noindent This leads to a simple form for the condensate density: $n(\mathbf{r}) = |\Psi(\mathbf{r})|^2 = [\mu - V(\mathbf{r})]/g$. The density thus simply follows the spatial profile of the trapping potential, up to the boundary at $V(\mathbf{r}) = \mu$ since $n(\mathbf{r}) \geq 0 $. 

Here we assume the dipole trapping light is red-detuned (attractive), with a fixed light sheet confining atoms in the vertical direction with trap frequency $\omega_z$, and transverse harmonic trapping with trap frequency $\omega_r$ provided by the projected light pattern. Modeling the local trap as a waveguide, a condensate in the three-dimensional regime will have a chemical potential scaling as $\mu\propto\omega_r\omega_z$. Thus, within the relevant energy scale, the chemical potential $\mu$ follows the local average optical intensity as $\mu\propto I_0^{1/4}$. For a typical condensate, the chemical potential is on the order of tens of nK, whereas the optical potential is typically much larger. Since the density of the condensate closely follows the optical potential, small perturbations in the optical field can result in significant density fluctuations across the condensate. These aspects highlight that the precision of the optical projection is a key consideration when implementing configurable optical potentials. This requirement is met with various trade-offs and advantages for the different technologies, which are described in subsequent sections.

\subsubsection{Spatial and Temporal Trapping Resolution}
The spatial resolution of the projection system can be determined through the scalar diffraction treatment in Sec.~\ref{sec:Diffraction_ints}, this calculation, along with the effective aperture of the system, determined by the input optical beam, determines the minimum feature size, or resolution element, of the projected potential. For beam deflectors, this will typically  be the focused spot size achieved by sending the deflected beam through the final focusing optical element. For projection of an SLM, the achieved resolution element is typically determined by the effective numerical aperture of the overall optical system, determined relative to the exit pupil (see \figo{fig:SLMTwopaths}).

The size of the resolution element is an important consideration in the design of a configurable trapping system. First, it clearly determines the minimum feature size that can be achieved with the system. Secondly, for techniques that result in the generation of discrete beam locations, such as rapid scanning of a beam deflector through a sequence of points, the size of the resolution element determines the separation between two projected points that still results in a smooth potential. This spacing can be conservatively estimated through the Sparrow criterion~\cite{sparrow1916spectroscopic,den1997resolution,trypogeorgos2013precise}, which minimizes the second derivative of the intensity variation between two adjacent resolution elements, resulting in an overall smooth potential.

When scanning a deflector controlled optical beam, or using pulse-width modulation of a DMD-SLM to produce time-averaged potentials, temporal resolution is likewise a key design parameter. As discussed in the previous section, temporal modulation must be fast, on the scale of trap frequencies, and must not lead to kinetic energies sufficient to escape the trap. Since the devices have a fixed bandwidth, resulting in a limited temporal discretization of a given trapping sequence, a balance between the spatial and temporal resolution must be found depending on the requirements of the desired potential. High-resolution time-averaged patterns require a small imaging system PSF, leading to a larger number of points needed to paint a potential in the deflector case, or a larger number of dithered DMD-SLM patterns to achieve the same smoothness in the potential (see Sec.~\ref{sec:DMDs} for further details). For a given optical intensity control, greater spatial resolution is thus possible at the expense of temporal resolution, and vice-a-versa. 

When implementing phase-modulating SLMs in the Fourier plane, the smoothness of the trapping potential can be effected through the production of speckle and optical vortices, resulting in 100\% local intensity modulation of the optical potential, that can be the result of iterative Fourier transform algorithms (ITFAs) that are often used to calculate phase-based holograms~\cite{aagedal1996theory,senthilkumaran2005vortex}. Improvements to these algorithms for application to cold atoms that mitigate these effects have been developed~\cite{gaunt2012robust,pasienski2008high}, and are discussed in further detail in Sec.~\ref{sec:SLMs_details}.

\subsubsection{Optical Trapping Depth}
Typical optical trapping configurations using the three technologies of AODs, direct-imaged SLMs, and SLMs in the Fourier plane are shown in \figo{fig:SLMTypes}. These configurations differ widely in their light efficiency, which, along with the detuning $\Delta_{ge}$ of the laser from the resonant transition and the numerical aperture of the optical system determine the achievable trap depth.

Beam deflection devices such as AODs or EODs provide the highest optical trap depths, as the crystalline materials exhibit high transmission ratios and very high optical damage thresholds, and are thus suitable for trapping thermal atoms prior to evaporation to BEC~\cite{barrett2001all,roy2016rapid}. However, for very high power applications, the TeO$_2$ material commonly used to manufacture AODs can exhibit thermal lensing~\cite{sparks1971optical,bendow1973optics}. This leads to a power-dependent shift of the focal position of the optical trap~\cite{hansen2013production}, although the effects may be mitigated through careful optical design~\cite{simonelli2019realization}, or choice of modulator materials with lower index of refraction dependence on temperature, such as quartz or flint glass. 

The diffraction efficiency of an AOD at its center frequency, and exactly meeting the Bragg condition can surpass 95\%. However, these conditions will not be satisfied simultaneously for all the deflection angles, and so overall efficiency will be slightly reduced. Phase array oscillator designs, which modify the angle of the acoustic wave as a function of drive frequency can provide a more uniform diffraction response across the tuning range. However, as the number of points making up the time-averaged pattern increases, the available trapping power will be divided across the number of points. For time-averaged trapping of BECs, power limits of AODs have not typically been a significant limiting factor. Instead, time-averaged traps are limited by the bandwidth of the scanned AOD, reducing the number of possible points that can be defined while still maintaining a time-averaged trap. 

The optical trapping depth achievable with SLMs similarly depends on the numerical aperture of the system, but will also be limited by the damage threshold of the devices, which is much lower than beam deflection devices. DMD-SLMs have a typical damage threshold of 25~W/cm$^2$~\cite{texas_instruments_laser_power_2012}, but care must be taken for thermal management of the device so that one only illuminates the reflective mirrors of the DMD-SLM, and thermal contact is made to the back of the DMD-SLM device. Furthermore, the fill factor of the DMD-SLM device (typically $>90\%$) and transmission losses, result in a reflectance from the device of $\sim 80\%$. However, as described in detail in Sec.~\ref{sec:DMDs}, the mirror array realizes a two-dimensional diffraction grating, which results in a wavelength dependent efficiency, which along with variance in the mirror angle results in a maximum efficiency of $\sim30\%$ -- $55\%$ in the primary diffraction order. In both the direct and Fourier plane implementations, the overall light throughput is reduced as the `off' mirrors dump the illuminating light. This results in reduced optical power available for trapping, given by the ratio of `on' mirrors to `off' mirrors, so the overall efficiency of a DMD-SLM in the direct imaging mode will be $\leq55\%$. With a DMD-SLM in the Fourier plane generating holograms based on linear diffraction grating efficiencies of $1\%$ -- $2\%$ have been reported~\cite{zupancic_ultra_2016}. For direct imaging, a further reduction in trapping intensity may result from overfilling the SLM aperture, as shown in \figreft{fig:SLMTypes}{b}, to achieve more uniform illumination from the input Gaussian beam.

Transmission and reflection LCD-SLMs can demonstrate much more promising performance, accommodating large illumination intensities $>400$ W/cm$^{2}$, and realizing first-order diffraction efficiencies exceeding 90\%~\cite{meadowlarkSLM}. 

\subsubsection{Modulator Bandwidth}

Optical deflectors such as AODs and EODs are the largest bandwidth devices, with center frequencies typically in the hundreds of MHz to GHz range, and modulation bandwidths of tens to hundreds of MHz [n.b. that EODs can either be resonant or broadband; for a broadband EOD there is no well-defined center frequency]. For AODs, the limitation of these devices is determined by the acoustic mode's propagation across the input beam diameter, and the number of points building up the scanning pattern. Typical scan frequencies, for patterns consisting of $\sim25$ points, are on the order of 10~kHz - 50~kHz, depending on the particular material and sound mode used in the device. For further details see Sec.~\ref{sec:AOD}.

For DMD-SLMs mirror switching speeds are on the order of 20~$\mu$s, but may be controller limited; full frame refresh rates can thus vary between 9 - 50 kHz. These rapid switching rates enable time-averaging in these devices. Furthermore, the DMD-SLMs can realize quasi-static patterns, although flicker noise is present as the mirrors are periodically cycled to avoid permanent locking of the mirrors in the on/off position. This flicker occurs at the refresh rate, and results in a mirror settling period of 20~$\mu$s~\cite{klaus2017note:}. This flicker of the image can lead to undue heating of the cold atoms, the effect that is more pronounced in lighter atoms. This flicker can be mitigated by modulating the refresh clock to prevent unwanted refresh~\cite{klaus2017note:}, although care must be taken to avoid permanent latching of the mirrors if too much time passes between refreshes. For further details see Sec.~\ref{sec:DMDs}.

Liquid crystal based SLMs also suffer from flicker, where the modulation of the light varies in time due to the change in orientation of the crystals under the influence of the time-varying applied field. A number of types of liquid crystal will thus slowly drift in their local orientation due to relaxation in an applied field. To maintain a consistent average orientation the applied field is dithered resulting in \emph{phase flicker}, typically on the order of 120~Hz~
\cite{lizana2008time}. The engineered refresh in the applied field will occur at about twice the frame rate due to criteria imposed by Nyquist sampling. However, this is not always going to be the case as some manufacturers try to correct for this phenomenon. Furthermore the flicker effects can be minimized through techniques such as reducing the temperature of the SLM~\cite{garcia2012flicker}. Further details are discussed in Sec.~\ref{sec:SLMs_details}.

\subsubsection{Computational Complexity}

Computational complexity is very low when SLMs are used in imaging mode and when frequency sweeping AODs. When used in a Fourier imaging mode, SLMs can require significant computation to generate the desired field in the image plane, particularly if an iterative algorithm is used to calculate the required hologram~\cite{zupancic_ultra_2016,gaunt2012robust,pasienski2008high}. However, devices in the Fourier plane come with the advantage that the phase of the field may also be controlled. This property arises because the optical transfer function transforms the field of complex numbers just like the Fourier transform does, and can be harnessed to correct for wavefront aberrations, for example.

\subsubsection{Hardware Selection}

Among the various technologies, the optimal choice of a spatial light modulator will be specific to the particular application. Some of the primary considerations when implementing a device, with regards to the above more detailed discussion, are summarized in the following paragraphs. Additional information is included in Sec.~\ref{sec:conclusion}, where a general guide to device selection is provided.\\

\noindent\textit{Desired geometry} -- A variety of geometries are required for different applications. Simple single Gaussian traps, and 1D and 2D arrays of such Gaussian traps, are suited to multifrequency-driven AODs or EODs, with or without time-averaging. Fourier plane SLMs are also suitable for producing arrays of Gaussian traps~\cite{nogrette_single-atom_2014}. More complex potentials are achieved with SLMs, either in the Fourier plane or direct imaging. Direct imaging is most suitable for producing hard-walled trapping potentials, such as uniform discs~\cite{gauthier_direct_2016,ville_loading_2017} or boxes~\cite{ville_sound_2018,luick_ideal_2020,kwon2020strongly}, as in the absence of halftoning methods (c.f Sec.~\ref{sec:DMDs}), the binary nature of the potential is directly transferred to the resulting potential.\\

\noindent\textit{Trap depth} -- The required trap depth is application dependent, and varies with the trapping wavelength and device efficiency. While AODs, EODs, and LCD-SLMs have efficiencies exceeding 90\%, and direct imaged DMD-SLMs have efficiencies of $\sim55\%$, DMD-SLMs in the Fourier plane typically will have efficiencies on the order of 4\%. This inefficiency can be offset through the use of more closely detuned light.\\

\noindent\textit{Time-averaged vs. static trapping} -- Time averaging requires sufficiently fast modulation of the trapping potential. This is achievable with AODs, EODs, and DMD-SLMs, and will also depend on the mass of the trapped particles~\cite{klaus2017note:}. \\

\noindent\textit{Trapping wavelength} -- While EODs and AODs can operate with a broad range of optical wavelengths, depending on device design and optical coatings, SLMs may be more wavelength specific. In particular, the diffraction efficiency of DMD-SLMs depends on the selected wavelength, for a given mirror pitch, see Sec.~\ref{DMD_Diffraction_Theory}. \\

\noindent\textit{Aberration correction} -- Correction of wavefront aberrations requires an SLM in the Fourier plane of the optical system, in the absence of static beam correction methods or other adaptive optics schemes.\\

\noindent\textit{Temporal and spatial resolution} -- Spatial resolution can be estimated by the Gaussian optics analysis of the optical system that focuses the light on the atoms. This can be determined by the condition of the diffraction limit for a plane wave of a specific wavelength input to the optical system, but care must be taken to consider optical aberrations since these will reduce the resolution. In terms of producing highly complex optical patterns, directly imaged SLMs are typically superior to other techniques. Temporal resolution is determined by the access time of the device, and modulation frequencies can range from $\sim 1$~kHz for LCD-SLMs, to $\sim 20$~kHz for DMD-SLMs, to $\sim 10$~kHz to $\sim 10$~GHz for AODs and EODs, respectively. For time averaged traps using AODs and EODs, where a beam is rapidly scanned, the requirement of maintaining the time-averaged conditions means that the access time is divided by the number of points in the painted pattern~\cite{bell2016bose,bell2018phase}.


\section{Beam Deflection Devices}
\label{sec:AOD}

Interactions between photons and phonons in optically transparent dielectric materials facilitate the control of an optical trapping beam's power, frequency, location and polarization. Acousto-optic modulators (AOMs) and acousto-optic deflectors (AODs) describe physically analogous devices which leverage these interactions, and are named according to their intended applications. Modulators primarily control the optical intensity and frequency \cite{schwenger2012high,aom2005,chang2008design}, while deflectors emphasize the deflection angle for optical beam steering and trapping applications, as shown in \figreft{fig:SLMTypes}{a}.

AODs induce Bragg diffraction to effectively deflect an incident optical beam. Using an acoustic traveling-wave, the input beam, typically a Gaussian TEM$_{00}$, is diffracted; the acoustic frequency determines the beam deflection angle and resulting trapping location after transmission through a focusing lens. Optical traps have been formed by driving the AODs with multiple frequencies simultaneously, and using rapid frequency modulation to form time-averaged optical fields~\cite{henderson2009experimental,Schnelle:08,bell2016bose}.

Electro-optical deflectors (EODs) are an alternative to AODs that use electro-optic materials and electric fields to directly steer an optical beam. The applied field perturbs the crystalline structure, resulting in a change to the refractive index \cite{Valentine08}. 

Optical deflectors typically modulate the beam along one direction. Optical traps extending across the two-dimensional plane at the lens focus can be realized with two deflection elements in close sequential series, permitting deflection of the incident beam along two orthogonal directions. Commercial bi-directional scanners normally co-house these deflectors within one device. Some modes of operation additionally enable bi-directional steering using a single integrated acousto-optic or electro-optic cell; examples include the longitudinal mode AODs and index gradient EODs discussed in later sections.

\subsection{Deflection Theory}

Acousto-optic interactions are just one example of the photo-elastic effect, which changes the optical properties of a material in response to mechanical strain. Acoustic running waves within a transparent material may thereby induce a refractive index grating from which the incident optical trapping beam will diffract. AODs are constructed to precisely control this diffraction grating by sandwiching the optical material between one or more piezoelectrical transducers and an acoustic damping block; these three components are collectively called a Bragg cell.

Electro-optic interactions may additionally change the optical properties of a material using an applied electric field. The Pockel and Kerr effects constitute the first linear and second quadratic orders of the multi-pole expansion for electric field vector $\vect{E}$, respectively \cite{ROMER2014}. While most materials exhibit a weak  Kerr effect, the Pockel effect only occurs in materials which lack inversion symmetry. Simple EODs use these interactions to homogeneously modulate the refractive index of materials held between high voltage plates. The applied voltage determines the optical angle of refraction through planar interfaces.

Gradient index EODs, constructed using Potassium Tantalate Niobate, produce larger deflection angles with smaller applied voltages, compared to surface refraction based devices \cite{High_Res_2009}. These devices exploit the Kerr and space-charge effects; the latter effect creates a spatially varying charge distribution by injecting electrons through the cathode \cite{doi:10.1063/1.2357335,doi:10.1063/1.2949394}. The resulting electric field gradient forms a spatial refractive index gradient which acts to continuously deflect the beam.

\begin{figure*}[t]
\includegraphics[width=\textwidth]{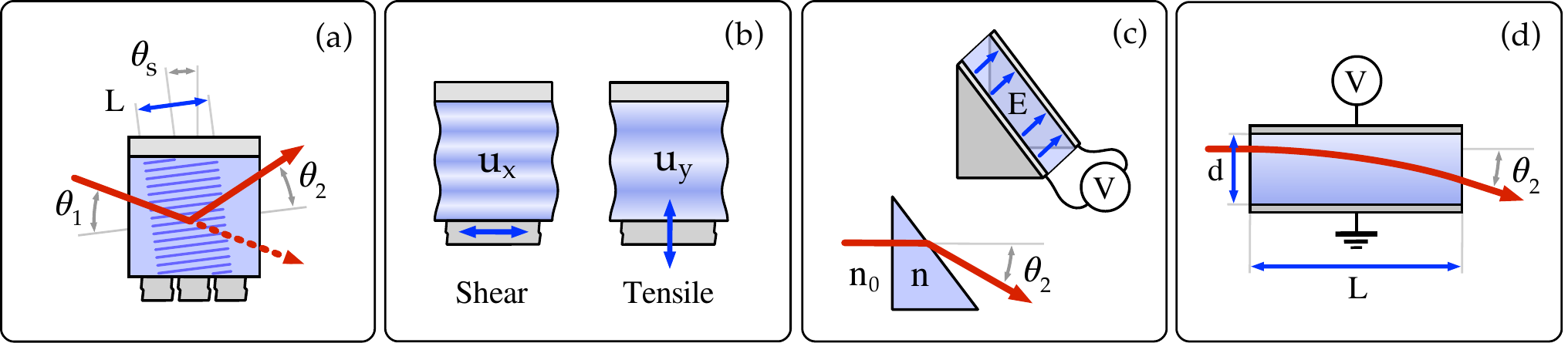}
\caption{Schematic representation for refractive index modulation schemes which enable several commercially available deflector technologies, and which are suitable for developing scanning beam based atom trapping applications.
(a)~AODs consist an optically transparent dielectric crystalline medium sandwiched between one or more piezoelectric oscillators and an acoustic absorption material. These elements generate traveling acoustic waves from which the incident optical beam diffracts. Phased array devices steer acoustic wave angle $\theta_s$ using multiple oscillators to preserve the Bragg condition across a broader deflection range $\theta_2$. (b)~Commercial AODs commonly induce transverse or longitudinal mode oscillations associated with shear and tensile strain waves respectively; these differently affect the optical polarization of the diffracted beam.
(c)~Interface refraction EODs homogeneously modulate the bulk refractive index of a transparent material between two capacitor plates using electrical potential $V$.
(d)~Gradient refraction EODs induce an inhomogeneous charge distribution within the Potassium Tantalate Niobate (KTN) crystal, which induces an electric field gradient $\partial E/\partial y$ and corresponding refractive index gradient \cite{ROMER2014}.}
\label{fig:DeflectorMode}
\end{figure*}

\subsubsection{Anisotropic Dielectric Media}

The starting point for understanding the interactions between light and acousto-optic or electro-optic materials is classical electromagnetism. Optical beams are treated using their principal electric $\vect{E}(\vect{r},t)$ and magnetic $\vect{H}(\vect{r},t)$ vector field components; these physically mediate the forces between charged objects. The supplementary electric $\vect{D}(\vect{r},t)$ and magnetic $\vect{B}(\vect{r},t)$ vector flux densities additionally quantify how the primary fields evolve through electric charge and current free spatial regions. Two constitutive relations between the electric $\vect{E}\leftrightarrow\vect{D}$ and magnetic $\vect{H}\leftrightarrow\vect{B}$ fields, and Maxwell's equations, govern the interactions between the optical trapping fields and the materials used to engineer beam deflectors \cite{AomTheory}. The relevant Maxwell's equations are
\begin{align}
\label{eqn:AnisoOpticEffect}
\nabla \times \vect{H}&=\frac{\partial\vect{D}}{\partial t}~, \hspace{4ex}\nabla \cdot \vect{D} =0~,\nonumber\\[2ex]
-\nabla \times \vect{E}&=\frac{\partial\vect{B}}{\partial t}~,\hspace{4.1ex}\nabla \cdot \vect{B} =0~.
\end{align}

Optical beam deflectors are commonly manufactured using linear anisotropic dielectric materials.  Interactions between the trapping beam and these materials are governed by the dielectric constitutive relationship between $\vect{E}\leftrightarrow\vect{D}$. This relationship may in general be described using the impermeability tensor $\vect{\zeta}$ and scalar vacuum permittivity $\epsilon_0$, using 
\begin{align}
\vect{\zeta}\vect{D} = \epsilon_0 \vect{E}~.
\end{align}

Common dielectric materials which are non-magnetic and are not optically activated are generally described using symmetric second order impermeability tensors, where $\zeta_{ij}=\zeta_{ji}$. The principal coordinates $x_i$ and the associated elements $\zeta_i=\zeta_{ii}$ are found using diagonalisation, and the refractive indices $n_{i} = \sqrt{1/\zeta_{i}}$ satisfy,
\begin{align}
\sum_{ij} \zeta_{ij} x_i x_j =  \sum_i \frac{x_i^2}{n_i^2} =1 ~.
\end{align}

Beam deflectors physically function by modulating the impermeability tensor to establish spatially and temporally varying refractive indices within an active material. The operation of these devices is therefore understood by considering in further detail how straining a material or applying an external electric field modulates the material impermeability.

\subsubsection{Impermeability Modulation}

AODs induce strain waves within suitable crystalline materials, from which the incident optical trapping beam is diffracted into one or more spatial orders. Piezoelectric transducers convert electrical modulations into molecular displacements $\vect{u}$, which then define the strain tensor $\vect{S}$. The relationship between the mechanical strain and the modulated optical properties of a specific material are characterized using the strain-optic tensor $\vect{p}$, and
\begin{align}
\label{eqn:AcoustoOpticEffect}
\zeta_{ij}(\vect{S}) &= \zeta_{ij}(0) + \sum_{kl} p_{ijkl} \hspace{0.2ex} S_{kl}~, \nonumber\\[1ex]
S_{kl} &= \frac{1}{2} \left( \frac{\partial u_k}{\partial x_l} + \frac{\partial u_l}{\partial x_k} \right)~.
\end{align}

EODs instead modulate the optical properties of suitable materials using an applied electric field $\vect{E}$. Those materials without and with inversion symmetry generally exhibit the Pockel and Kerr effects most strongly, respectively. The electric field and optical qualities of a specific material are related using the linear $\vect{q}$ and quadratic $\vect{r}$ electro-optic tensors, and
\begin{align}
\label{eqn:ElectroOpticEffect}
\zeta_{ij}(\vect{E}) &= \zeta_{ij}(0) + \sum_{k} q_{ijk} \hspace{0.2ex} E_{k} + \sum_{kl} r_{ijkl} \hspace{0.2ex} E_{k} E_{l}~, \nonumber \\[1ex]
q_{ijk} &= \frac{\partial \zeta_{ij}}{\partial E_k}~, \hspace{5ex} r_{ijkl} = \frac{1}{2} \frac{\partial^2\zeta_{ij}}{\partial E_k \partial E_l}~.
\end{align}

\subsubsection{Acousto-Optic Deflection}

The operation of AODs can be better understood if we neglect the crystal anisotropy, polarisation effects and assume that the photoelastic interaction region is sufficiently narrow that diffraction occurs from a stationary grating \cite{AomTheory}. With these assumptions, the acoustic wave fronts effectively constitute reflectors with length L, as shown in \figreft{fig:DeflectorMode}{a}. Let $\{\lambda_1,\theta_1,k_1\}$ represent the incident optical wavelength, angle and wave-vector and $\{\lambda_s,I_s,k_s\}$ the acoustic wavelength, intensity and wave-vector; $\mathcal{M}$ the crystal photoelastic parameter; and $n$ the bulk refractive index. The  first-order diffraction efficiency then becomes
\begin{align}
\label{eqn:AODEfficient}
\eta_{_\mathrm{AOD}}\hspace{-0.5ex} &\propto \sin^2\hspace{-0.7ex}\left( \hspace{-0.6ex}  
\frac{2\pi n L\lambda_s}{\lambda_1^2}\sqrt{\hspace{-0.4ex}\frac{\mathcal{M}\hspace{-0.1ex}I_s}{2}}
\hspace{0.8ex} \text{sinc}\hspace{-0.7ex}\left[\hspace{-0.4ex}\frac{\left(\hspace{-0.2ex}2 k_1 \sin\hspace{-0.2ex} \theta_1\hspace{-0.5ex}-\hspace{-0.4ex}k_s\hspace{-0.2ex}\right)\hspace{-0.5ex}L}{2\pi}\hspace{-0.3ex}\right] \hspace{-0.6ex}\right)~.
\end{align}

\noindent The maximum optical diffraction efficiency occurs for $2 k_1 \sin \theta_1\hspace{-0.5ex}=\hspace{-0.3ex}k_s$. Consistent with three wave-mixing, this condition conceptually simplifies the operation of AODs. The diffracted optical frequency $\omega_2$ and the deflection angle $\theta_2$ result from energy and momentum conservation. The diffracted power $P_2$, can be controlled, up to the input power $P_1$, by modulating the acoustic power $P_s$, below saturation $P_0$. Given sound velocity $\nu_s$, AODs are controlled using
\begin{align}
P_2 &\approx  P_1 \sin^2\hspace{-0.7ex}\left( \hspace{-0.5ex}\frac{\pi}{2} \sqrt{\frac{P_s}{P_0}} \hspace{0.2ex} \right)~,\nonumber\\[0.5ex]
\theta_2 &= \arcsin\hspace{-0.5ex}\left(\frac{\lambda_1 \omega_s}{4\pi\nu_s}\right)~,\nonumber\\[0.5ex]
\omega_2 &= \omega_1 \pm \omega_s ~.
\label{eqn:AODBragg}
\end{align}

Since the relative alignment of the AOD and incident optical beam are mechanically fixed, standard AODs with a fixed acoustic grating angle $\theta_s$ ensure that $\theta_1 \neq \theta_2$ for deflection angles about some center. The diffraction efficiency and deflected optical power then appreciably vary across broader scanning patterns, modulating the depth of planar optical potentials. Smoother potentials are achieved using phased array piezoelectric oscillators, shown in \figreft{fig:DeflectorMode}{a} \cite{Wang17,Pieper:83}. Phased array AODs control the relative phases between the adjacent oscillators to tune $\theta_s$, and maintain the nearer to optimal Bragg condition $\theta_1\approx\theta_2$ across a broader bandwidth $\omega_s$. 

Beyond the scalar grating model, the anisotropy of the modulation crystal and the strain wave polarization may additionally affect the trapping beam polarization. While optical potentials formed using far-detuned fields are commonly unaffected by the trapping beam polarization, the experimental integration of several optical fields often requires a specific polarization; combining trapping and imaging light using a polarizing beam cube is one example. The longitudinal and transverse sound modes, associated with the tensile and shear oscillations, respectively, are illustrated in \figreft{fig:DeflectorMode}{b}. While longitudinal mode AODs may preserve the optical polarization of the diffracted beam, transverse mode AODs behave like half-wave plates \cite{ALIPPI1973397}.

\subsubsection{Electro-Optic Deflection}

EODs deflect an incident optical trapping beam using refraction, rather than diffraction \cite{ROMER2014}. Refraction either occurs along the planar interfaces between two materials with different bulk refractive indices, or continuously through materials within which a spatially varying refractive index has been established.

Interface refraction based EODs are manufactured by sandwiching an electro-optic material between high-voltage capacitive plates; shown in \figreft{fig:DeflectorMode}{c}. Modulating the electrical potential $V$ creates the spatially homogeneous electric field $E$, which alters the bulk refractive index $n(E)$. The beam deflection angle $\theta_2$ is then simply governed by Snell's law and the precise geometry of the crystal or crystals manufactured.

Gradient refraction based EODs similarly consist an electro-optic material between high voltage capacitive plates. However, the selection of specific materials which additionally exhibit the space-charge effect continuously deflects the incident optical beam through the crystal; see \figreft{fig:DeflectorMode}{d}. Using Potassium Tantalate Niobate, charge injected via the cathode creates a spatially varying electric field which, combined with the Kerr effect, creates an approximately linear refractive index gradient. Provided the necessary stringent temperature requirements have been satisfied, the beam deflection angle is
\begin{align}
\theta_2 &= \frac{-\alpha \hspace{0.2ex} n^3 \epsilon_0^2\hspace{0.2ex} \epsilon_r^2\hspace{0.2ex} V^2 L}{d^3}~,
\label{eqn:EODGrad}
\end{align}
where $n=2.2$ and $\epsilon_r=3\times10^4$ are the refractive index and relative permittivity at $V=0$, respectively; $L$ is the crystal length; $d$ the crystal thickness; $\epsilon_0$ the free-space permittivity; and the constant $\alpha=0.153~\text{rad}\hspace{-0.2ex}\cdot\hspace{-0.2ex}\text{m}^4\hspace{-0.2ex}/\hspace{0.1ex}\text{C}^2$.

EODs offer many advantages over AODs. Near unity deflection efficiencies are achievable, limited by reflection and absorption losses. The deflection angle does not depend on the wavelength of the optical beam as the index dispersion is relatively constant, unlike AODs. EODs also overcome the aberrations associated with acoustic non-linearity \cite{Woody18}. The EOD random access response is limited by the capacitive response time of the electro-switching circuit, rather than the AODs speed of sound. However, the deflection angles normally achieved using EODs are smaller.

Despite these advantages, EOD's have had much more limited usage than AODs in cold atom trapping. For commercial optical systems, they have been used for precision beam steering, and laser data storage and recording, with improved stability and frequency response over mirror beam steering. For scientific applications, their  primary use has been  steerable optical tweezers in biological systems~\cite{LaserTweezerBiology,padgett2011holographic,Curtis2002}.

\subsection{Multiple Beam Optical Traps}

Optical potentials that extend across larger areas or include regions of smoothly varying depth require multiple overlapping beams. In practice, aligning and combining large numbers of independent trapping beams is increasingly difficult. Beam deflectors can simplify the creation and alignment of several adjacent optical beams. Rectangular arrays of optical beams can be generated using a multiplexed driving scheme, which drives the deflector at multiple frequencies simultaneously. Time-averaged scanning methods are used to enable the formation of more flexible trapping geometries.

Time-averaged AOD optical potentials have demonstrated the formation of planar matterwave circuits and waveguides \cite{henderson2009experimental,bell2016bose,1367-2630-17-9-092002,PhysRevLett.111.205301} for studies involving persistent currents and phase slips \cite{PhysRevLett.110.025302}. AODs have facilitated the assembly of defect free atomic arrays \cite{Barredo2016,PhysRevLett.122.203601,PhysRevA.102.063107}. AOD controlled optical tweezers with two \cite{Roberts:142,PhysRevA.90.051401} and three \cite{ThesisChrisholm2018,barredo_synthetic_2018} spatial dimensions have also enabled the formation of neutral atom based quantum logic gates \cite{PhysRevLett.123.170503} and vortex generation \cite{PhysRevLett.80.3903,PhysRevLett.84.806,CreateVortex}.

Here we outline some of the additional considerations relevant when forming complex optical potentials using beam deflectors. In most cases, we wish to create uniform depth optical potentials across an extended area.

\begin{figure*}[t]
\includegraphics[width=\textwidth]{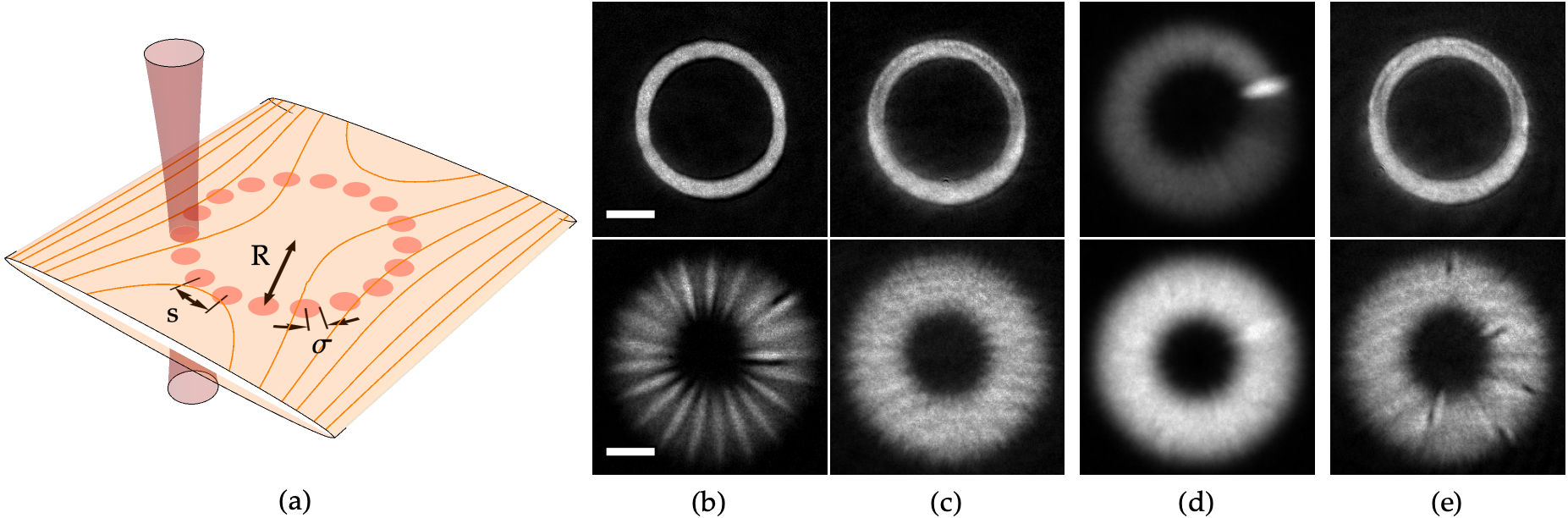}
\caption{Optical potentials created using an acousto-optic deflector. Summary visual reference which illustrates the experimental technique and signatures characteristic of the predominate issues affecting the implementation of beam deflector based optical potentials.
(a)~Schematic representation of the typical optical trapping beam configuration, here realizing a ring shaped (toroidal) geometry. Approximately planar confinement against gravity is provided by the elliptically focused Gaussian sheet beam, shown in orange. The orthogonal scanning beam, shown in red, provides transverse confinement.
(b)~Absorption images showing signatures characteristic of an insufficient number of discrete scanning points around the ring's circumference. While the trapped density distribution captured following \SI{1}{\milli\second} time-of-flight (TOF) shows no observable signatures of azimuthal nonuniformity, residual corrugations appear as regular density fringes following \SI{20}{\milli\second} TOF; the beam waist $\sigma=\SI{26.5}{\micro\meter}$ and spacing $s=0.8\hspace{0.1ex}\sigma$. 
(c)~The condensate density distributions observed for feedforward corrected ring potentials with spacing $s=0.6\hspace{0.1ex}\sigma$.
(d)~Despite achieving uniform density, the scanning beam introduces phase variation revealed as a density signature using a technique known as rotationally accumulated mean-density images (RAMDI)~\cite{bell2018phase}. The two frames correspond to RAMDI TOF images of the condensate shown in panel (c) for two scanning frequencies: $f_s=\SI{0.5}{\kilo\hertz}$ (top) and \SI{6.25}{\kilo\hertz} (bottom). 
(e)~Care must be taken to avoid unwanted excitations if the trap is dynamically evolved -- here uneven loading of the condensate into the ring trap results in quantized vortices being excited. 
The ring potential radii $R\approx\SI{80}{\micro\meter}$ and the scale bars have \SI{50}{\micro\meter} length. Images were captured using \SI{1}{\milli\second} TOF (top) and \SI{20}{\milli\second} TOF (bottom) in frames (b-c,e). Both images in frame (d) were captured using \SI{20}{\milli\second} TOF.}
\label{fig:AOD_Atoms}
\end{figure*}

\subsubsection{Density Based Feedforward}
\label{subsub:AODFeedforward}

Optical potentials using steered beams generated from an acousto-optical deflector, either in multiplexed or time-averaged operation, generally require some level of beam intensity correction, particularly when a uniform and smooth potential is desired. This is due to a number of factors. Firstly, the efficiency of the device is not constant with driving frequency, resulting in a variable beam intensity for different points across the deflection range.
Secondly, due to the comparatively weak confinement along the propagation axis, additional trapping fields are used to provide confinement in that axis, and trap atoms within the steered beam's focus. Typically this would be through the use of a cylindrically focused optical sheet potential, which has a broad transverse Gaussian waist oriented perpendicular to the scanned potential and provides planar confinement against gravity \cite{davidson1995long,Schnelle:08,henderson2009experimental}; see \figreft{fig:AOD_Atoms}{a}.

The resulting trapping potential reflects both the beam intensity variation with deflected angle, as well as the residual curvature of the Gaussian sheet, defined by the transverse waist and Rayleigh length. As noted above in Sec.~\ref{SpatialTrapUniform}, for condensates in the Thomas-Fermi regime, the atom density closely follows the resulting trapping potential. Thus, the atom distribution itself is a sensitive probe of local optical potential, and can be observed through imaging the resultant cold atom cloud. Through iterative feedforward correction of the optical potential, and repeated observation of the atom density, the trap can be corrected until the desired density distribution is achieved \cite{bell2016bose}. 

The effectiveness of density based feedforward using AODs has been ultimately limited by the measurement resolution of the imaging system, rather than the control resolution of the trapping system \cite{bell2020PhD}. The discrete pixel depth of the camera used for absorption imaging constrained the experiments measurement sensitivity to changes in atomic density. By comparison, the optical intensity control was limited by the amplitude resolution of the function generator. While the smallest incremental changes in the optical intensity could be measured using optical means, the resulting changes in the trapping potential were not detectable using the atoms. Other imaging techniques are therefore likely to be important considerations in the development of future systems. In particular, non-destructive imaging methods foreseeably enable the real-time optimization of dynamical processes \cite{PhysRevA.80.013614,PhysRevA.82.043632,Wigley:16}.

\subsubsection{Spatial Resolution Criteria}

Optical potentials with smooth spatial variation may be formed using the combination of suitably overlapping Gaussian beam components \cite{trypogeorgos2013precise}. Since additional beams add experimental complexity, the least number of beams necessary to achieve the intended spatial control, which corresponds to the greatest beam spacing, is preferred. 

To evaluate the minimum optical beam spacing needed to form homogeneous optical potentials, we consider the energy scales within a simple trapping geometry which involves two adjacent beams. Disconnected potentials are formed when the constituent beam centers are separated by several beam waists. Decreasing the separation between the trapping beams reduces the amplitude of the potential barrier between the centers. Connected potentials are formed once the barrier potential is comparable to the interaction energy of the trapped ensemble, the chemical potential. Aggregate traps formed using the larger beam numbers require still smaller beam spacing to form potentials which negligibly corrugate the atomic density distribution.

Releasing the trapped atoms and destructively imaging the density distribution following free-expansion often provides greater insight regarding the trapped state \cite{PhysRevLett.77.5315,PhysRevA.54.R1753}. The \SI{20}{\milli\second} time-of-flight (TOF) image in \figreft{fig:AOD_Atoms}{b} shows regular azimuthal density corrugations associated with potential corrugations around the time-averaged ring; these density signatures were obscured following \SI{1}{\milli\second} TOF. Here the scanning points were discretely spaced by $s=0.8\hspace{0.2ex}\sigma$, where $\sigma$ represents the $1/e^2$ Gaussian beam waist. Spacing of order $s\approx0.6\hspace{0.2ex}\sigma$ avoid these density signatures by suitably smoothing the potential, while minimizing the number of points \cite{bell2020PhD}; see \figreft{fig:AOD_Atoms}{c}.

Discrete AOD scan patterns are experimentally formed using frequency shift keying (FSK) function generators. Simple scanning patterns such as rings or lines are similarly achieved using continuously frequency chirped signals. Continuous scanning approaches circumvent the issues associated with selecting point spacing $s$. However, parameterizing the signals necessary to continuously scan more complex multidimensional patterns is prohibitively difficult in general. FSK function generators therefore provide greater experimental flexibility when developing AOD systems. These arguments are largely irrelevant for EOD systems, since the signal amplitude controls the beam deflection angle, rather than its frequency.

\subsubsection{Multiplexed Operation of AODs}

Driving an AOD using a single continuous frequency experimentally controls the deflection angle and power of an incident optical trapping beam. This can be extended to multiplexed signals which include several frequencies, generating multiple beams from one incident beam input. Provided the waveform generator enables the individual control over the component wave amplitudes and frequencies, then the trapping beam positions and powers are individually controllable. Linear potentials, 1D trap arrays and runaway optical evaporation have been demonstrated using multiplexed driving schemes \cite{PhysRevA.90.051401}.

In 2D, there are additional challenges which preclude the complete and arbitrary control over the potential landscape. For one, using a dual axis AOD deflector, the geometry is limited to a rectangular array of trapping beams. Additionally, since the input beam is sequentially diffracted along orthogonal directions, the amplitude of any frequency component within the waveform controlling one diffraction axis affects the intensity of an entire line of diffraction orders oriented along the orthogonal direction. This inherent property of dual axis deflectors prevents the individual control of the beam intensities necessary to correct spatially asymmetric trap nonuniformities using density based feedforward.\\

\subsubsection{Time-Averaged Operation of AODs}

Time-averaged, or painted optical potentials, offer a greater degree of freedom in the geometries that can be configured, with modulation of the driving frequencies providing individually addressable control over amplitude and beam position.  

Time-averaged traps are formed when the scanning frequency exceeds the rate required by the trapped atoms; these requirements are described in Sec.~\ref{sec:TimeAvg}. Modulating the beam power and position together enables local control of the time-averaged optical trap. These capabilities enable adjustments of the potential based on observations of the atomic density, and the preparation of nearly arbitrary density distributions using density based feedforward.

The minimum scanning frequency required for effective time-averaging is equally determined by the response of the trapped atoms and the intended application. While unperturbed density distributions are readily achieved using modest scanning frequencies, substantially greater frequencies are required for phase sensitive applications; further details are provided in Sec.~\ref{sec:PhaseMicromotion}. These considerations have been explored using AOD time-averaged ring potentials, which revealed the existence of uneven imprinted phase distributions \cite{bell2018phase}. Upon release, these non-uniform phase distributions were seen as the density signatures shown in \figreft{fig:AOD_Atoms}{d}. These results motivate the development of more rapidly scanning EOD based time-averaged optical potentials.

The dynamical density and phase distributions of condensates trapped within time-averaged optical rings are very similar to those within stirred ring potentials \cite{0953-4075-49-23-235301}. Provided another toroidal trap confines the atoms, repulsive barriers, analogous to the trapping beam used to form attractive time-averaged potentials, may be used to stir the trapped atoms and form persistent currents \cite{PhysRevA.91.033607,PhysRevA.91.063625,0953-4075-46-9-095302,PhysRevA.80.021601,PhysRevLett.110.025302}. Within these systems, metastable quantum phase transitions between circulation states are mediated by incompressible excitations \cite{PhysRevA.79.063616}, motivating studies into the relationship between the current and phase distributions \cite{PhysRevX.4.031052,PhysRevA.92.033602,PhysRevA.81.033613,0953-4075-49-23-235301,PhysRevA.94.063642}.

\subsubsection{Dynamical Trapping Sequences}

The final comment here relates to the minimization of unwanted excitations, such as vortices, sound, and breathing modes, in preparing trapped condensates in configurable optical potentials. This is particularly relevant during evaporative cooling, where, for instance the dipole sheet trap depth as well as the trap depth of the configurable optical potential (from e.g. AOD, DMD-SLM) are simultaneously lowered, resulting in complex dynamics in the potential minima.  Possible consequences of a non-ideal evaporation ramp in a time-averaged ring trap are shown in \figreft{fig:AOD_Atoms}{e}, where unwanted vortices were generated, and observed in the condensate following TOF absorption imaging~\cite{PhysRevA.61.023605}.  Particular care should be exercised to dynamically correct for the total potential during evaporation or any dynamical change to the optical potential, and this may be achieved by observing the atomic density at multiple intervals throughout the sequence.

\subsection{Technical Considerations}

Several additional considerations are important when evaluating which commercial deflector would optimally satisfy the requirement of the intended application. These include the homogeneity of the diffraction efficacy across broader deflection regions, suitable methods for correcting the residual corrugations detected across the optical potentials formed, and the physical characteristics which constrain the ultimate spatial and temporal trapping resolutions.

\subsubsection{Deflection Efficiency}

Deflection angle dependent transmissivity introduces optical intensity and trapping depth variations across the optical potential. Since the incident optical trapping beam is mechanically fixed relative an AOD, the maximal diffraction efficiency is achieved for some central deflection angle where $\theta_1=\theta_2$. Scanning the deflection angle to form optical traps changes the diffraction condition and increases the proportion of optical power diffracted into other unused orders. EODs use refraction to redirect a single output beam and are therefore immune to these intensity modulations.

One method for overcoming these changes involves the steering of the acoustic diffraction grating angle as the acoustic wave frequency changes. Directional control over the acoustic traveling wave angle through the AOD is achieved using phase array piezoelectrical oscillators. These devices incorporate several adjacent mechanical oscillators which are driven with an equal frequency but variable relative phases. The relative phases between these oscillators control the acoustic wave propagation angle, preserving $\theta_1\approx\theta_2$.

Several other considerations affect the useful optical intensity diffracted into the optimally aligned order and available for trapping atoms. The most important of these considerations involves the selection of a device which suits the optical wavelength intended for trapping atoms. For any AOD, the maximal diffraction efficiency is achieved using the manufacturer defined central optical wavelength and acoustic driving frequency. How the diffraction efficiency responds to changing the grating frequency depends on the optical wavelength. AODs which incorporate phased array transducers are therefore only able to preserve the optimal grating angle and diffraction efficiency for the central wavelength.

While the phased array acoustic wave steering may improve the Bragg condition, for example, optimizing the $+1$ diffraction order, this must narrow the useful effective bandwidth of the opposite -1 order. Operators would be required to adopt the $+1$ diffraction order in this example and accept the corresponding $\omega_2=\omega_1+\omega_s$ detuning. The frequency of other trapping beams around the system, or alternatively the input frequency $\omega_1$ may need to be adjusted to avoid spatial interference with other trapping beams.

\subsubsection{Hardware Selection}

The required spatial and temporal trapping resolutions are important considerations when comparing commercial beam deflectors. One advantage of multiplexed AOD systems is the spatial and temporal resolutions are largely separable. The spatial resolution is determined by the focused spot size. Greater spatial resolution is achieved by increasing the beam waist prior to being focused; see \eqnreft{eqn:AtomPolariseMany_GaussWaist}{}.

Magnifying the beam waist after the deflector inversely scales the deflection angle and proportionately reduces the trapping area.  Alternatively, a deflector with a larger active aperture accepts an incident beam with a larger initial waist which increases the spatial resolution without reducing the trapping area.

Improving the optical trapping spatial resolution using large active area AODs to enlarge the trapping area cannot be done without compromise. Increasing the active area increases the time required for the running acoustic wave to traverse the optical beam diameter within the crystal. The access-time required for acoustic wave changes to propagate across the optical beam and change the diffraction conditions determines the temporal resolution of the trapping regime. The spatial resolution using any specific AOD cannot be improved without sacrificing either the trapping area or temporal resolution.

This inherent compromise between the temporal and spatial resolutions is important when engineering AOD controlled time-averaged optical potentials with large trapping areas. The larger the trapping area the larger the number of scanning points required to generate optical potentials with the same spatial resolution. Given the scanning point time is limited by the AOD access-time, it is useful to develop practical metrics for comparing the relative merits of commercial devices.

Since the temporal resolution of trapping points are independent of the number of points when a multiplexed AOD driving scheme is adopted, for these applications greater emphasis should be placed on increasing the spatial resolution by maximizing the unit-less deflection range parameter $\alpha_x$. Alternately, when developing time-averaged trapping applications, greater emphasis should be given to maximizing the temporal resolution, by maximizing the scanning frequency parameter $\alpha_t$. Given the device specific active radius~$R_s$, sound velocity $\nu_s$ and deflector bandwidth $\Delta_s$, these comparative metrics are 
\begin{align}
\label{eqn:AODCompare}
\alpha_x = \frac{\Delta_sR_s}{\nu_s}~,\qquad \alpha_t = \frac{\nu_s}{R_s}~.
\end{align}

\section{Digital Micromirror Devices}
\label{sec:DMDs}

\begin{figure}[t!]
\includegraphics[width=1.0\columnwidth]{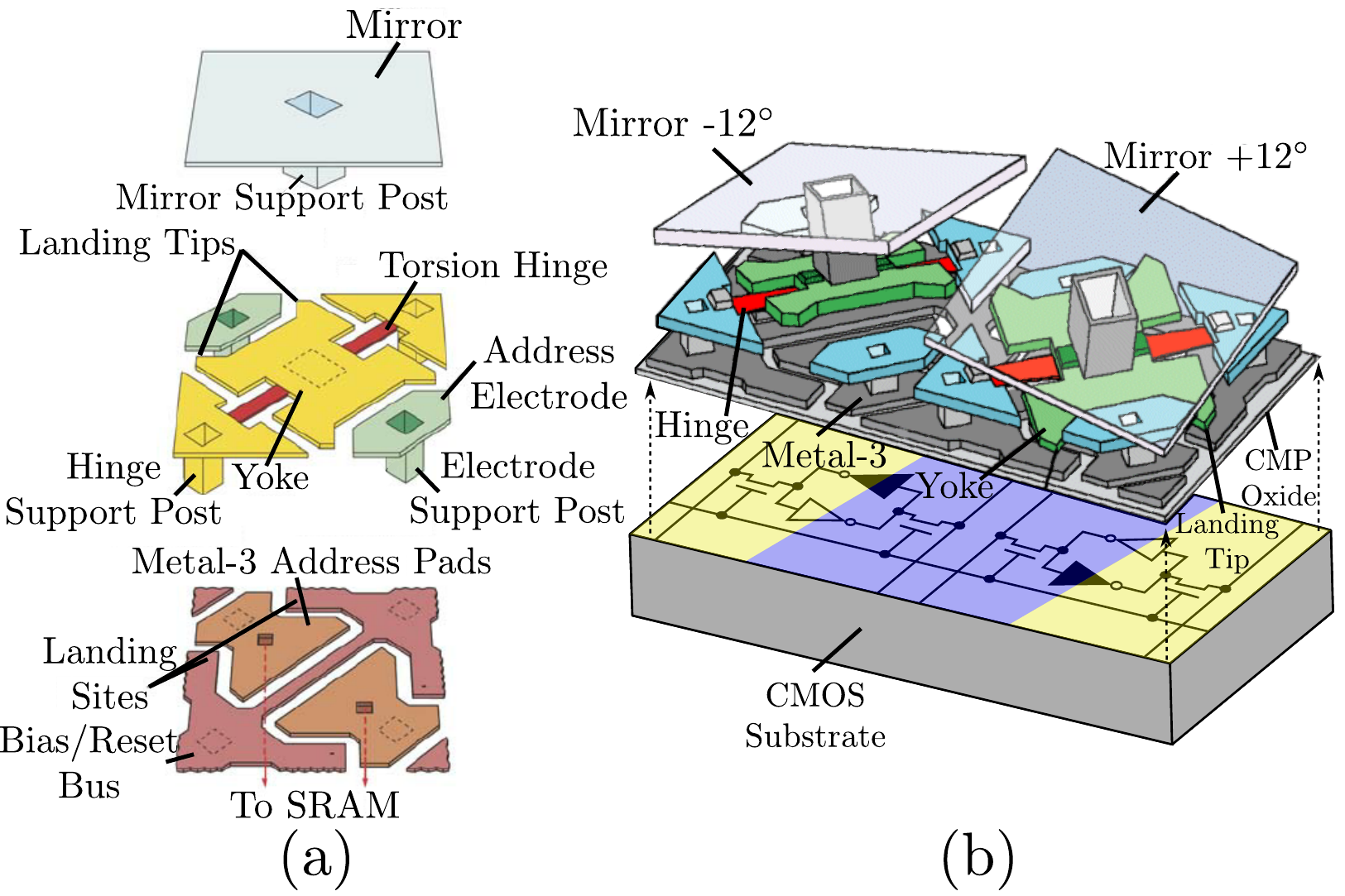}
\caption{Representation of the physical implementation of the individual mirrors making up the two-dimensional DMD-SLM array (a) Layer view of DMD-SLM pixel. (b) Two DMD-SLM pixels in oppositely tilted states with underlying CMOS circuit in opposite logic states represented. Courtesy of Texas Instruments.}
\label{fig:DMDPhyscialStructure}
\end{figure}

Digital micromirror devices (DMD-SLMs) are micro-opto-electro-mechanical systems (MEMS) which are typically used to spatially modulate light beams using tiny electromechanical mirrors~\cite{dudley2003emerging}. The device is trademarked by Texas Instrument and was first created by their Fellow Emeritus Dr. Larry Hornbeck in 1987. DMD-SLMs are most commonly used in projectors where they modulate the spatial light field to project different images onto a screen. A DMD-SLM typically consists of a two-dimensional rectangular array composed of hundreds of thousands to millions of mechanical mirrors, arranged in a grid pattern where the mirrors can be individually addressed and rotated electronically, see \figo{fig:DMDPhyscialStructure}.

Mechanically, each individual mirror is mounted on a yoke which is suspended by torsion hinges connected to posts joined to the substrate. The torsion hinges allow for the rotation of the yoke, which in turn rotates the mirror about the torsion axis. The maximum rotation of the mirror is mechanically fixed by landing tips on the yoke which make contact with the landing site on the metal substrate layer at certain angles limiting the rotation of the micromirrors to $\pm$ 12$^{\circ}$-17$^{\circ}$, often referred to as the on/off state of the mirror. Landing tips and sites are designed so as to limit the surface area contact of the on/off state to prevent Van der Waals forces from permanently latching the mirror in a particular state. The hinges can be oriented such that the mirror rotation axis is about the diagonal (orthogonal) which are referred to as the diamond (square) hinge operation mode. The mirror array is typically behind an anti-reflection coated optical window with $>96$\% double pass transmission efficiency. This window prevents dust from damaging the array, while also containing an inert gas to protect the mirror surfaces.

Electrodes are connected onto both sides of the rotation axis of the mirror. They hold diametrical potentials to create a field that attracts the mirror/yoke assembly on one side and repel it on the other. These off-center forces provide the torque needed to rotate the mirror into the desired state. The polarity of the bias is stored in a complementary metal oxide semiconductor (CMOS) static random access memory (SRAM) fabricated on the underlying wafer and connected to the biasing electrodes. These registers can be programmed and accessed individually by the DMD-SLM controller board. Once the mirror has landed it must be released before it can be switched to a new state. This allows for the pre-loading of the next image pattern into the CMOS memory, enabling  near instantaneous switching  of the DMD-SLM mirror array state once the next image has been pre-loaded, taking 0.5~$\mu$s to 5~$\mu$s depending on the device.

Optically, each of the mirror elements are highly reflective with a typical reflectively of $\sim$89\% over the visible region~\cite{texas_instruments_DMD_2019}. A simple picture of the operation of the DMD can be considered by examining a beam of light that travels perpendicular to the hinge rotation axis, with twice the mirror tilt angle from the DMD-SLM array normal. Given that the DMD mirrors are tilted towards the incident light, the light beam will simply reflected along normal direction to the DMD plane (the `on' position). If the mirror is instead tilted away from the incident light (the `off' position), it is reflected at four times the mirror tilt angle from the plane normal. By setting up the imaging system to only image the light which gets reflected normal to the DMD-SLM, the amplitude of the projected light field can be spatially tailored. The mirror on state fill factor is $\sim 92$\% which represents the ratio of the array area which the tilted mirrors cover. 

However, this simple picture neglects the diffractive nature of the DMD-SLM mirror array. The array of mirrors act as a blazed diffraction grating, which will limit the reflection efficiency depending on the diffraction efficiency at the desired reflection angle. DMD-SLM diffraction theory will be explored in depth in the next section.

For a more complete introduction to the basics of DMD-SLMs and their operation see~\cite{lee2013dmd, hornbeck1995digital, hornbeck1996digital}, for more in-depth information about the exact nature of the electrostatic rotation of the mirror see~\cite{hornbeck1990deformable}, for comparison of DMD-SLM performance against other popular projection technology see~\cite{hornbeck2001dmd}, for an explanation overview of the packaging and testing of the devices see~\cite{mignardi2000digital} and for an overview of emerging and current applications of DMD-SLM technology see~\cite{sampsell1994digital, dudley2003emerging, bansal_digital_2013}.

\subsection{DMD-SLM diffraction theory}
\label{DMD_Diffraction_Theory}

\begin{figure*}[t!]
\includegraphics[width=0.75\textwidth]{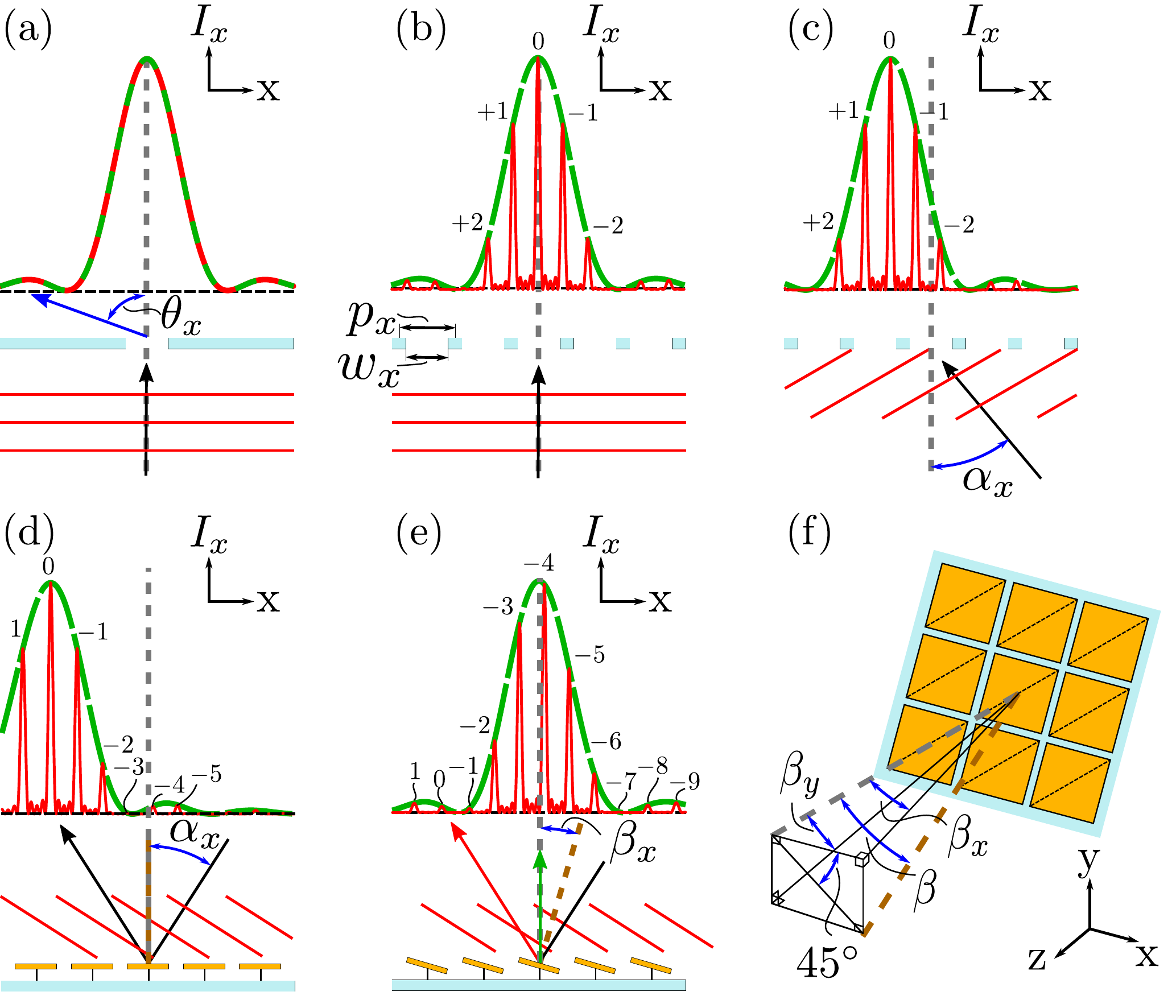}
\caption{Fundamentals of DMD-SLM diffraction grating from single slit to 2D blazed diffraction grating. (a) The diffracted light intensity from a single slit aperture with diffraction angle ($\theta_{x}$). The red lines represent the incident wavefronts, with the black arrow representing the directionality of the incident wavevector. The red lines on the intensity plots represent the actual intensity pattern and the green lines represent the single slit diffraction envelop. The dashed gray line is the diffraction grating normal. The single slit diffraction intensity pattern is given by \eqnreft{eq:SingleDiffractionSlit}{}. (b) Diffraction from multiple slits (N$_{x}$ = 5) with slit width ($w_{x}$) and slit pitch ($p_{x}$). Each slit produces a diffraction pattern with a slightly offset peak. The interference of these patterns forms the red line interference pattern with the constructive interference order points labeled. The multiple slit diffraction intensity pattern is given by \eqnreft{eq:MultipleDiffractionSlit}{}. (c) Tilted incident wavefront with incident angle ($\alpha_{x}$) leads to shifted interference order and diffraction peak. The diffraction intensity pattern is given by Eqs.~(\ref{eq:MultipleDiffractionSlit})~\&~(\ref{eq:MultipleDiffractionSlitIncidentAngle}). (d) Reflective diffraction grating which is shown to be equivalent to a transmission diffraction grating of same slit pitch and width. The brown dashed line represents the normal of the reflective elements. (e) Diffraction pattern for the case of tilted reflective elements ($\beta_{x}$). The red arrow represents the position of the 0$^{th}$ constructive interference order which is located at $\theta_{x} = \alpha_{x}$ and the green arrow the position of the diffraction envelop peak which is located at $\theta_{x} = \alpha_{x} - 2\beta{x}$. Tilting the reflective elements decouples the constructive interference points from the diffraction peak. The diffraction intensity pattern for tilted mirrors is given by Eqs.~(\ref{eq:MultipleDiffractionSlit})~\&~(\ref{eq:MultipleDiffractionTiltedMirror}{}). (f) Most Texas Instruments DMD-SLMs use a square lattice of square mirrors with diagonal representing the tilting axis (black dashed line). Assuming the diffraction grating normal is taken to be the z axis and the square lattice vectors form our \{x,y\}-Cartesian basis. The mirror tilt angle ($\beta$) can be decomposed into a \{x,y\} tilt component with $\beta_{x} = \beta_{y}$. And each axis diffraction pattern is independent and can individually be understood from (a) - (e) and their combined diffraction intensity pattern is given by Eqs.~(\ref{eq:MultipleDiffractionSlit})~\&~(\ref{eq:MultipleDiffractionDMD}).}
\label{fig:DMDBlazeGrating}
\end{figure*}

DMD-SLMs act as a two dimensional classic optical diffraction grating when all of the mirrors are in their `on' state. Every single micromirror acts as a diffraction slit which creates a diffraction envelop of light,  and the multiple light patterns interact resulting in an interference pattern. To understand diffraction from a DMD-SLM surface, one must first understand the diffraction from a single slit, the diffraction from multiple slits, the diffraction from a blazed optical grating, and generalize the concept to two-dimensions. A quick overview of these concepts will be given in the context of diffraction from a DMD-SLM. 

Each of the mirrors on the DMD-SLM acts as a diffraction slit. The diffraction from a single slit in the Fraunhofer limit for light under normal incidence is depicted in \figreft{fig:DMDBlazeGrating}{a}, and is given by
\begin{equation}
I = I_{0}\hspace{0.25ex} \textrm{sinc}^2\hspace{-0.25ex}\left(\phi_x\right),\qquad \phi_x = \frac{k\hspace{0.25ex}w_x}{2}\sin{\theta_x},
\label{eq:SingleDiffractionSlit}
\end{equation}
where $k$ is the spatial angular frequency (wavenumber) of the wave ($k = 2\pi/\lambda$), $I_0$ is the incident wave intensity, $w_x$ is the slit width, and $\theta_{x}$ is the measurement angle relative to the slit normal. 

As mentioned earlier, DMD-SLMs are made up of multiple mirrors which each act as diffraction slits and which leads to a typical multi-slit interference pattern shown in \figreft{fig:DMDBlazeGrating}{b}. The green dashed line represents the intensity envelop due to single slit diffraction, \eqnreft{eq:SingleDiffractionSlit}{}, and the red solid line depicts the interference intensity pattern given by
\begin{equation}
I = I_{0}\hspace{0.25ex} \textrm{sinc}^2\hspace{-0.25ex}\left(\phi_x\right)\left[\frac{\sin{\left(N_{x} \gamma_x\right)}}{\sin(\gamma_x)}\right]^2,
\label{eq:MultipleDiffractionSlit}
\end{equation}
\noindent where

\begin{equation}
\gamma_x = \frac{k\hspace{0.25ex}p_x}{2}\sin{\theta_x},
\end{equation}
with $p_x$ the mirror grating pitch, $N_x$ the number of slits contributing to the interference pattern, $\gamma_{x}$ parameterizing the interference pattern, and $\phi_x$ is defined in \eqnreft{eq:SingleDiffractionSlit}{}. The location of the maxima of the interference pattern are labeled $n$ and their location is given by
\begin{equation}
\theta^{\text{max}}_n = \arcsin{\left(\frac{2n\pi}{k\hspace{0.25ex}p_x}\right)} \qquad \forall \left\{n \in \mathbb{Z} \hspace{0.75ex} \& \hspace{0.5ex} \left|n\lambda\right| \leq p_x \right\}.
\label{eq:Interference_Order_Maximum}
\end{equation}

One degree of control when diffracting light from the DMD-SLM is the incident angle of the light relative to the normal ($\alpha_x$). Shifting $\alpha_x$ shifts the position of the diffraction intensity peak and the $0^{th}$ order interference peak geometrically as illustrated in \figreft{fig:DMDBlazeGrating}{c}. The intensity pattern is then given by \eqnreft{eq:MultipleDiffractionSlit}{}, but with the replacements

\begin{equation}
\begin{split}
\gamma_x &= \frac{k\hspace{0.25ex}p_x}{2}\left(\sin{\theta_x}-\sin{\alpha_x}\right), \\
\phi_x &= \frac{k\hspace{0.25ex}w_x}{2}\left(\sin{\theta_x}-\sin{\alpha_x}\right).
\end{split}
\label{eq:MultipleDiffractionSlitIncidentAngle}
\end{equation}

The normal mode of operation for a DMD-SLM is for the mirrors to be tilted by a manufacturer tilt angle relative to the DMD-SLM surface. The tilting of the reflection surface affects the position of the maximum of the diffraction envelop, green arrow in \figreft{fig:DMDBlazeGrating}{e}, which obeys the reflection law with respect to the reflection mirror surface normal, shown with a brown dashed line in the figure. The position of the $0^{th}$ order interference peak (red arrow) is given by reflection of the incident beam with respect to the diffraction grating normal (gray dashed line), and is therefore independent of the reflection surface tilt. The diffraction envelop intensity maximum is thus no longer tied to the location of the $0^{th}$ order interference peak.

Figure \figreft{fig:DMDBlazeGrating}{e} accounts for the tilting of the reflective elements, as is the case for DMDs. The intensity pattern is again described by \eqnreft{eq:MultipleDiffractionSlit}{}, but with the replacement
\begin{equation}
\phi_x = \frac{k\hspace{0.25ex}w_x}{2}\left[\sin{\left(\theta_x+\beta_x\right)}-\sin{\left(\alpha_x-\beta_x\right)}\right];
\label{eq:MultipleDiffractionTiltedMirror}
\end{equation}
$\beta_x$ describes the mirror tilt angle. The interference pattern, dictated by $\gamma$, is unaffected by the mirror tilt. However, the position of the intensity envelop changes and is described through the modification to $\phi_x$.

The full DMD-SLM consists of a 2D square array of mirrors which rotate about their diagonals as shown in \figreft{fig:DMDBlazeGrating}{f}. The diffraction from this array can be thought of as being made up of a blazed diffraction grating along both axes. The resulting intensity pattern is thus the product of two one dimensional intensity patterns, \eqnreft{eq:MultipleDiffractionSlit}{}, with $\phi_x$ defined by \eqnreft{eq:MultipleDiffractionTiltedMirror}{}, giving the overall DMD-SLM diffraction intensity of 
\begin{equation}
I\hspace{-0.5ex} = \hspace{-0.5ex} I_{0}\hspace{0.25ex} \textrm{sinc}^2\hspace{-0.5ex}\left(\phi_x\right)\textrm{sinc}^2\hspace{-0.5ex}\left(\phi_y\right)\hspace{-0.5ex}\left[\frac{\sin{\left(N_{x} \gamma_x\right)}\sin{\left(N_{y} \gamma_y\right)}}{\sin(\gamma_x)\sin(\gamma_y)}\right]^2 ,
\label{eq:MultipleDiffractionDMD}
\end{equation}

\noindent where
\begin{equation}
\begin{split}
&\gamma_i = \frac{k\hspace{0.25ex}p_i}{2}\left[\sin{\theta_i}-\sin{\alpha_i}\right], \\
&\phi_i = \frac{k\hspace{0.25ex}w_i}{2}\left[\sin{\left(\theta_i+\beta_i\right)}-\sin{\left(\alpha_i-\beta_i\right)}\right].
\end{split}
\end{equation}
\noindent The index $i \in \left\{ x,y \right\}$ indicates the two orthogonal axes of the DMD-SLM surface. 

\begin{figure*}[t!]
\includegraphics[width=\textwidth]{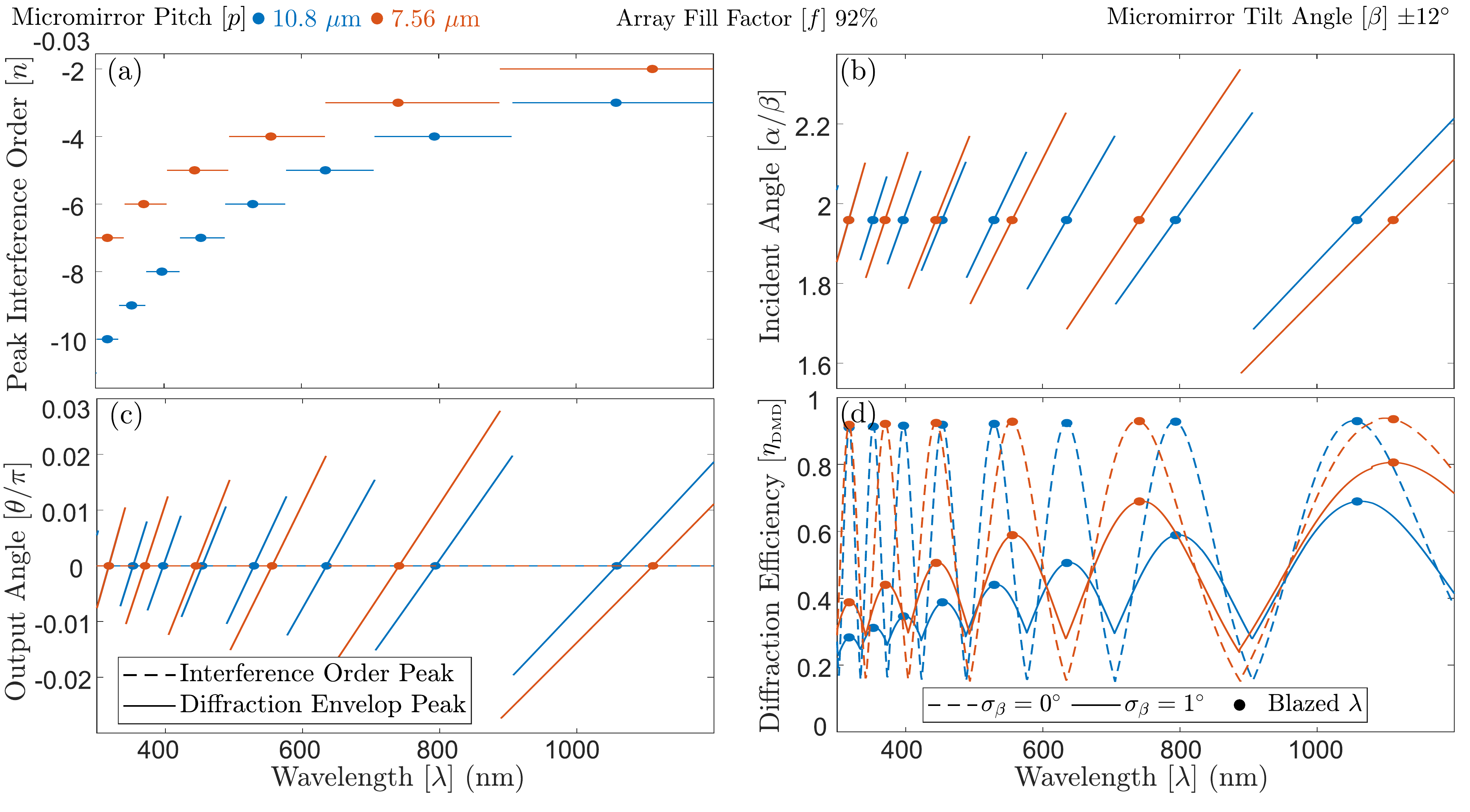}
\caption{Diffraction parameters when direct imaging a Texas Instruments {\color[rgb]{0,0.447,0.741}DLP9500} ($p = 10.8$~$\mu$m) and {\color[rgb]{0.851,0.325,0.098}DLP9000} ($p = 7.56$~$\mu$m) chip using single order. Both chips have a micrometer tilt angle of $\beta = \pm 12^{\circ}$ and an array fill factor of 92\%. (a) Diffraction efficiency given by \eqnreft{eq:DiffractionEfficiencyWithUncertainty}{} assuming no micromirror grating tilt uncertainty ($\sigma_{\beta} = 0^{\circ}$) are shown using dashed line and using manual specified micromirror tilt uncertainty ($\sigma_{\beta} = 1^{\circ}$) are shown using solid lines. The wavelengths at which the diffraction gratings are blazed which can be calculated using \eqnreft{eq:BlazedWavelengthCondition}{} are shown using solid points. (b) The output angle about the micromirror tilt diagonal $\theta$ at which the interference order for our projection optics (dashed line) and the peak of the diffraction envelop (solid line) are located. The interference peak is always aligned at $\theta = 0$ by design since that is where the diffraction limit of the imaging system will be optimized. (c) The angle of the incident beam required to align the closest diffraction order with the diffraction grating normal calculated using \eqnreft{eq:IncidentAngleToMaximizePowerOutput}{}. (d) The corresponding interference orders which line up with the grating axis can be calculated using \eqnreft{eq:OptimalInterferenceOrder}.}
\label{fig:DMDDiffractionEfficiency}
\end{figure*}

The DMD-SLM diffraction efficiency for a plane wave can be calculated from \eqnreft{eq:MultipleDiffractionDMD}{}. In the limit that the number of diffracting elements in the grating becomes large $\lim_{N_i\rightarrow \infty} \sin(N_i \gamma_i)/\sin(\gamma_i) = \delta[\hspace{-1.5ex}\mod\hspace{-0.5ex}(\gamma_i,\pi)]$. For most DMD-SLMs, $N_x$ and $N_y$ are relatively large which means that the diffraction efficiency of each of the constructive interference peaks can be calculated using
\begin{equation}
\begin{split}
\eta_{_\text{DMD}}&(n_x,n_y,\alpha_x,\alpha_y,\beta) = \hspace{0.25ex} A\hspace{0.5ex}\textrm{sinc}^2\hspace{-0.5ex}\left(\phi_x\right)\textrm{sinc}^2\hspace{-0.5ex}\left(\phi_y\right),  \\ 
\phi_i &= \frac{k\hspace{0.25ex}w_i}{2}\left[\sin{\left(\theta_{n,i}^{\textrm{max}}+\beta_i\right)}-\sin{\left(\alpha_i-\beta_i\right)}\right],
\end{split}
\label{eq:DiffractionEfficiency}
\end{equation}
where A is the normalization factor, calculated below in \eqnreft{eq:NormalizationDiffractionModes}{}, and $\theta_{n,i}^{\textrm{max}}$ is the location of the constructive interference peak given by
\begin{equation}
\begin{split}
\theta_{n,i}^{\textrm{max}} &= \arcsin{\left[\frac{2n_i\pi}{kp_i}+\sin{\alpha_i}\right]} \\ \forall &\left\{n_i \in \mathbb{Z} \hspace{0.75ex} \& \hspace{0.5ex} \left|\frac{n_i\lambda}{p_i}+\sin{\alpha_i}\right| \leq 1 \hspace{0.75ex} \& \hspace{0.5ex}  i \in \left\{ x,y \right\}  \right\}.
\end{split}
\label{eq:ConstructiveInterferenceAngle}
\end{equation}

When using the DMD-SLM in the Fourier plane, one wants to maximize the light intensity which is directed in the order $n_i$ to be projected onto the atoms. This is achieved by solving for $\phi_i = 0$ in \eqnreft{eq:MultipleDiffractionDMD}{} for both coordinates and is also known as the blazed grating condition. For most wavelengths $\lambda$ and diffraction orders there will exist a blazed angle which gives near perfect diffraction efficiency.

In the case of direct imaging, an image of the DMD-SLM needs to be projected onto a plane with normal given by the projection objective optical axis, for this to be the case the DMD-SLM normal also needs to be parallel to the  optical axis. Adding the requirement that the aberrations be minimized adds the constraint that the DMD-SLM should be centered on the optical axis. This leads to an optimal measurement angle of $\theta_i = 0^{\circ}$. If the DMD-SLM normal is not parallel to the projection optical axis the final projection objective will also need to be tilted relative to the atom plane, greatly complicating the optical system when trying to achieve a diffraction limited imaging system.

The requirement that $\theta_i = 0^{\circ}$ leads to $\alpha_i = -\arcsin{\left(n_i\lambda/p_i\right)}$. The blazed angle is approximately given by $\alpha_i \approx 2\beta_i$ which can be used to find the diffraction order which maximizes the diffraction efficiency for a given wavelength and mirror tilt angle:
\begin{equation}
n^{\text{max}}_i=-\Bigl\lfloor\frac{p_i}{\lambda}\sin{2\beta_i}+\frac{1}{2}\Bigr\rfloor,
\label{eq:OptimalInterferenceOrder}
\end{equation}
and plotted in \figreft{fig:DMDDiffractionEfficiency}{a} for two common DMD-SLM devices. Once $n^{\text{max}}_i$ is determined, the incident angle which maximizes the diffraction is given by 
\begin{equation}
\alpha_{i}^{\text{max}} = -\arcsin{\left(\frac{\lambda}{p_i}n^{\text{max}}_i\right)},
\label{eq:IncidentAngleToMaximizePowerOutput}
\end{equation}
and is plotted in \figreft{fig:DMDDiffractionEfficiency}{b} as function of the DMD-SLM mirror tilt angle.

One may now ask which wavelengths maximize the direct imaging optical power of a given DMD-SLM. This is done by solving for wavelengths which satisfy the blazing condition, as well as the desired $\theta_i = 0^{\circ}$, giving
\begin{equation}
\lambda_i = \frac{-\sin{\left(2 \beta_i\right)} p_i}{n_i},
\label{eq:BlazedWavelengthCondition}
\end{equation}
where $\beta_i$ is the tilt angle of the micromirror along the coordinate $i$. Since DMD-SLM mirrors typically rotate about their diagonal, as depicted in \figreft{fig:DMDBlazeGrating}{f}, $\beta_x = \beta_y = \arctan{\left[\tan{\beta}/\sqrt{2}\right]}$, where $\beta$ being the tilt angle of the micromirrors about the diagonal. For square pixels, where $p_x = p_y$, this leads to $\lambda_x = \lambda_y$, meaning the blazed wavelengths along both diffraction axis are the same. \figurereft{fig:DMDDiffractionEfficiency}{c} shows the output angle along the diagonal $\theta$ of the interference order, which maximizes the power in the direct imaging order (aligned to $\theta = 0^{\circ}$ by design) and the envelop output angle that is depicted in \figreft{fig:DMDBlazeGrating}{e}. As can be seen in \figreft{fig:DMDDiffractionEfficiency}{d}, only at the wavelengths satisfying the blazed condition do both the envelop maximum and interference order overlap. This leads to a local maximum in the direct imaging diffraction efficiency for these wavelengths. 

The diffraction efficiency for a given $\lambda$, $\beta$ and diffraction order ($n_x$, $n_y$) in the direct imaging case is therefore given by
\begin{equation}
\eta{_{_\text{DMD}}}(n_x, n_y, \beta, \lambda) = \frac{\eta_{_\text{DMD}}(n_x,n_y,\alpha_x^\text{max},\alpha_y^\text{max},\beta)}{\sum\limits_{\hspace{0.5ex}n'_x}\hspace{-0.5ex}\sum\limits_{\hspace{0.5ex}n'_y} \eta_{_\text{DMD}}(n'_x,n'_y,\alpha_x^\text{max},\alpha_y^{\text{max}},\beta)},
\label{eq:EfficiencyWavelengthNoUncertainty}
\end{equation}
where $n'_{x(y)}$ range over every possible interference order that fulfill the conditions $\left|\theta_i\right| \leq \pi/2$  and  $\left|\alpha_i\right| \leq \pi/2$, with $n_i^{\text{max}}$ given by \eqnreft{eq:OptimalInterferenceOrder}{}, $\alpha_i^{\text{max}}$ given by \eqnreft{eq:IncidentAngleToMaximizePowerOutput}{}, and $\eta_{_\text{DMD}}(n_x,n_y,\alpha_x,\alpha_y,\beta)$ is given by \eqnreft{eq:DiffractionEfficiency}{}. For small tilt angles $\beta$
\begin{equation}
\sum_{n_x}\sum_{n_y} \eta_{_\mathrm{DMD}}(n_x,n_y,\alpha_x^{\text{max}},\alpha_y^{\text{max}},\beta) = A \left(\frac{p_i}{w_i}\right)^2,
\label{eq:NormalizationDiffractionModes}
\end{equation}
which means the diffraction efficiency of the modes along the diagonal $n_x = n_y = n$ is given by 
\begin{equation}
\eta_{_\text{DMD}}(n, \beta, \lambda) = \left(\frac{w}{p}\right)^2 \textrm{sinc}^4 \hspace{-0.5ex} \left(\pi w\left[\frac{n}{q} \hspace{-0.25ex} + \hspace{-0.25ex} \frac{\sqrt{2} \beta}{\lambda}  \right] \right),
\label{eq:DiffractionEfficiencyDiagonalModes}
\end{equation}
where the small angle approximation was used extensively to simplify the equation, $w$ is the micromirror square width, and $\beta$ is defined in \figreft{fig:DMDBlazeGrating}{f}. This equation gives the diffraction efficiency for the diagonal order $n$ when it is aligned to $\theta = 0$ for direct imaging.

Micromirror deflection angle uncertainties ($\sigma_{\beta}$) associated with the manufacturing tolerances may furthermore appreciably modify the experimental DMD-SLM diffraction efficiency at a given wavelength. This can be accounted for using
\begin{equation}
\eta'_{_\text{DMD}}(\beta,\lambda,\sigma_{\beta}) = \int_{-\infty}^{\infty} d\kappa \, g(\kappa,\beta,\sigma_{\beta}) \hspace{0.3ex} \eta(\kappa,\lambda),
\label{eq:DiffractionEfficiencyWithUncertainty}
\end{equation}
where $\eta'_{_\text{DMD}}$ is the optical efficiency of the maximum order accounting for the tilt uncertainty of the micromirror and can be evaluated using \eqnreft{eq:EfficiencyWavelengthNoUncertainty}{}, or using equations~(\ref{eq:DiffractionEfficiencyDiagonalModes}) combined with (\ref{eq:OptimalInterferenceOrder}). The micromirror tilt probability distribution $g$ is given by 
\begin{equation}
g(\kappa,\beta,\sigma_{\hspace{-0.3ex}\beta}) = \frac{1}{\sigma_{\hspace{-0.3ex}\beta}\sqrt{2\pi}} \exp\left( -\frac{1}{2}\frac{(\kappa-\beta)^2}{\sigma_{\hspace{-0.3ex}\beta}^2}\right).
\label{eq:MicromirrorTiltAngleProbabilityDistribution}
\end{equation}
The optimal diffraction order, incident angles and diffraction efficiency of the DLP9500 and DLP9000 are plotted as a function of wavelengths in \figo{fig:DMDDiffractionEfficiency}.

Equation~(\ref{eq:DiffractionEfficiencyWithUncertainty}) only approximates the diffraction efficiency of the DMD-SLM, $\eta_{_\text{DIFF}}$, but the efficiency is further reduced by the single-pass window transmission efficiency, $\eta_{_\text{SPWT}}$, the efficiency of the array due to non unitary fill factor, $\eta_{_\text{FF}}$, and the mirror reflectively efficiency, $\eta_{\text{MR}}$. This all combines to give a DMD-SLM efficiency of $\eta_{_\text{DMD}} = \eta_{_\text{SPWT}} \times \eta_{_\text{FF}} \times \eta_{_\text{DIFF}} \times \eta_{_\text{MR}} \times \eta_{_\text{SPWT}}$.

\subsection{DMD-SLM imaging implementations}
Direct imaging of DMD-SLMs has been used to perform a variety of experiments in cold atoms. Described below are the main ways in which DMD-SLMs are used experimentally along with cold atoms experiments which used these techniques.

\subsubsection{Direct Imaging}
Direct imaging is the projection of the spatially modulated light pattern produced by the DMD-SLM onto the atomic plane where it is used to shape the potential experienced by the atoms, in the case where phase modulation is not required~\cite{gauthier_direct_2016,ville_loading_2017}. 

It is a straightforward technique which requires little computational power. The propagation of light from individual pixels can be calculated separately leading to an illumination profile and the atomic potential proportional to the uploaded pattern, convoluted with the point-spread function of the imaging system~\eqnreft{equ:PSFconvolution}{}. The drawback of the direct imaging method is that aberrations in the optical path experienced by the illumination from the different pixels leads to phase shifts which will affect the performance of the imaging system with no possibility of being compensated for using the SLM. This means that in the case of direct imaging particular care needs to be taken in the design of the imaging system to achieve the required resolution level. 

Assuming that the desired resolution has been achieved using a high-performance imaging system, then direct imaging is perfectly suited for the creation of large, and complex structures atomic trapping potentials. The limitations of direct imaging become apparent when trying to create small features such as beams to individual address lattice sites in cold atom microscope. For small features on the order of the PSF of the optical system, few mirrors contribute to the feature which limits the control over the shape and amplitude of the addressing feature. Second, since only a small portion of the mirrors are `on', the ratio of the diffracted light decreases drastically, resulting in a low power efficiency. The creation of small spatially distributed features is usually better served by using the DMD-SLM in the Fourier plane where the inverse scaling of the Fourier transform can be taken advantage of.

\begin{figure}[t!]
\includegraphics[width=\columnwidth]{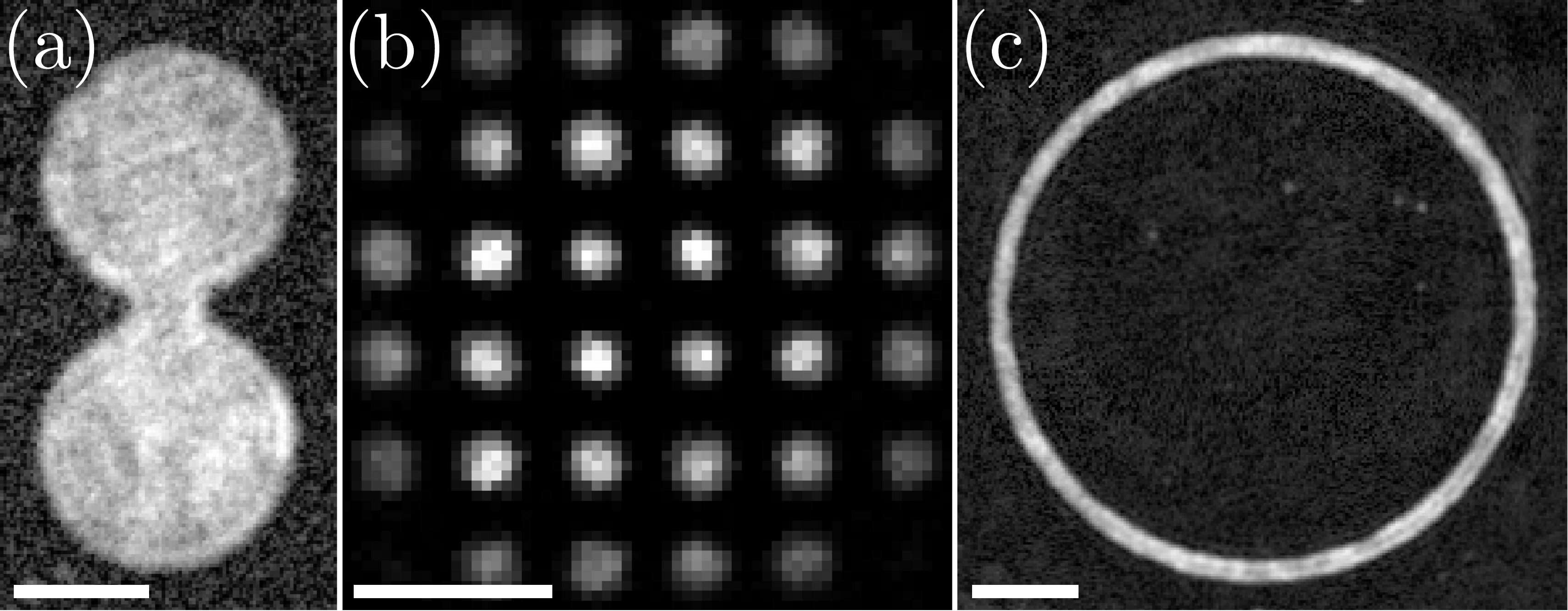}
\caption{Imaged atomic density distribution of traps created using direct imaging of DMD-SLMs. (a) A dumbbell geometry of two reservoirs connected by a channel. (b) A square lattice of BECs with 10~$\mu$m spacing formed using the projection of a halftone DMD-SLM pattern. (c) A ring-shaped BEC with a 110~$\mu$m diameter and 10~$\mu$m radial width. The scale bar on each image indicates 20~$\mu$m. Permissions requested for \textit{Roadmap on Atomtronics}, submitted for publication in \textit{AVS Quantum Science}.}
\label{fig:DMDAtomtronicsGeometries}
\end{figure}

However, direct imaging is a powerful technique and allows for the creation of near arbitrary 2D trapping potentials. Some examples are shown in \figo{fig:DMDAtomtronicsGeometries}, including a dumbbell trap consisting of two reservoirs connected by a channel with tunable reservoir size, channel length/width which can be used for studies of superfluid transport [\figreft{fig:DMDAtomtronicsGeometries}{a}]; a square lattice with 10~$\mu$m period for which the period can be dynamically changed, formed using halftoning [\figreft{fig:DMDAtomtronicsGeometries}{b}]; a ring-shaped BEC of diameter 110~$\mu$m with 10~$\mu$m radial width, where the diameter and width are easily tunable [\figreft{fig:DMDAtomtronicsGeometries}{c}]. This versatility in the possible potentials using direct imaging, combined with the dynamic capabilities of the device has allowed for simulating the expansion of the universe~\cite{eckel_rapidly_2018}, generating various energy scale turbulent states~\cite{gauthier_giant_2019,johnstone_evolution_2019,stockdale_universal_2020}, creating tunable velocity solitons~\cite{fritsch_creating_2020}, studying superfluid transport through a mesoscopic channel~\cite{gauthier_quantitative_2019}, creating dense slabs of cold atoms~\cite{corman_transmission_2017}, studying the relaxation dynamics of the recombination of multiple condensates after independent evolution~\cite{aidelsburger_relaxation_2017}, generating sound and studying its propagation~\cite{ville_sound_2018}, and providing radial confinement for the first realization of a two-dimensional homogeneous Fermi gas~\cite{hueck_two-dimensional_2018}. Two DMD-SLMs have been used together for shaping the potential for the realization of an ideal Josephson junction in an ultracold two-dimensional homogeneous Fermi gas~\cite{luick_ideal_2020}, for the generation of sound using phase imprinting to study of the propagation and quantum limited damping of sound in a two-dimensional homogeneous Fermi gas~\cite{bohlen_sound_2020}, to shape the potential during the observation of superfluidity in a strongly correlated two-dimensional homogeneous Fermi gas~\cite{sobirey_observation_2020}, and to flatten and shape a Gaussian trapping potential for realizing a Fermi–Hubbard antiferromagnet with long-range order in 2D optical lattice experiment~\cite{mazurenko_cold_2017}.

\subsubsection{Halftoning}
\label{sec:Haltoning}
Halftoning is a technique that was used to produce gray-scale images for newsprint and relies on the imaging system not being able to resolve the individual binary elements that create the image (in the case of a newspaper the observer's eye). For newsprint, either black ink spots are deposited at a given location, or the white background is left alone. To create regions that appear gray to the eye, the density of these spots are either increased or decreased to create regions of which appear darker or lighter. This can be done by either increasing the size of the ink dots or by increasing the number of said points in a given region. This technique easily maps to DMD-SLMs since they are similarly binary by nature. Here the imaging system is the projection system, and depending on the magnification and resolution of the direct imaging system, the micromirror size in the image plane can be modified such that multiple mirrors contribute the illumination at a spatial location through their point-spread function. This results in a situation similar to the newspaper where multiple modulation element contribute to the illumination of a given spatial location which allows for the control of the light amplitude by changing the ratio of on/off elements around a given spatial location~\cite{liang_1.5_2009}.

Usually for most imaging systems, the magnification of the system is fixed and changing it requires changing lenses in the projection system. On the other hand, a simple filter iris in the Fourier plane of the projection system allows for control of the limiting aperture of the system, by limiting the solid angles which can travel through the system~\cite{liang_1.5_2009}. The spatial resolution's upper limit will be given by the non iris-limited imaging system, but reducing the spatial resolution will broaden the spatial contribution of each micromirror, allowing finer control over the light amplitude. However, this comes at the cost of the inability to create sharp features. 

Given a grayscale image, there are multiple possible methods which can be used to convert the image into a binary halftone image. The simplest method is to normalize the image by the highest intensity point and spatially assign mirrors states using the normalized image as a probability map. This method creates sub-optimal patterns since the states of two adjacent mirrors where the light level should be halved are uncorrelated. These half intensity regions should optimally form a periodic checkerboard pattern which are highly correlated to increase the halftone fidelity with the original grayscale image. Another method is to use an error diffusion where the state of already assigned neighbor mirrors is taken into account when assigning the state of the next mirror~\cite{dorrer_design_2007}; these methods are sometimes alternatively called dithering \cite{hocevar_reinstating_2008} or halftoning \cite{kimoto_halftone_2013}. Generally, these algorithms start in one corner of the image, where the state which minimizes the local error is assigned. The residual local error is propagated to the adjacent mirrors and the process is repeated over the full array. This redistribution of the local error to adjacent sites leads to a correlation between sites, leading to a reduction of the global error. Modest improvements to trap shaping~\cite{gauthier_direct_2016} and uniformity when creating linear density gradients~\cite{PhDGauthier2019} where observed in BECs when utilizing error diffusion methods, as opposed to probability distribution methods.

An example halftone lattice is depicted in \figreft{fig:DMDAtomtronicsGeometries}{b}. The technique has been used to create a wide range of atomic potentials, to perform Bragg spectroscopy in a shaken optical lattice~\cite{ha_roton-maxon_2015}, imprint phonon modes on persistent current states and detect the current through Doppler shift measurements~\cite{kumar_destructive_2016}, and in 1D systems to create arbitrary potentials~\cite{tajik_designing_2019}, with the help of error correction algorithms which are discussed below in Sec.~\ref{sec.DMDFeedback}.

\subsubsection{Time-averaging}
\label{sec:TimeAveraging}

\begin{figure*}[t!]
\includegraphics[width=\textwidth]{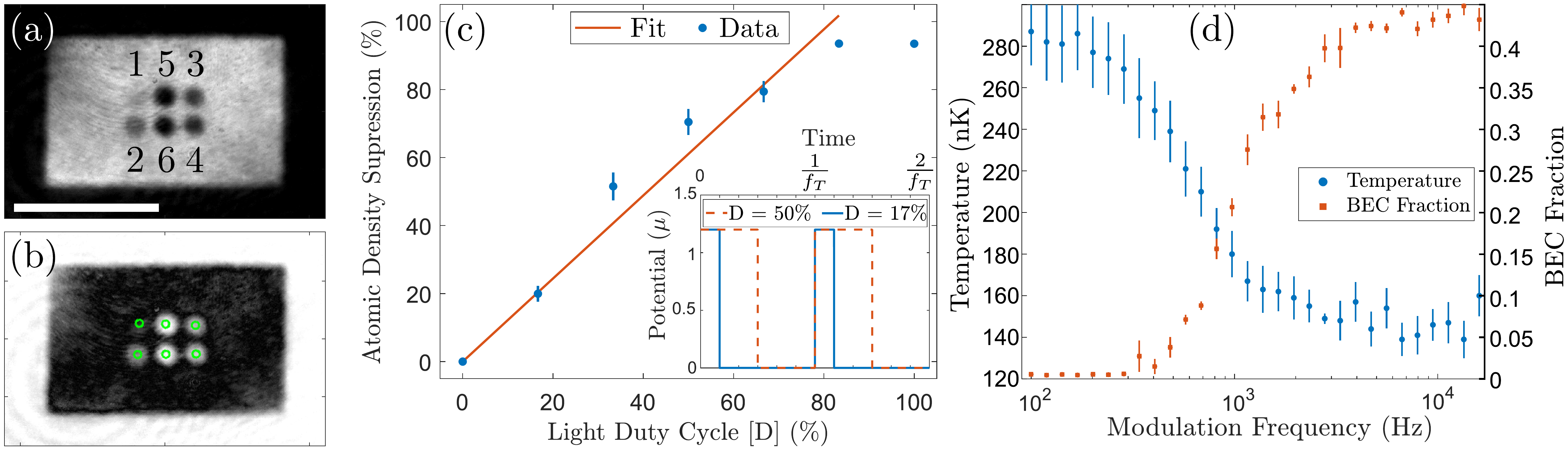}
\caption{(a) Grayscale blue-detuned time-averaged pattern applied to the atoms, averaged over 10 sequential images. Each of the numbered patterns correspond to the fractional duty cycle of 2.75~kHz time-averaging frequency $f_T$, with 6 being the maximum, corresponding to a repulsive potential barrier of $1.2~\mu$. The scale bar represents 50~$\mu$m. (b) Images used for analysis of the gray levels. The grayscale image was subtracted from the average atomic background with no barriers present. The mean density and standard deviation were calculated over the indicated circles. (c) Gray levels achieved through this process. A best fit least squares regression line (LSRL) is indicated. As the optical potential of point 6 exceeds the condensate chemical potential it was excluded from the fit. Image background and untrapped atoms result in an apparent maximum density suppression of 96\%. Inset shows example pulse width modulations, corresponding to patterns 1, the minimum pulse, and 3, a 50\% duty cycle. The time-averaging frequency $f_T = 2.75$~kHz and DMD-SLM frame refresh frequency $f_r = 16.5$~kHz divisions are indicated. (d) Turning on and off barrier 3 only, with varying frequency, for a total modulation time of 500~cycles provides an estimate of the heating rate from mirror switching. Rates above $\sim3$~kHz appear to have a negligible heating effect over this modulation period. Adapted with permission from \cite{gauthier_direct_2016} \textcopyright{} The Optical Society.}
\label{fig:TimeAvgeraging}
\end{figure*}

As discussed in Secs.~\ref{sec:Introduction} and \ref{sec:AOD}, time-averaged potentials can be achieved for atomic dipole potentials through rapid modulation of the optical field~\cite{Schnelle:08, henderson2009experimental, bell2016bose}. Using a DMD-SLM, the potential experienced by the atoms can be changed faster than the rate at which the atoms can react. They then experience an effective time-averaged potential described by \eqnreft{eq:Time_Averaged_Potential}{}.

Time-averaging can also be used for the creation of grayscale potentials with the DMD-SLM. This is done by controlling the fraction of the cycle each micromirror spends in the `on' state, see~\figo{fig:TimeAvgeraging}. Assuming a device with a refresh rate of $f_{r}$ and $N$ images being time-averaged together, the time-average frequency becomes $f_{T} = f_{r}/N$. The mirror can be on from $0$ up to $N$ frames which allows for time-averaged grayscale potential with $N$ levels. The maximum number of frames will depend on the level of heating of the atoms that is acceptable for a given application, which is also highly dependent on the mass of the atomic species for which the time-averaged trap is being created. This method is fully compatible with the halftoning method previously discussed, meaning that $M$ tones of halftoning combined with $N$ time-averaged images leads to $N \times M$ tones for the time-averaged grayscaled potential.

As was discussed previously in Sec.~\ref{sec:ScanningFrequencyRequirements}, kinetic energy can be gained by the atoms due to non-uniform phase imprinting, which induces micromotion. The steady-sate phase evolution of the groundstate wavefunction conforms to the time-averaged potential. Non-uniform perturbations of the phase distribution away from the steady state solution are given by \eqnreft{eqn:phaseDensEqns}{} while looking solely at the perturbations relative to the time-average potential $V(\mathbf{r},t) \rightarrow V(\mathbf{r},t) - \overline{V}(\mathbf{r},t)$. In essence, this treats the error in the individual frames relative to the time-averaged potential as a perturbation that evolves the phase relative to its steady state value. Any non-uniformity in the difference will lead to a phase gradient and micromotion of the atoms given by \eqnreft{eq:CondensateVelocityField}{}. The higher the instantaneous velocity, the weaker the assumption of the kinetic energy term in the GPE being irrelevant leading to energy transfer and eventually heating. 

There are a few strategies which can be employed to minimize energy transfer during time-averaging. The most obvious is to increase the frequency at which the frames are time-averaged, as this will reduce the maximum relative phase gradient imprinted. The second temporal method is to put frames with opposite gradient errors next to each other in the time-averaged sequence of frames. This will reduce the maximum velocity achieved by the atoms thereby reducing micromotion and heating. 

An additional method is to pick the spatial distribution of the on and off mirrors in such a way so that it minimizes the RMS gradient between the time-averaged potential and the potential created by the individual frames. This ensures that a minimal amount of anomalous phase is imprinted during each frame. There are two main approaches to accomplish this. The first is a judicious placement of the mirrors in the `on' and `off' state so as to minimize the difference between the time-averaged potential and the instantaneous potential~\cite{PhDGauthier2019}. The second method is to minimize the maximum possible gradient in the created potentials by reducing the spatial resolution of the system, which leads to decreased phase gradient imprinting by sacrificing the possibility for sharp features. This is useful for target time-averaged potentials which are smoothly varying~\cite{bell2020PhD}, but the optimal point-spread function trade-off between sharp features and low heating rate needs to be determined on a case by case basis. Combinations of each of the above methods can be used in conjunction to minimize heating and micromotion. One parameter can be used to reduce the requirements on another, for example, the smaller the instantaneous erroneous phase gradient imprinted, the lower the time-averaging frequency requirement will be. In the limit that each of the frames is equal to the time-averaged potential the time-averaging frequency will tend towards zero. 

\begin{figure*}[t!]
\includegraphics[width=\textwidth]{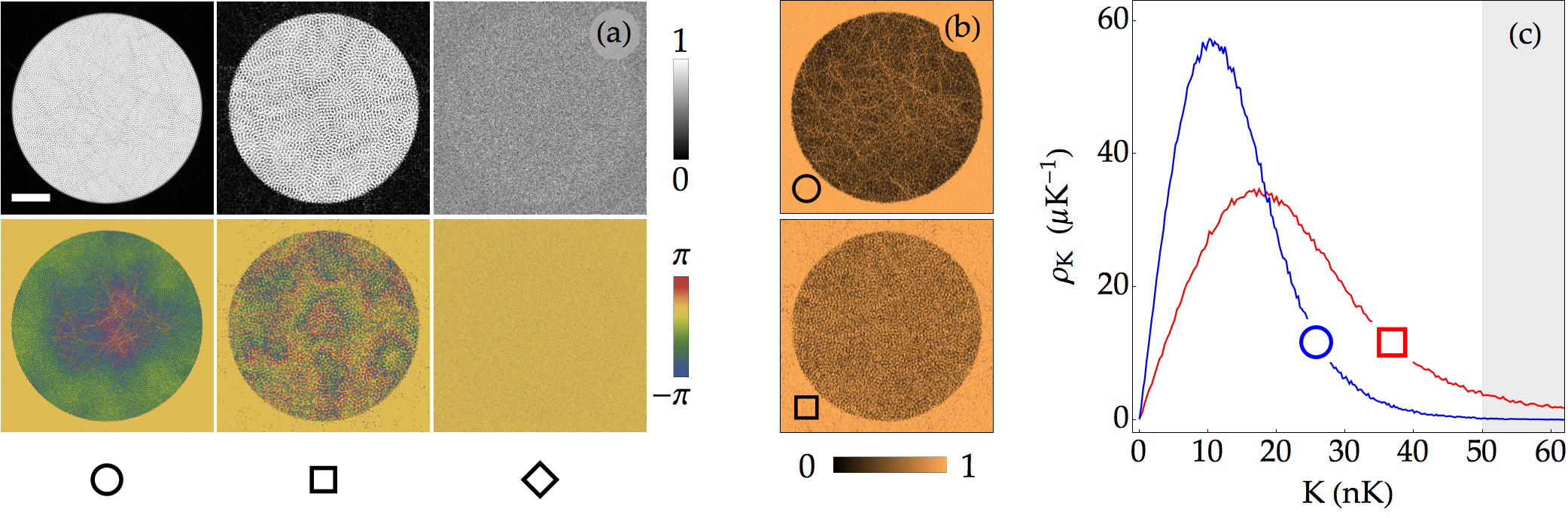}
\caption{Numerical determination of the requirements for time-averaged confinement using scanned planar reservoir potentials. The repulsive DMD-SLM potential has an R = 45~$\mathrm{\mu}$m radius, 100~nK depth and the simulated optical system has a 0.65~$\mathrm{\mu}$m FWHM spatial resolution. Time-averaged potentials of 50~nK (50 \%) depth are created which are made up of 10 random halftone DMD-SLM patterns. (a) Condensate density (top) and phase (bottom) distribution for frequencies $f_T$ = 2.222~kHz (circle), 1.068~kHz (square) and 0.514~kHz (diamond). The scale bar has 20~$\mathrm{\mu}$m length. (b) The instantaneous kinetic energy spatial distribution $T(\vect{r},t)$ given by \eqnreft{eqn:KenSingle}{}. (c) The corresponding kinetic energy probability distributions, with energies exceeding the time-averaged potential depth shaded. Reproduced with  permission from \cite{bell2020PhD}.}
\label{fig:DMDTimeAverage}
\end{figure*}

\figureo{fig:DMDTimeAverage} shows numerical GPE modeling of the density and phase distributions, and the kinetic energy probability distribution, of a 50\% half toned potential for different time-averaging frequencies. The halftoning pattern is uniformly random, meaning the probability of any mirror being on is 0.5. Three different time-averaging frequencies are considered, 0.514~kHz, 2.222~kHz, and 1.068~kHz. The lower time-averaging frequency of 0.514~kHz generates too much micro motion to trap the atoms. As for 2.222~kHz and 1.068~kHz, the expected decrease in the kinetic energy spectrum mean energy is observed as the time-averaging frequency increases, due to suppression of the anomalous phase imprint. Atoms which have a kinetic energy above the trapping depth will tend to be lost from the trap resulting in heating-induced losses.

Due to the imprinted phase, \eqnreft{eqn:phaseDensEqns}{}, being independent of atomic mass, but the velocity field being inversely proportional to the mass, \eqnreft{eq:CondensateVelocityField}{}, the transferred kinetic energy due to phase imprinting has an inverse dependence on mass. This means that the time-averaging frequency needed to prevent heating will be dependent on the atomic species mass. Insignificant heating was found when time-averaging above 3~kHz for trapping of $^{87}$Rb~\cite{gauthier_direct_2016}; on the other hand, when tapping atoms with $^{6}$Li technical noise at $f = 9.5$~kHz caused by DMD-SLM mirror refreshing reduced the lifetime of the atoms by more than a factor of $\sim 80$~\cite{klaus2017note:}. Complex time-averaged optical traps have been demonstrated using DMD-SLMs by ~\citet{gauthier_direct_2016}.

\subsubsection{Binary Fourier plane holograms with DMD-SLMs}
\label{sec:BinaryFourierImaging}
The general aspects of Fourier imaging are presented in Sec.~\ref{Sec:Fourier_Transform_General}. Here we describe the specific considerations that come from using a binary amplitude mask device, a DMD-SLM, to encode the calculated complex-valued optical field onto an incident wave front to obtain the desired trap.

In a direct approach there are two main ways to represent a complex-value field on the DMD-SLM. The first is by redefining the zero such that it represents a 50\% DMD-SLM light level~\cite{lerner_shaping_2012}; this way the real value of the complex field lies in [0, 1] range. This mapping is then combined with halftone techniques such as the Floyd-Steinberg error diffusion algorithm to create a grayscale map of the hologram~\cite{holland_benchmarking_2018}. The drawbacks of this technique are the reduction in the amplitude range control by a factor of two and inclusion of a background signal due to the offset. Another approach to this problem is to use phase control, where negative values represent a $\pi$ phase shift~\cite{zupancic_dynamic_2013, Goorden2014Jul}.

\begin{figure}[t!]
\includegraphics[width=0.85\columnwidth]{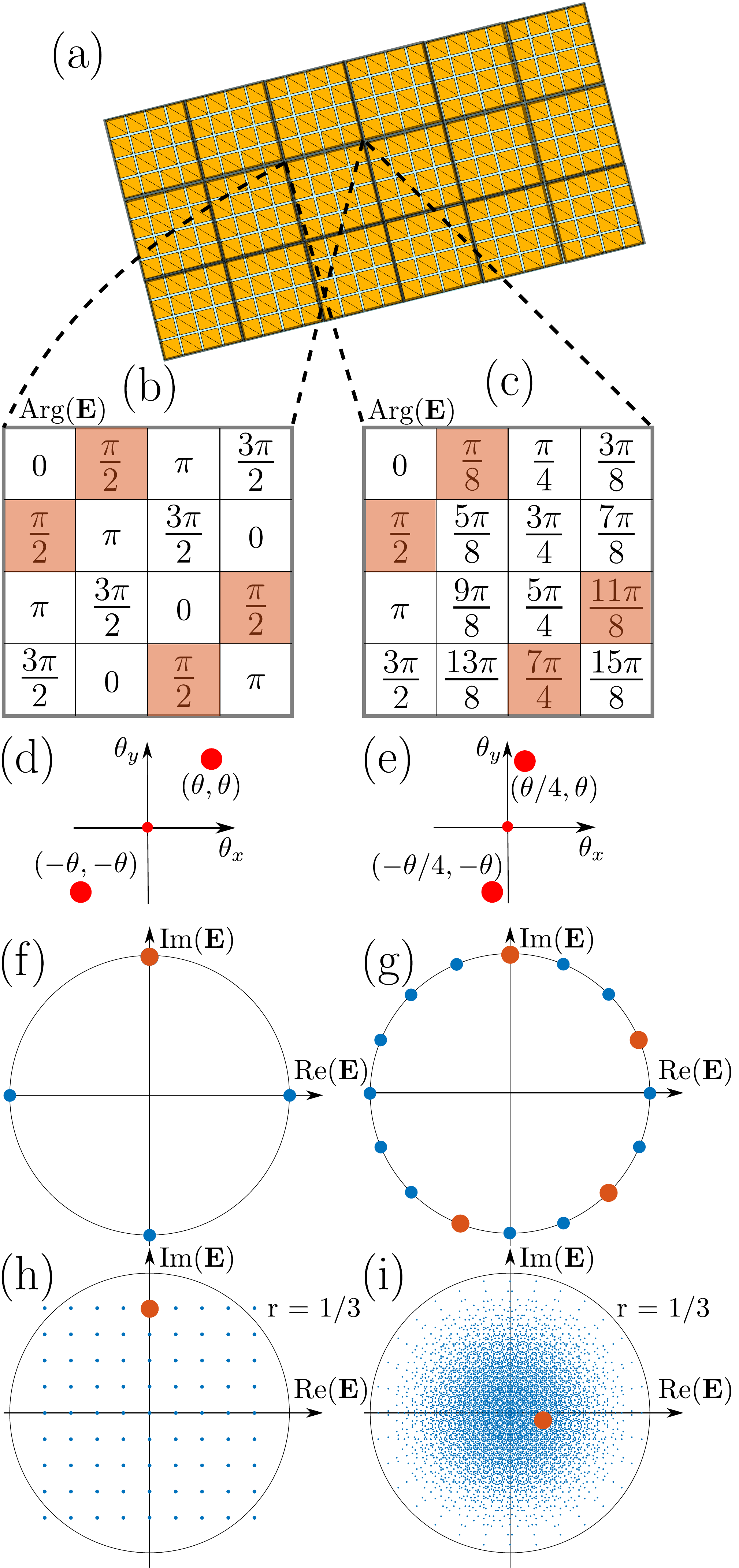}
\caption{DMD holographic superpixels decomposition~\cite{Goorden2014Jul}. (a) DMD micromirror array with effective holographic superpixels depicted inside black edged boxes for $H = 4$. (b,c) Relative electric field phases for each individual micromirror making up the holographic superpixel. The pixel phase mapping in (b,c) depends on their sampling angle which are depicted in (d,e) for the two respective cases. (d,e) Imaged diffraction order angle (large red dots), relative to DMD diffraction order (small center red dot) for (b,c), respectively. The angle is given by $\theta = \theta^{\text{max}}_{n,1}-\theta^{\text{max}}_{n,0} \approx \lambda/Hp$, with $\theta^{\text{max}}_{n,h}$ given by \eqnreft{eq:Holographic_Order_Location}{}. To obtain the staggered phasor of (c) the sampling must be done off-diagonal at $\theta_x = \theta/H = \theta_y/H$  (f,g) Imaginary plane projections of the electric field resulting from individual pixels depicted in (b,c), respectively, with the large orange dots corresponding to the shaded pixels. (h,i) All possible 81, 6561 unique holographic electric fields resulting from binary combinations of pixels from (b,c) respectively, with large orange dots representing the field resulting from the sum of the shaded pixels in (b,c).}
\vspace{-1em}
\label{fig:DMDFourierPhasorMapping}
\end{figure}

One might assume that using a binary amplitude control device precludes fine phase control for the correction of errors in the incident light field and in the transmission system. In actuality, fine control can be realized through holography, which takes advantage of destructively interfering holographic diffraction orders. Holographic diffraction orders are those formed by conceptually combining micromirrors into larger superpixels made up of $H \times H$ micromirrors, artificially increasing the spacing of the diffraction grating, see \figreft{fig:DMDFourierPhasorMapping}{a}. The holographic orders will be located at angles
\begin{equation}
\theta_{n,h}^{\textrm{max}} = \arcsin{\left[\frac{2(n+h/H)\pi}{kp}+\sin{\alpha}\right]},
\label{eq:Holographic_Order_Location}
\end{equation}
where all the parameters are identical to \eqnreft{eq:ConstructiveInterferenceAngle}{}, with the exception of the holographic order $h\in [\lceil -H/2 \rceil,\lfloor H/2 \rfloor]$, and superpixel length/width $H$.

The holographic orders are suppressed by destructive interference between the diffraction orders associated with opposing pixels within the $H \times H$ groupings. For these holographic orders, spatial locations of the micromirrors are associated with different path lengths and therefore electric field phases, which permits phase control. Imaging the first holographic diagonal order at $\theta_x = \theta_y = \theta_{n,1}^{\textrm{max}}$, shown in \figreft{fig:DMDFourierPhasorMapping}{d}, associates each pixel in the holographic superpixel with
the electric field phase depicted in \figreft{fig:DMDFourierPhasorMapping}{b}. These give four possible electric fields in the imaginary plane, \figreft{fig:DMDFourierPhasorMapping}{f}.
Binary combinations of holographic pixels can form 81 unique electric fields for the holographic superpixel, depicted in \figreft{fig:DMDFourierPhasorMapping}{h}, provided the individual DMD micromirrors cannot be resolved. 

To create a more diverse electric field using the same number of pixels, the light from the off diagonal diffraction order located at $\theta_x = \theta_y/H = \theta_{n,1}^{\text{max}}/H$, as proposed by \citet{Goorden2014Jul} and shown in \figreft{fig:DMDFourierPhasorMapping}{e}, can be imaged instead. This different sampling angle leads to a uniformly increasing electrical field phase without phase wrapping over the holographic superpixel, \figreft{fig:DMDFourierPhasorMapping}{c}, and in the imaginary plane, see \figreft{fig:DMDFourierPhasorMapping}{g}. This leads to a drastic increase in the possible electric field phasors that can be generated by the same holographic superpixel, depicted in \figreft{fig:DMDFourierPhasorMapping}{i}, changing from 81 to 6561 for the example case of $H = 4$ shown. This method is equivalent to creating a $H_x^2 = H_y^2 = H^2$ holographic pixel and sampling both in the first order $h_x = h_y = 1$, but treating them as $H$ separate holographic superpixels along the x-direction, since each superpixel has the complete $[0, 2\pi]$ phase mapping. A subtlety worth noting is that the location of the zero global phase shifts with period $H$ from holographic superpixel to holographic superpixel when traveling along the finer phase sampling direction.

Irrespective of the mapping method, holographic superpixels ultimately sacrifice spatial resolution to obtain phase and half toned amplitude control from a binary amplitude array. The larger the holographic superpixels are, the finer the phase and amplitude control becomes at the cost of a lower spatial resolution of the overall reconstructed light field, which means that local phase error will no longer be correctable due to the integration of the phase error over the holographic superpixel. The trade-offs between finer local phase correction and better phase control for a DMD-SLM have been investigated in~\cite{zupancic_dynamic_2013}, but will ultimately depend on the imaging system being built.

To obtain the maximum theoretical power that can be attained in an holographic diffraction order, all of the constructively interfering pixels in that order must be turned on, so that $0 \leq \text{Arg}(\bf{E}) < \pi$, and all the other pixels are turned off. The maximum theoretical diffraction efficiency for a given holographic order $h$ is $\eta_{H} = \textrm{sinc}^2\left(h\pi/2\right)/4$. Phase holography cannot be performed in the $0^\textrm{th}$ order since all light diffracted from the grating interferes constructively in this order. The best order which allows for amplitude holography and maximizes the diffraction efficiency is the $\pm 1^{\mathrm{st}}$ order, which both have maximum diffraction efficiencies of $1/\pi^2\approx 10\%$, irrespective of the holographic superpixel size. 

Due to imperfections in the manufacturing of the DMD-SLM, the micromirror plane tends to be distorted which means that phase error in the holographic field ensues. For Fourier plane holography, it is important to characterize these distortions and correct for them~\cite{zupancic_dynamic_2013}. A local phase correction is added to the final value of the complex valued calculated light field to correct for the induced aberrations. It can also correct for overall aberrations of the optical system.

Algorithms exist that will take advantage of the freedom of the electric field phase at the atom plane, due to the intensity being the important metric for trapping. These degrees of freedom allow for the optimization of the output intensity profile through iterative means, achieving the desired intensity profile with smooth features and a high diffraction efficiency. The smoothness is important due to small-scale imperfections leading to phase defects detrimental to atomic interferometry~\cite{leanhardt_propagation_2002, fortagh_surface_2002, leanhardt_bose-einstein_2003, esteve_role_2004}. These imperfections can also adversely affect the study of quantum many-body physics through Anderson localization~\cite{damski_atomic_2003}. For binary representations of a hologram, a direct binary search is one method which can be used to optimize the hologram representation on the DMD-SLM~\cite{michael_digital_1987,legeard_multicriteria_1997}. Due to its high-computational cost, direct search algorithms have fallen out of fashion and instead iterative Fourier transform algorithms (ITFA) are usually used. ITFA consist of projecting a guess modulated field through the system using fast Fourier transforms and then modifying the guess through successive iterations, based on the error between the target and calculated fields. The commonly used variants of this algorithm are the Gerchburg-Saxton (GS) and Adaptive-Additive (AA) algorithms~\cite{wyrowski_diffraction_1990, ripoll_review_2004, andrews_structured_2011}. For cold atoms which are trapped in a plane, where no three-dimensional shaping of the trap is necessary, a more robust approach has been shown to be mixed-region amplitude freedom method (MRAF)~\cite{pasienski2008high, holland_benchmarking_2018}. Iterative methods to beam shaping are an important component to phase based SLMs and are discussed in more detail in  Sec.~\ref{sec:IterativeSolutionToBeamShaping}.

Fourier plane imaging of a DMD-SLM was used to set the initial motional state of the atoms in an optical lattice in a study of strongly correlated quantum walks~\cite{preiss_strongly_2015}. It was used to initialize atoms on a sub-region of an optical lattice to study of the thermalization of quantum many-body systems through entanglement~\cite{kaufman_quantum_2016}. Adding a second DMD-SLM to  project a quasi-periodic disorder potential allowed for the probing of entanglement in a similar many-body system~\cite{lukin_probing_2019}. It was used to restrict the evolution of atoms in a lattice to a $N \times 2$ sub-region and explore the interacting Harper–Hofstadter model in the two-body limit~\cite{tai_microscopy_2017}. It was used to create a mesoscopic lattice with tunable number of sites to explore the emergence of band structure in fermionic transport~\cite{lebrat_band_2018}. This technique was also used to create Bragg beams with tunable relative incident angles which allowed for the probing of the excitation spectrum of rotons in dipolar Bose gases~\cite{petter_probing_2019}.

\subsubsection{Potential Correction Methods (feedback)}
\label{sec.DMDFeedback}
Cold atoms, due to their isolated nature, and the precision measurement and manipulation techniques available, are a versatile system where long time quantum evolution can be coherently observed~\cite{bloch_many_2008,langen_ultracold_2015}. Precise control of the potential is required when trying to gain access to previously unobserved phenomena. Homogeneous systems can unveil phenomenons suppressed by inhomogeneous densities~\cite{rauer_recurrences_2018,gauthier_giant_2019,johnstone_evolution_2019,stockdale_universal_2020} or masked~\cite{gaunt_bose_2013,mukherjee_homogeneous_2017,hueck_two-dimensional_2018}. 

To precisely control the final trapping potential, it is sometimes impossible to know \emph{a priori} the exact shape of the potential which needs to be projected, due to optical aberrations in the projection system, or the unknown quality of the laser beam illuminating the SLM. To circumvent these factors, one can use feedback which relies on the direct and/or indirect measurement of the final potential, along with the control over the input potential, to correct the projected potential until the target intensity is achieved.

To measure the final potential two main methods can be used, the direct and/or indirect method. The direct method consists of direct imaging of the DMD-SLM shaped light pattern somewhere else in the system from which the projected shaped light pattern can be feedback. In the indirect method, the atomic density distribution is imaged from which the underlying potential can be approximated.

The direct method requires either the implementation of an independent imaging camera which is used exclusively to image the light intensity pattern or control of the atomic imaging path to be able to focus both, the atoms resonant light and the trapping far-detuned light, on the same camera. This method is quick since it does not rely on the duty cycle of the experiment for feedback which is usually the limiting factor for the speed of the feedback in the indirect method. A drawback of the direct method is correcting for imperfections in the projected potential. Any imperfections due to the vertical/radial trapping method will needed to be accounted for independently. 

The indirect method is usually done using the atomic imaging system. Once the atomic distribution is imaged a mapping is required to extract the underlying potential from the measured atomic density, this is usually done using Thomas-Fermi approximation~\cite{PhDGauthier2019,tajik_designing_2019}, but more involved methods such as GPE ground state finding with numerical optimization can be used. The main drawback of this method is that the rate a which feedback can be performed is limited by the experimental sequence. 

Common to both indirect and direct feedback methods is that aberrations in the re-imaging system appear as features that need to be corrected, but will instead degrade the potential in the atom plane if accounted for.

Combining both methods is possible such that the majority of the correction of well known aberrations and target potential are done using direct method and fine tuning is performed using an indirect method.

Once the actual potential is measured using indirect or direct methods, a mapping is needed to map the spatial position between the measurement and DMD-SLM mirrors. This can be done using a fixed linear mapping~\cite{tajik_designing_2019}, by estimating the geometric transformation that best aligns measured potential and the expected potential~\cite{PhDGauthier2019}. The advantage of geometric transformation estimation is that it will make the feedback relatively insensitive to spatial drift between the imaging system and projection system. A down side is the extra computational time required and the added noise that comes from a transformation that varies from shot to shot which might slow down the convergence rate.

With mapping and measurement, the error of the potential relative to the target potential can be calculated. Taking advantage of halftoning and possibly time-averaging, one can lower or increase the local potential where needed. Many different approaches can be taken to choose which mirrors to flip. They usually fall under two broad classes, probabilistic~\cite{PhDGauthier2019} or deterministic~\cite{tajik_designing_2019} approaches. 

Feedback using DMD-SLMs has been done in one-dimensional system which allows for the sacrifice of spatial resolution along the direction perpendicular to the trap direction to increase the number of halftone gray-levels available. This is due to multiple mirrors along the low resolution axis contributing to the light amplitude at the atom location. This method has been achieved using a combination of magnetic micro-traps generated using atom chips and direct imaged DMD-SLM to create and correct imperfections in a range of one-dimensional trapping potentials~\cite{tajik_designing_2019}. Feedback has also been achieved in 2D system using halftoning and time-averaging to create a variety of trapping potentials~\cite{PhDGauthier2019}.

\subsection{Technical Considerations}
The factors to consider when choosing a DMD-SLM are the number of mirrors in the array, the switching frequency, and the diffraction efficiency of the array for the wavelength used for projection. 

The larger the feature size one is trying to achieve, the more important the diffraction efficiency and the damage threshold, typically 25~W/cm$^2$~\cite{texas_instruments_laser_power_2012}, since more power is required to achieve similar potential depths in larger patterns. The same is true of the detuning, the farther the detuning, the more power is needed to maintain the same potential depth. Although most of the projection efficiency comes from the DMD-SLM such as the diffraction efficiency, mirror reflectivity, window transmissivity and mirror array fill-factor, one should also consider the transmission efficiency of the projection system itself which comprises of multiple lenses/mirrors and transmission through the vacuum glass. As with every application, it is good practice to include a safety factor of $\sim 2$, since it is easier to turn down the laser power than to increase it. It is worth noting that the stated diffraction efficiency in DMD-SLM manual is usually for broadband visible lasers with a lens collecting light from multiple diffraction orders. For most cold atoms applications one is interested in narrow band single order diffraction efficiency which can be approximated using \eqnreft{eq:DiffractionEfficiencyWithUncertainty}{} for direct imaging and is shown for DLP9500 \& DLP9000 in \figo{fig:DMDDiffractionEfficiency}.

The number of micromirrors the DMD-SLM has is important, in the Fourier plane it will determine the frequency resolution and range which can be used to create the image. For a low number of mirrors the user will need to decide between large spatial extent or sharp features. But with more mirrors it is possible to obtain both large spatial extent with sharp features, or the extra resolution can be leveraged to create sharper features. For direct imaging, the higher the number of mirrors the more gray levels will be available when performing halftoning, leading to more precise sculpting of the potential without the need to sacrifice the sharpness of the projected features. If insufficient number of mirrors reside within a resolution element of the imaging system (PSF), then trade offs between the number of halftone levels and the sharpness of the features through the PSF will ensue; determining the best compromise will be application dependent.

The importance of the switching rate will be application dependent. For cases where time-averaging is considered, a higher rate will lead to the possibility of more time averaged frames, or to the same number of averaged frames but displayed at a higher frequency leading to a lower heating rate. In the case where phase imprinting is the goal, being able to imprint on a time scale that is short compared to the response of the atoms is going to be important, so the smaller the atomic mass of the trapped species the faster the switching rate requirement will be to meet this criterion. For applications where switching will be used to generate dynamics in the system, a safe margin is for the switching rate to be an order of magnitude faster than the frequency along which the dynamics is being generated (see Sec.~\ref{sec:ScanningFrequencyRequirements}). If all other requirements are met, faster switching rates are nonetheless preferable. For applications where the atomic mass is small, the ability to produce truly static potentials which do not heat up the cloud is paramount, and a procedure which bypasses the DMD-SLM clock, latching the mirrors and preventing kHz switching noise, is described in \cite{klaus2017note:}.

Some less critical considerations are the number of frames the controller can upload, since it is not possible for computers to upload faster than the rate at which they can be displayed. This means the frames to be displayed must be uploaded to the controller before the sequence of images are displayed. The larger the number of frames, the longer the sequence of unique images which can be displayed by the DMD-SLM. Controllers usually come with a sequencer which allows the user to define how the images stored in the controller memory will be displayed. This enables the programming of triggers, loops, and waits which allows the same image to be displayed at multiple different times during a sequence, for external control of the image display sequence, and for control of image display duration.

One of the factors to consider in terms of thermal power management is the thermal load placed on the DMD-SLM by the light, as some will be absorbed due to the array fill factor, the partial reflectively of the mirrors and the overfilling of the array to flatten the illumination. To prevent over-heating of the device, even when operating below the damage threshold, care must be taken to actively cool the DMD-SLM array through the thermal back plate.


\section{Liquid Crystal Devices}
\label{sec:SLMs_details}

\begin{figure} 
\includegraphics[width=0.99\columnwidth]{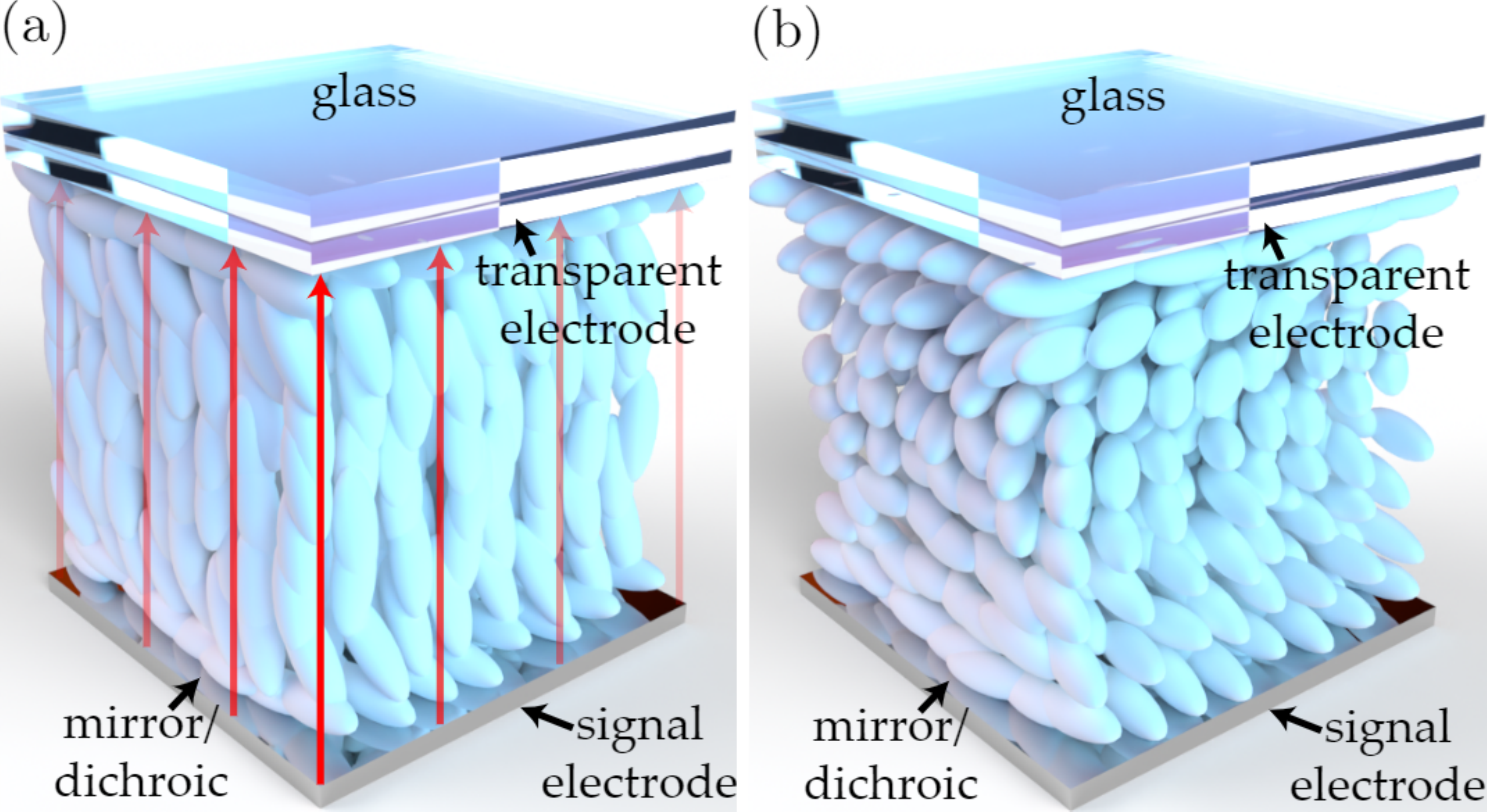}
\caption{Two examples of electrically addressed nematic liquid crystal cells. In reality, nematic liquid crystals are not organized in layers as suggested here (though smectic LCs are), but are distributed in a disordered state to fit the available space. The long axis of the ellipsoids represents the long axis of the liquid crystal molecules. (a) Nematic liquid crystals align with electric fields (red arrows). In the case of parallel at surface alignment, the crystal rotates between a birefringent and non-birefringent state for normally incident light. (b) The twisted nematic liquid crystal is shown in its resting state.}
\label{fig:slmcells}
\end{figure}

The use of liquid crystal SLMs for beam shaping and control has been driven in part by significant advances in technical capability and price reductions from the development of consumer display technology \cite{Barnes1989Nov}, e.g. digital clocks and television. Here we give a general introduction to the use of liquid crystal SLMs for light shaping, and a brief review of key literature discussing their applications.
To explain the detailed operation of a liquid crystal spatial light modulator, we must discuss the function of liquid crystals and more specifically their ordering.
Liquid crystals are room temperature liquids consisting of anisotropic molecules with the ability to form crystal-like structure due to the complex interplay between electromagnetic and hydrodynamic forces \cite{Stephen1974}. There are several distinct crystal-like alignments or phases that arise from the free energy of each liquid. The local measure of the orientation of the long axis of a liquid crystal is called the director. Several varieties of liquid crystals can be categorized by their resting structures, including: nematic (thread-like), smectic (layered), and cholsteric (chiral, twisted). Several of the configurations for nematic liquid crystal SLMs are determined by their interaction with substrates which fundamentally affects their usability and method of operation \cite{deBougrenetdelaTocnaye1997Mar, Khoo1993}. There are three commonly considered cases \cite{Stephen1974,Khoo1993}:
\begin{enumerate}
    \item perpendicular---where the crystal orients its long axis away from both surfaces,
    \item parallel---where the crystal orients its long axis with the surfaces [\figo{fig:slmcells}(a)],
    \item twisted---where in addition to the parallel case  the crystal  rotates from one parallel orientation to another [\figo{fig:slmcells}(b)].
\end{enumerate}
The ambient condition equilibrium alignment of a liquid crystal determines the effect that external electromagnetic forces have on the director. The result of modulating an electric and magnetic fields is that for particular angles of incidence the overall birefringence of the liquid crystal can be modified.

\begin{figure}
\includegraphics[width=0.99\columnwidth]{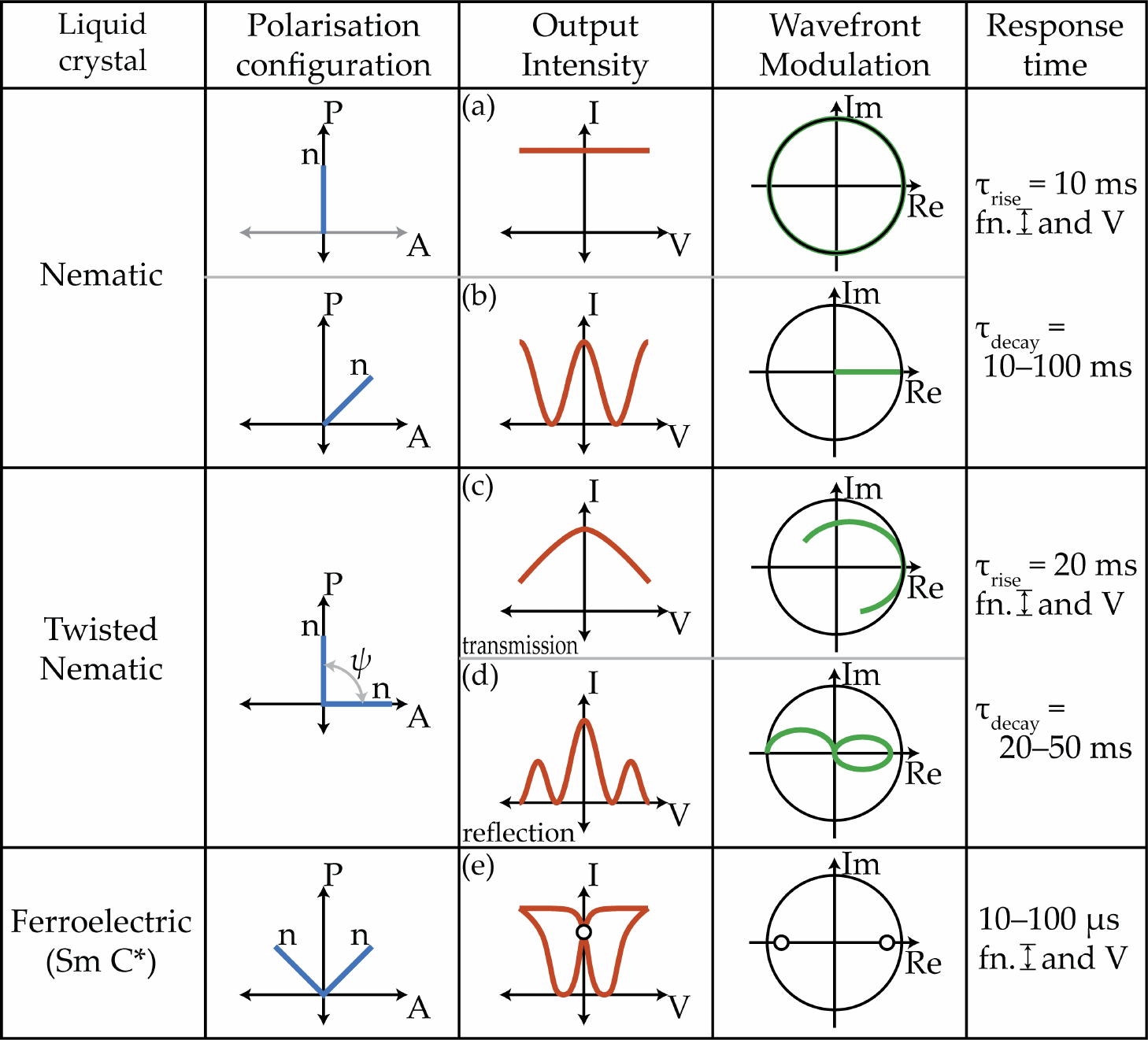}
\caption{The properties of liquid crystals in devices discussed in this review acting under voltage bias. (a) Nematic liquid crystal devices are typically operated using a single polarization state of light. (b) With incident light with polarization at $\pi/4$ to the long axis nematic liquid crystal modulate the light level when coupled with an orthogonally oriented analyzer. (c) Twisted nematic liquid crystals act more like variable waveplates with the polarization and phase being coupled together. (d) In reflection, twisted nematic liquid crystals get one complete twist instead of one half in transmission. They are thicker and are capable of accessing more states of polarisation and phase. (e) Chiral smectic C liquid crystals have two stable orientations and respond quickly between the two states under bias. Adapted with permission from \cite{deBougrenetdelaTocnaye1997Mar} \textcopyright{} The Optical Society.}
\label{fig:SlmTable}
\end{figure}

Most LCD-SLMs act as controllable polarization switches. \figo{fig:SlmTable} shows several types of liquid crystal displays that have been used. The variables in the figure are: the liquid crystal rise time $\tau_{\rm rise}$, fall time $\tau_{\rm decay}$, voltage $V$, irradiance $I$, polarizer state $P$, analyzer state $A$, thickness $\updownarrow$. A polarizer/analyzer optical system can be used to modulate radiant power between the polarization states even with a nematic liquid crystal device, \figo{fig:SlmTable}(b). A twisted nematic device on the other hand will have correlated changes in phase and amplitude, \figo{fig:SlmTable}(c). \figo{fig:slmcells} shows two potential configurations of liquid crystal used in a SLM: (a) A parallel aligned nematic liquid crystal with the electrodes in the `on' state, and (b) a twisted nematic liquid crystal in a relaxed, `off', state. In the first case, the nematic liquid crystals have their long axis aligned with the electric field between the signal electrodes and the transparent coverslip electrode. In the second case, the crystals spiral in a helix due to alignment with the surface coating. The director of the crystal will become more aligned with the field when the signal electrode is in the `on' state, changing the path length difference and polarization in a similar manner to rotating a waveplate.

A majority of the devices discussed in this section are controlled using direct electrode bias of the liquid crystal. We would like to point out use of a device occupying the transition between optically and electrically addressed modulators. An electrically addressable projector illuminating a photoconductive layer to bias a parallel aligned liquid crystal device was used to demonstrate precise positioning of single trapped atoms \cite{Bergamini2004Nov}.

Liquid crystal SLMs (LCD-SLMs) have great flexibility in controlling light. They can \emph{potentially} come with a few drawbacks that were foreseen in early experiments and techniques developed for cold atom experiments using electrically driven elements \cite{Boyer2004Sep}. Ferroelectric Liquid crystals move easily and are thus very sensitive to changes in the driving field. The frequency range and strength of the field affect the intensity stability and both are highly device dependent. More recent experiments with nematic type liquid crystals do not have significant intensity changes \cite{Haase2017}. Indeed, the LCD-SLM used there was very similar to that used in \cite{garcia2012flicker} which noted small intensity fluctuations, but the systems used different back plane drivers. In either case, cooling the liquid crystal will increase the effective mass and thus increase the time for the liquid crystal to respond to an applied field. Decay time to the resting state is also increased and that may be undesirable for some dynamic applications.  As we will show throughout these sections, these limitations are technical concerns and do not seem to limit the successful use and numerous applications of liquid crystal devices to cold and single atomic systems.

We outlined an example of complex modulation of light using the Fourier transform in Sec.~\ref{Sec:Fourier_Transform_General}. 
An AOD performs dynamical control of atoms and the generation of complex optical potential landscapes using a high-speed time scan. By contrast, a SLM can create these kinds of optical potentials after transformation through diffraction. It is thus a spatial counterpart to the temporal multiplexing from AODs. We will also show how liquid crystal SLMs may be used in imaging applications.

\subsection{Beam shaping with liquid crystal SLMs}
\label{sec:LCSLM_BeamShaping}
We will briefly discuss the usage of a SLM to shape the phase, amplitude and polarization of the beam of light.

The complex amplitude of a polarized paraxial beam of light at some fixed plane ($u,\,v$), traveling in the $w$-direction, can be represented with a function:
\begin{equation}
\label{equ:slmone}
\vec{U}(u,v)=A_u(u,v)e^{i \Phi_u(u,v)}\hat{u}+A_v(u,v)e^{i \Phi_v(u,v)}\hat{v}
\end{equation}
where $\vec{U}$ is the complex time-harmonic electric field, $A$ are amplitudes, and $\Phi$ is the phase with subscripts in either vector component. This expression represents a fundamental laser mode when the complete wavefunction is an eigenfunction of the Fourier transform operator. SLMs change the momentum of the wavefunction by modulating the amplitude, phase or polarization over its surface. The design of a device will usually mean that the path length differences produced by the liquid crystals can be represented as a linear operator or transfer function \cite{deBougrenetdelaTocnaye1997Mar} acting on a particular subspace of an optical field.

\subsubsection{Fundamental principles of Fourier mode of operation}
\begin{figure}[t!]
\includegraphics[width=0.99\columnwidth]{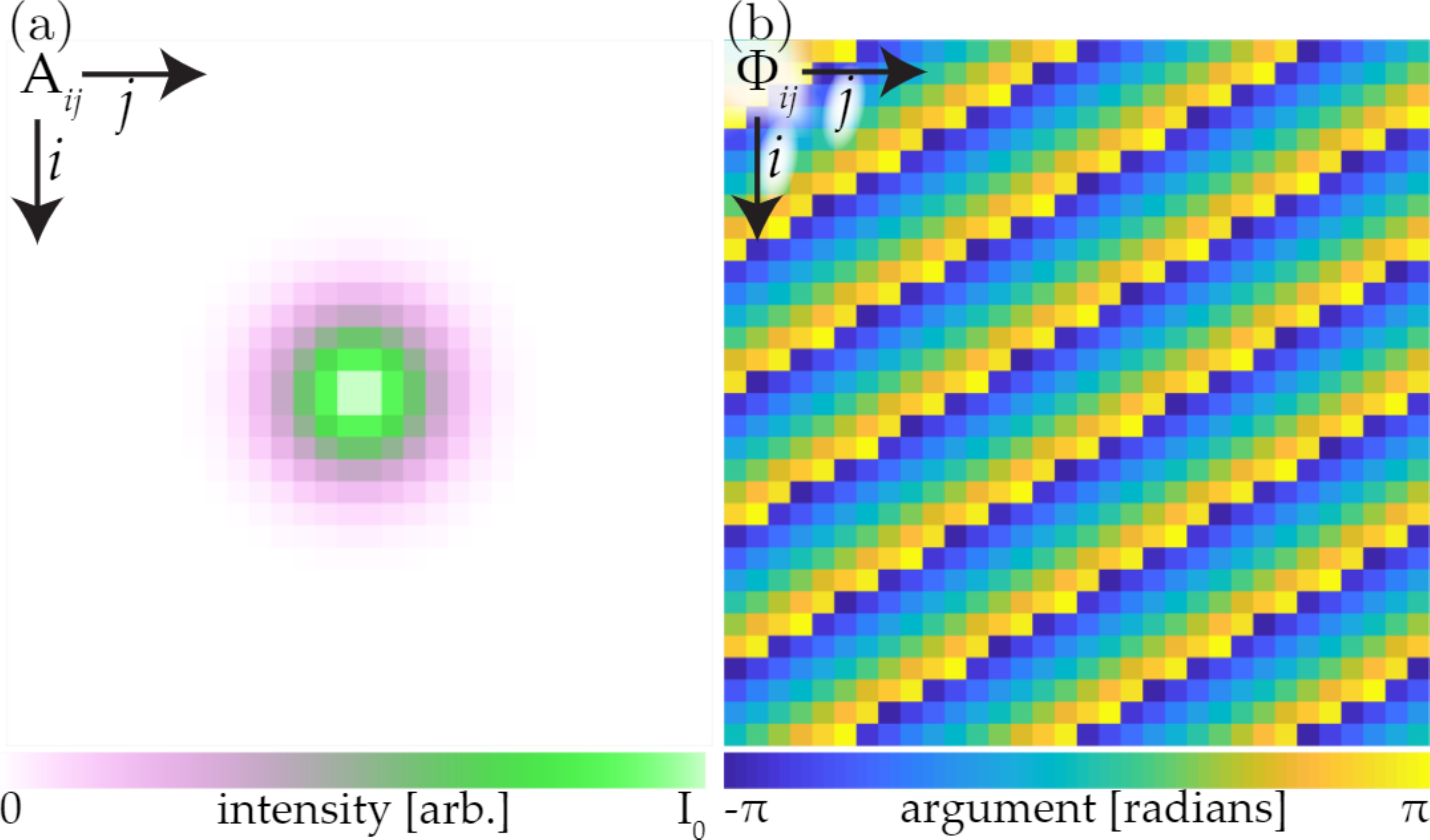}
\caption{Discrete approximation of a pixelated SLM with 32$\times$32 elements. (a) Amplitude of the incident beam on each element. (b) Imprinted phase designed to shift the beam in space.}
\label{fig:cheapslm}
\end{figure}

There are many examples in the literature of modulating a light field in order to project light to an image plane (near-field) from its Fourier conjugate (far-field). For a discrete element liquid crystal SLM capable of phase-only modulation, using one polarization component, the following discrete approximation to \eqnreft{equ:slmone}{} is assumed:
\begin{equation}
U_{ij}=A_{ij}e^{{i}\Phi_{ij}},
\label{fig:slmphase}
\end{equation}
where ${i,j}$ are the element index on the SLM, and $A_{ij}$ is complex mode amplitude of light striking the SLM, and $\Phi_{ij}$ is the phase shift caused by the SLM. Assuming that $A_{ij}$ is a real valued Gaussian mode profile then the result is that the discrete Fourier transform will have a point field with a Gaussian PSF. Some trivial examples of this mode of operation arise from considering the Fourier shift when $\Phi_{ij}=\mathbf{k}_{ij}\cdot \mathbf{r}$, where $\mathbf{k}_{ij}$ is a linear function of element position. In cylindrical coordinates, a shift along the beam propagation direction is, $k_z=\mathbf{k}\cdot \hat{z}= k\left[1-(k_{\rho}/k)^2\right]^{1/2}$, and thus achieve focusing/defocusing and in effect gain the ability to move the focal spot along the direction of beam propagation in addition to the transverse directions (2D discrete shift pictured in \figo{fig:cheapslm}).

\begin{figure}[t!]
\includegraphics[width = 0.85\columnwidth]{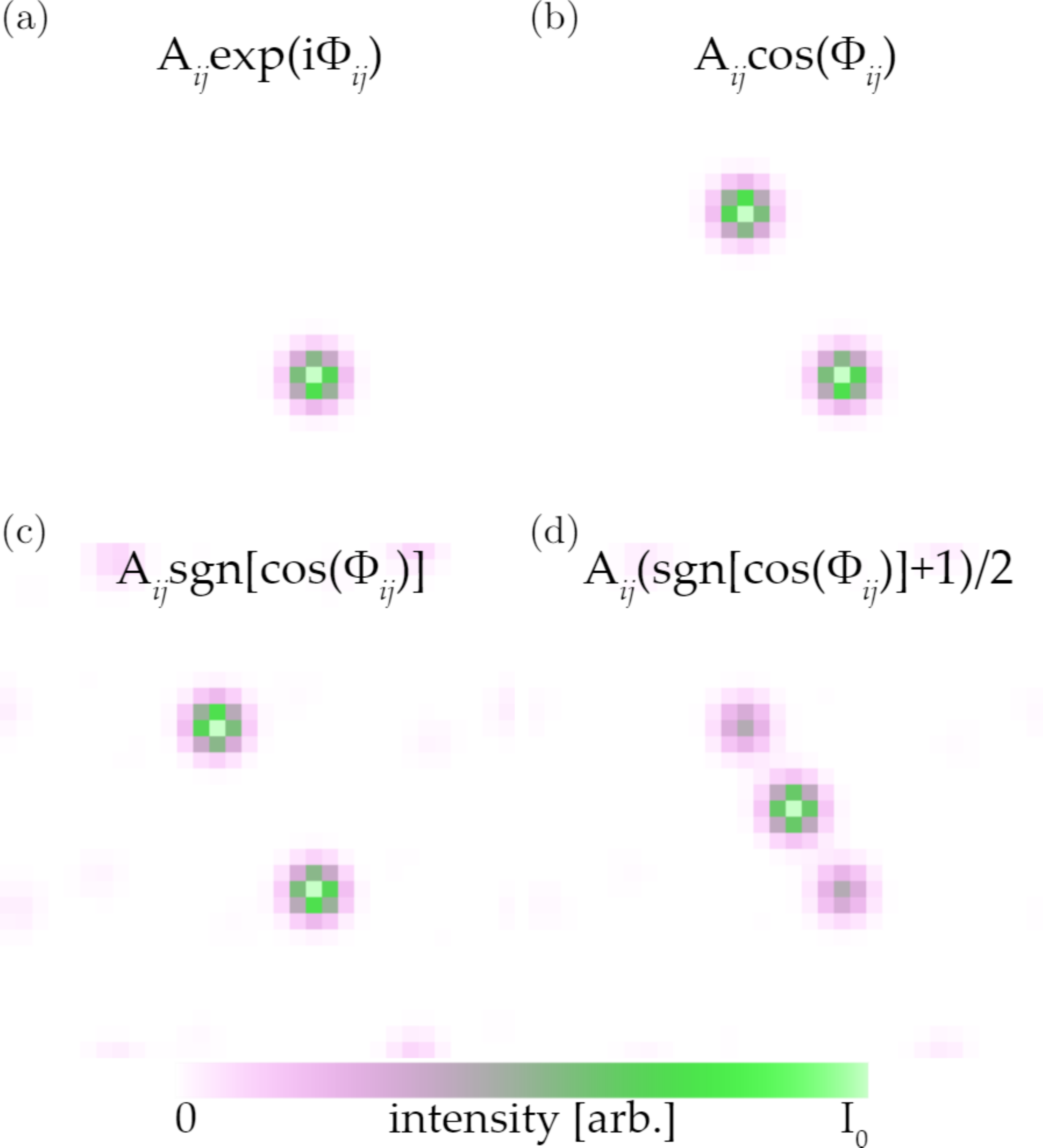}
\caption{The result of performing a DFT on variations of \figo{fig:cheapslm}. 32$\times$32 element realization of the labeled functions (a)--(d). $A_{ij}$ are amplitude at the device and $\Phi_{ij}$ is the phase for each pixel.}
\label{fig:cheapslm2}
\end{figure}

The same principle applies for other types of modulation. Take, for example, the following potential alternatives shown in \figo{fig:cheapslm2}; the panels respectively represent control over the unity magnitude complex ($-\pi<\arg e^{i\Phi_{ij}}<\pi$), reals ($-1<\mathrm{Re}\,e^{i\Phi_{ij}}<1$), binary phase ($[0,\pi]$), and binary amplitude ($[0,1]$). All produce at least the desired spot, but at the expense of the addition of high order harmonic distortions. Spatial filtering in the optical system is required to remove the harmonics to allow only the target mode to traverse the optical system with a corresponding reduction in diffraction efficiency. It is important to keep in mind that, even if a device has the ability to modulate a full wavelength shift (or more), these devices are typically operated using digital control electronics and thus the finite number of levels must be considered for an accurate mapping between any simulations, iterative optimizations, and the reality of a physical device projecting light. We discuss this further in the technical considerations of this section.

\subsubsection{Amplitude efficiency}
There are several kinds of related diffraction efficiencies to consider for light shaping with a SLM. The first arises due to the physical constraint associated with diffraction when projecting light between two planes with a thin pixelated device such as a SLM, which is essentially a variation of \eqnreft{eq:SingleDiffractionSlit}. The second is caused by the changes to the light field that the pixels of the device are capable of making, for example, the maximum phase shift produced by the device, absorption, reflection, duty cycle or phase stroke. We will introduce this  in a device/engineering independent manner here, before we describe other effects such as fringing in technical considerations, Sec.~\ref{sec:fringing}. 

\begin{figure}[t!]
\includegraphics[width=0.99\columnwidth]{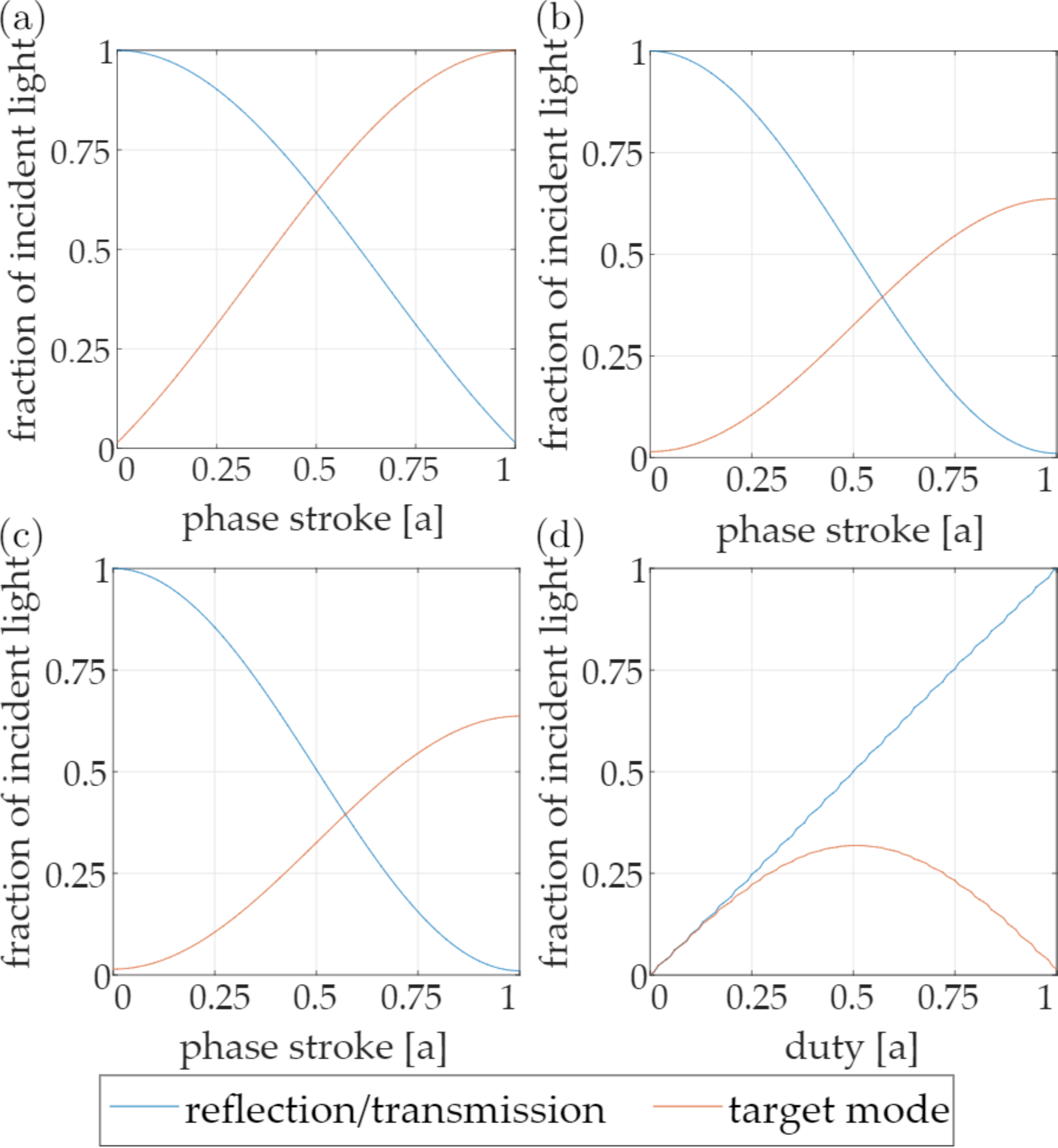}
\caption{Fraction of incident field amplitude in either the reflection/transmission (blue) or targeted diffracted mode (red) with stroke for higher resolution versions of functions corresponding to those in \figo{fig:cheapslm2}.}
\label{fig:diffreff}
\end{figure}

How much light can be diverted into particular modes? A Fourier transform of a non-eigenfunction will result in a new function with a distribution of power in the Fourier plane that is different at the device. This means that patterns may be mixed into the pattern displayed on the SLM to precisely control amplitude and phase at the expense of loss to the surrounding environment. The maximum amplitude that can be diverted to a particular pattern of amplitude and phase in a single step process is determined by the overlap between the Fourier transform of the ideal target mode and the transfer function and pattern on the beam shaping device \cite{Stilgoe2016May}. This simultaneously explains how one can precisely control fields, but also how efficiently this process can be achieved. As shown in \figreft{fig:SLMTwopaths}{b}, a change in the light in one plane  can result in a delocalized effect in another. Proceeding logically from the position that a device doing nothing simply reflects or transmits the light impinging on it, one can determine how changes to the phase stroke for a particular displayed pattern alters the efficiency. One of the simplest of these would be the shifting function introduced earlier, on the examples shown in \figo{fig:cheapslm2}. We can find the diffraction efficiencies as a function of a parameter, $a$, scaling the stroke ranges in amplitude and phase, shown in \figo{fig:diffreff}.
The curves in \figo{fig:diffreff} in the limit of a continuous function are
\begin{align}
\label{equ:diffra}\sqrt{\eta_{_\text{SLM}}({a,l})}&=\mathrm{sinc}\,{\pi(a-l)},\tag{\figreft{fig:diffreff}{a}}\\[1ex]
\sqrt{\eta_{_\text{SLM}}({a,l})}&=(2/\pi)^l\cos^2{\pi(a-l)/2},\tag{\figreft{fig:diffreff}{b,c}}\\
\label{equ:diffrd}\sqrt{\eta_{_\text{SLM}}({a,l})}&=\begin{cases}a, & l=0\\(1/\pi)\sin{\pi a}, & l=1\end{cases}, \tag{\figreft{fig:diffreff}{d}}
\end{align}
where $l=0$ for the reflection and transmission channel and $l=1$ for the target mode. The last case, when all mirrors are on, no target light is produced, subject to the diffraction efficiency that was previously derived. At maximum, half of the light will simply be reflected. As a general rule, the more conditions that are needed to be met in the Fourier plane, the less light can be diverted into those target modes. Rigorously, a diffraction integral will need to be solved for every pattern generated by the device; the data shown in \figo{fig:diffreff} is nonetheless a good rule of thumb estimate. The total diffraction will be a product of the first kind of diffraction previously discussed and that described in this section.

\subsubsection{Reshaping light fields}
One of the classic examples of beam shaping is the production of patterns of isolated spots that can be dynamically re-arranged to move particles in optical tweezers. The ideal function for $N$-uniquely placed spots is  \cite{Reicherter1999May,Barredo2016,endres2016atom},
\begin{equation}
    T_{ij}=A_{ij}\sum_{n=1}^N e^{{i}\Phi^{(n)}_{ij}}, \label{equ:Tij}
\end{equation}
where $T$ is the ideal field at the device. The problem is that the beam shaping device will not typically be able to produce this as it is the result of multi-beam interference. There are a few ways to proceed for a full phase only SLM, equation (\ref{equ:Tij}) can be approximated as $A_{ij}\exp{\mathrm{i}(\arg{\sum_n^N e^{{i}\Phi^{(n)}_{ij}}})}$ as is often done with holographic optical tweezers \cite{Reicherter1999May,Grier2003}. The consequence of this approach is the harmonic distortion such as the faint `phantom' spots displayed in \figo{fig:cheapslm2} are produced. Let us label this harmonic distortion in the projection plane as $\tilde{D}$. Occasionally one of the diffraction orders will be a harmonic of another and the interference from the harmonic distortion will change the intensity at that spot. To avoid this outcome a harmonic distortion or dither can be engineered into the field to expose the target modes, $\tilde{T}$, either approximately using an analytical approach \cite{kirk1971shaping,davis1999encoding,vanPutten2008,liang2010high,Goorden2014Jul,Stilgoe2016May,Lenton2020}, or with an iterative process \cite{Yang1994Jan,Dufresne2001Mar,DiLeonardo2007Feb,pasienski2008high,gaunt2012robust,Boyer2004Sep,nogrette_single-atom_2014,Bowman2017May,Lenton2020}, which may be more effective for complex patterns. Conceptually, this can be written for the device as:
\begin{equation}
    U_{ij}=\mathcal{T}^{-1}_{ij}\left(\tilde{T}+\tilde{D}\right),
\end{equation}
where $\mathcal{T}^{-1}_{ij}$ is the propagator onto the $ij$-th element of the device. Ideally, the magnitude of the inner product of field moduli $|\tilde{T}|,\,|\tilde{D}|$ is near or at zero which demonstrates minimized harmonic distortion. This will typically be the case, but some of the methods discussed use the appropriate phase shift of distortion at the spot location to increase intensity.
\subsubsection{Analytical approximations to beam shaping}
The price paid from trying to reproduce a facsimile of a target wavefunction or beam shaping, $\tilde{T}$, with a device not capable of the exact transformation needed is loss, $\tilde{D}$. The general expression to produce a particular transformed scalar potential is:
\[
U_{ij}=A_{ij}H_{ij},
\]
where $H_{ij}$ is the transfer or modulation function of the beam shaping device.

There are several approximate techniques by which the transfer function can be modified to project $\tilde{T}$ \cite{kirk1971shaping,davis1999encoding,vanPutten2008,liang2010high,Goorden2014Jul,Stilgoe2016May}. Considerations of these types of functions will amount to the amplitude diffraction efficiencies shown in the equations of \ref{equ:diffra}{} -- \ref{equ:diffrd}. An SLM with the freedom of choosing a particular phase will have amplitude diffraction efficiency equal to the equation shown in \ref{equ:diffra}{}. We thus must apply the inverse of these functions to appropriately tune the harmonic distortion. The scheme used in \citet{davis1999encoding} is appropriate for nematic liquid crystal devices (phase-only) accounting for the mode structure of the incident light and aberration. This effectively amounts to applying the following phase:
\begin{equation}
    \phi_{ij}=(1-c_{ij})\arg{T_{ij}}-\arg{A_{ij}}-\arg{X_{ij}},
\end{equation}
where $c_{ij}=\mathrm{sinc}^{-1}(\pi N|T_{ij}|/|A_{ij}|)$, $N$ normalizes the maximum of the argument to $\pi$, and $X_{ij}$ is the aberration. Usefully, the term, $(1-c_{ij})\arg{T_{ij}}$ in general is: $(\arg{T_{ij}}-c_{ij}\arg{M_{ij}})$ where $M_{ij}$ is any phase pattern distinct from $T_{ij}$, e.g. a position shift, etc. 

The work of \citet{vanPutten2008} (and older work on superpixels by, e.g., \citet{Birch2000}) was more general, splitting up a twisted nematic liquid crystal device into sets of pixels capable of independent addressing of amplitude and phase by finding points where the complex modulation that results from twisted nematics cancels the real or imaginary part of the phasor. This concept was later applied to DMD-SLMs \cite{Goorden2014Jul}, and equivalently in \citet{zupancic_dynamic_2013,zupancic_ultra_2016,Stilgoe2016May} by changing the relative position of fringe patterns. Provided a good refresh rate, these device patterns can be dynamically altered; we will discuss this further in the technical considerations section.

A smooth Laguerre-Gauss-type ring potential with angular momentum was imprinted on a $^{87}$Rb BEC in the near-infrared to create blue and far-infrared light using four-wave mixing, demonstrating trans-spectral orbital angular momentum transfer \cite{Walker2012}. To get an efficient output, the appropriate phase matching needed to be attained between the different wavelengths of light. They used a SLM to shape the two pump beams to better match the Gouy phase shift between the two near-infrared beams, enhancing the probability of the two photon excitation of the BEC into the four state superposition.

Gaussian beams focus to a single bright spot. On the other hand, a bottle beam or dark spot beam is created by engineering destructive interference at the focus to trap neutral atoms with repulsive light \cite{Chaloupka1997,Ozeri1999}. This scheme can be implemented with a SLM \cite{Arlt2000,Xu2010} and used for trapping of single atoms \cite{Xu2010}. More recently, bottle beams have been used to create 3D optical trapping potentials for confining Rydberg atoms~\cite{barredo_three-dimensional_2020}. In this work, $^{87}\text{Rb}$ atoms were trapped with $1/e$ lifetimes of 222~$\mu$s in the 84$S_{1/2}$ state without cooling. In the same work they subsequently demonstrated coherent state transport between two trapped Rydberg atoms by driving Rabi oscillations (using the dipole blockade mechanism \cite{Lukin2001}). This kind of scheme can be used for the purposes of preparing and simulating the interactions of two qubit states, and potentially greater numbers of entangled states in the future.

\subsubsection{Iterative solutions to beam shaping}
\label{sec:IterativeSolutionToBeamShaping}
\begin{figure}[t!]
\includegraphics[width=0.99\columnwidth]{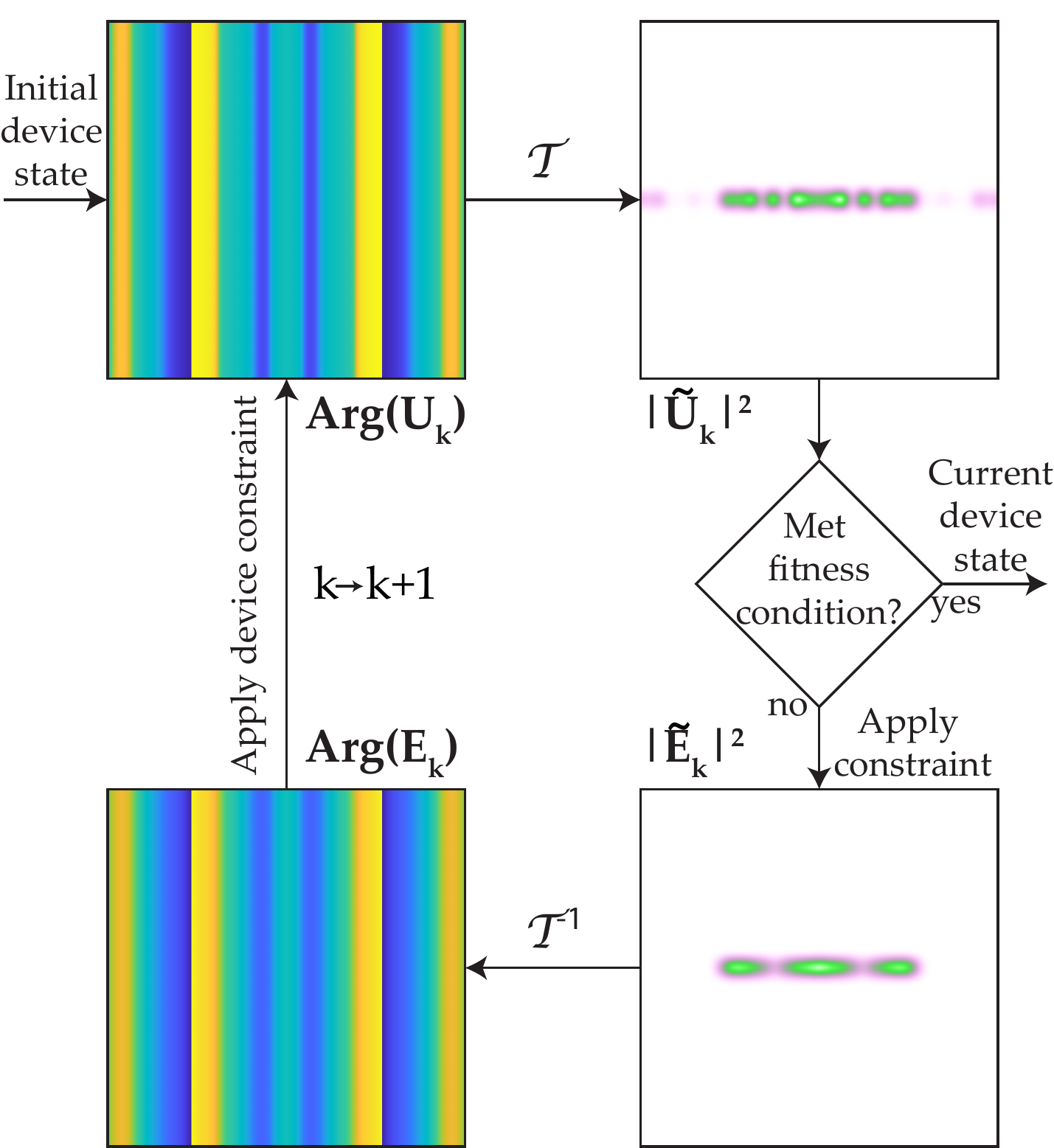}
\caption{An IFTA takes a guess of state, $U_k$, performs the transform to the projection plane, checks convergence to the ideal output state, if the convergence criterion is met the current device state is collected, otherwise the output constraint, $\tilde{E}_k$, is applied followed by the inverse transform and re-application of the device constraint (including finite phase/amplitude levels) for the next iteration. Alternate approaches based on minimization/maximization of cost functions may circumvent the bottom line of the figure and directly act on device elements.}
\label{fig:ifta}
\end{figure}

As mentioned earlier, spatial multiplexing can achieve smooth and flat optical potentials for trapping cold atoms. A na\"ive implementation of an iterative Fourier transform algorithm, IFTA, on a Gaussian laser beam will result in the production of phase singularities. Phase singularities can be introduced into light with an SLM with as little as three plane wave components \cite{OHolleran2006}. The production of a large field of phase singularities in the target plane can be considered equivalent to modulating the initial laser light with a diffuser \cite{Braeuer1991}. Instead, one would prefer to shape the initial Gaussian laser mode as if it were being transmitted with the smallest possible aberration or through an effective gradient index lens \cite{Rhodes1980,Han1983,Wang1993}. The removal of the singularities that particular filters can introduce is widely researched due to their importance for controlled beam shaping \cite{aagedal1996theory,senthilkumaran2005vortex,gaunt2012robust,Bowman2017May} and in particular to confine BEC in uniform potentials \cite{gaunt_bose_2013,navon_emergence_2016}. \figo{fig:ifta} shows a schematic representation of a typical IFTA. It is necessary, though tautologous to note that IFTAs or any other algorithm we discuss will only converge with particular device constraints, including the digitization of phase or amplitude shift levels, if the product of the incident beam mode and device modulation is capable of projecting the desired target pattern.

Convergence greatly differs with the circumstances of the device and optical system and thus care needs to be taken in designing the control loop. The classical Gerschberg-Saxton or Yang-Gu algorithms are suitable for phase retrieval \cite{Yang1994Jan}, which can approximate the phase pattern needed to optimize projected intensity \cite{Dufresne2001Mar,DiLeonardo2007Feb}, but not the generation of amplitude and phase. This and similar phase optimization techniques have been used to trap single atoms in 2D and 3D structures \cite{Bergamini2004Nov,nogrette_single-atom_2014,barredo_synthetic_2018}. A more meaningful investigation of optimization reveals that the quality of the solution is sensitive to the class of problem. A traditional IFTA (such as Gerschberg-Saxton or Yang-Gu algorithms), assumes that the problem being solved is convex, that is, has a simple minimum \cite{gill1981}. When the problem being solved is an array of resolvable spots it is highly likely that a minimum can be found. In general this is not the case. The algorithm is `greedy', it will dump as much light as possible into the target points. Convergence of this method can thus be poor for particular shapes and curves. This is the likely outcome for continuous line or ring patterns. The symmetry of the space is too high and many nearly degenerate local minimums can be found. Other schemes such as those in \citet{pasienski2008high,gaunt2012robust,Bowman2017May} are more lossy with regard to the light deflected to the target modes, but enable the production of smoother potentials as judged by their performance metrics. These schemes can create very uniform and reproducible illumination but at the cost of unavoidable dither noise ($\tilde D$). \citet{Bowman2017May} is based on conjugate-gradient minimization of a `cost' function that has good convergence properties that does not seem to have a strong dependence on the initial state.

\begin{figure}[h!]
\includegraphics[width=0.99\columnwidth]{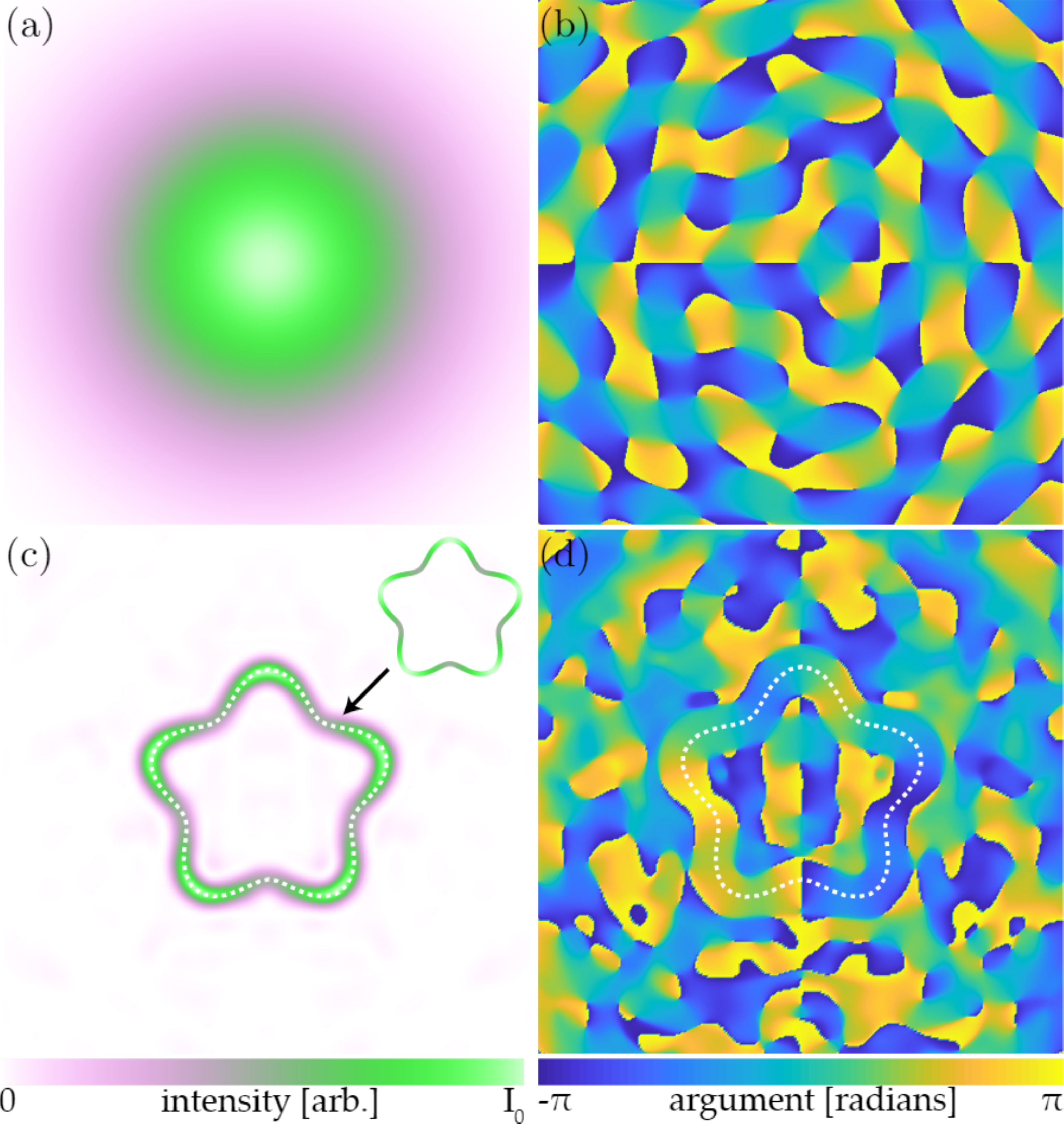}
\caption{Using a modification of the algorithm presented in \citet{DiLeonardo2007Feb} the phase pattern (b) is computed assuming an amplitude (a) to produce intensity and phases (c) and (d) in the plane of projection. The curve in the top right of (c) represents the shape and target intensity used in the algorithm. Good agreement between the shape and intensity can be observed.}
\label{fig:dileonardomethod}
\end{figure}

The weighted Gerschberg-Saxton method in \citet{DiLeonardo2007Feb} is straight forward to implement because it operates by updating the phase associated with the target modes (truncated Fourier series) only. There is an inherent cost to this as there are only a small number of degrees of freedom to modulate and so many ideal projection states may not be accessible to the algorithm so care should be taken when using it. It has the implicit advantage over a FFT method as spots can be generated at non-integer frequencies, which using a FFT would require an over sampling of points. The interesting feature of this kind of beam shaping is that the harmonic distortion amplifies the target modes. This conserves energy because the filtered signal must have a lower radiant flux than the input. \figreft{fig:dileonardomethod}{c} utilizes a modification to the weight function to also account for the amplitude in addition to having the points so close together their phases match up. It is important to choose a good starting configuration as it will enhance the chances that the algorithm converges to a more ideal result. The star pattern example has good convergence because it has low symmetry compared to a line or ring and light energy can be distributed away from the target pattern.

One of the clearest benefits of dynamical control of optical potentials is to transport BECs. For an SLM the concept is straightforward---create a dynamically addressable spot pattern to move, split or combine BECs. There are technical considerations to prevent unnecessary atom loss. There are a number of equivalent phase patterns that produce the same intensity spots at the same position with varying phases. Liquid crystals also require a finite time to move. The result is that light is scattered into the target state with time varying amplitude. In \citet{Boyer2004Sep} an algorithm to reduce the number of pixels changes when a ferroelectric Liquid crystal (chiral smectic C type) SLM was used to dynamically change multiple beam spots. The algorithm reduces this by finding patterns with minimal phase difference when transitioning between patterns. This process comes at the cost of also changing the phase of each spot. The subsequent experiment demonstrated transport of two and three well potentials containing Bose-Einstein condensates \cite{boyer_dynamic_2006}.

To address intensity flicker that can sometimes occur when dynamic changes to the SLM are made when trapping single atoms, a feedback scheme \cite{Kim2016}, and IFTA have been developed \cite{Kim2019}. Using these techniques high loading rates of $^{87}$Rb atoms and dynamical re-arrangement was demonstrated. Negligible atom loss was observed for 200~nm displacements. They also demonstrated movements of atoms over a 20~$\mu$m range. The ability to dynamically move traps with a LCD-SLM without high chance of loss is equally effective when compared to the creation of a large number of time-shared traps using feedback controlled AODs.

The correct model of the optical system is important because it directly impacts diffraction and hence distribution of speckle in the light field. A FFT without the use of an accurate point spread function results in a pure angular spectrum that will not accurately emulate the targeted output. The appropriate lens, transfer and point spread functions need to be included to minimize noise as these physical properties change the structure of speckle. These considerations are particularly important in an iterative approach when the projected light is a fine featured potential \cite{gaunt2012robust,gaunt_bose_2013}. Using shaped repulsive light, cylindrical `boxes' containing near uniform density BEC have been created \cite{gaunt_bose_2013}. This trapping configuration has been used to induce large length-scale excitation with a time varying magnetic field gradient and study the resulting turbulent cascade \cite{navon_emergence_2016}.

So far shaping has been discussed in this section with regard to beam shaping at a single plane. This highly constrains the degrees of freedom that may be modified and consequently lowers projection efficiency. Beam shaping in multiple planes is more flexible, but requires more optical elements in the beam path due to the extra operations. However, the results can be impressive with near perfect construction of beam modes using two-passes of an LCD-SLM utilizing an iterative algorithm on the first pass to change the amplitude profile and in a different plane correct the divergence and shape the phase \cite{jesacher2008near}. Multiple passes over successive beam shaping planes can effectively act as a volume hologram to sort beam modes \cite{Fontaine2019}. A consequence of this process is that beams with different output profiles are produced and can conceivably be used for any number of experiments including those on cold atoms.

\subsubsection{Direct imaging}
In a similar vein to the section on DMD-SLMs we will also highlight several uses of liquid crystal SLMs for imaging purposes. As twisted nematic liquid crystal SLMs rotate the polarization with the phase shift, a polarizing beam splitter can be used to split phase shifted light from the original beam. This enables direct imaging of a pattern of light similar to that achieved with DMD-SLMs. This particular technique was used to create and couple a source of cold atoms to a drain through a controllable disordered channel. This dumbbell trapping potential allowed for the measurement of Anderson localization in a two-dimensional system~\cite{white_observation_2020}. If we consider \figo{fig:SlmTable}(b) we see that even a phase only liquid crystal modulator can be used to change the amplitude of light when the `twist' of a twisted nematic crystal is emulated with the appropriate polarizer/analyzer combination.

Generalized phase contrast is a natural fit as a method for shaping light diffracted off a nematic liquid crystal device in the imaging mode as it can be used to project light into a wide variety of shapes \cite{Glueckstad2009,Lee2014}. Such methods were utilized to control the coupling between two potential wells and thus simulate a super current RLC circuit using BEC \cite{eckel2016contact}. With the appropriate choice of phase contrast filter the original laser mode shape can be transformed, for example from a Gaussian to flat top profile at high efficiency \cite{Banas2014} which can then be used for uniform trapping potentials.

\subsection{Technical Considerations}
At the heart of our discussion here is the use of various technologies for dynamic high-resolution control of ultracold atoms. Liquid crystal devices cover a range of modulation types including amplitude, phase and polarisation. However, no single device is equally capable of all kinds of modulation. Liquid crystals are viscoelastic and take time to re-orient and relax. Thus, care needs to be taken in choosing the correct device for a particular set of experiments. Here several physical effects are considered that may impact performance of these devices, and a few points to consider when operating them are described.
\subsubsection{Speed}
In early experiments on liquid crystal SLM control of BEC \cite{boyer_dynamic_2006}, technical limitations of the device and apparatus used meant that only a small sequence of trapping patterns could be used, and in some instances, up to half the BEC was lost due non-adiabatic movement of the harmonic potentials. Atom losses of less than $2\%$ were demonstrated if the gas could expand into the double or triple potential slowly by reducing the distance of travel of the potentials, effectively reducing the size of momentum kicks imparted to the gas and reducing the number of atoms lost over the edge of the potential in a large displacement. The device used here had fast update rates up to about 1~kHz which was achievable as the liquid crystals were chiral smectic C and only needed to rotate between easily accessible bi-stable orientations. Many of the available nematic type devices have liquid crystal rise times on the order of 20--25\,ms though relaxation times may be much longer because nematic liquid crystal motion is frustrated by neighbors. There are products which use high-voltage back planes which force the liquid crystal to move faster, down to about $\sim$3ms for several operating wavelengths. Given a large potential phase stroke and dynamic range over the back plane and fast GPU computation it may be possible to modify the behavior of some LCD-SLMs to attain millisecond phase level transition times for dynamic light projection \cite{Thalhammer2013Jan}.

\subsubsection{Calibration}
Optimal light projection efficiency with a phase plate of any sort relies on knowing the exact phase shift the plate induces. Liquid crystals cells exhibit non-linear response to bias, therefore each device needs to be calibrated. Furthermore, the optical path length difference will change with lab conditions such as temperature (as with most materials) and with different levels of illumination. Water cooling to stabilize response is an option available for a number of different models and makes of LCD-SLMs. Some devices are shipped with calibrations pre-loaded into the driving hardware and are convenient for general use. Alternatively, there are a large number of calibration techniques that can be found in the literature. We only pick a few (based only on being a classical technique) as so many other alternatives exist, and indeed, calibrations are fairly straight forward to develop given some knowledge of optics in the era of cheap and readily available components and computation. Classic methods include: Mach-Zehnder Interferometry~\cite{Konforti1988}, dual pinhole interference~\cite{Bergeron1995}, parametric fitting of a device model~\cite{Marquez2001}, and modulation of the stroke in using diffraction of binary-phase level (Ronchi) gratings~\cite{Zhang1994}. Manufacturers will also typically have recommended calibration methods for their devices.

\subsubsection{Fringing (cross-talk), and flicker}
\label{sec:fringing}
Liquid crystals can be modeled as continuous fluid acting under the influence of external fields. The forces acting on liquid crystal from the external field, in turn change the electric potential throughout the liquid crystal. Coupled with the fact that the field should leak between adjacent electrodes, this means that the phase difference created by the system consisting of signal electrode and liquid crystal is non-uniformly smoothed \cite{Bouvier2000,Apter2004} and this effect is magnified in thicker liquid crystal cells \cite{Linnenberger2006}. There is no guarantee that any part of the LC display reaches the actual set level. The practical outcome is that the diffraction efficiency can change even with a uniform phase-shift across the device, let alone for particular fine structural details of a pattern. Diffraction efficiency will change with a moved spot or pattern, even between two similar patterns. The accuracy can be somewhat improved by re-biasing the field between the adjacent elements over the discontinuity produced by taking the argument \cite{Linnenberger2006,Thalhammer2013Jan}. Another way this problem can be avoided is by compensating with a computational model to predict the phase retardation and light scattering on a per pixel level. Altogether this is a computationally intensive process. Progress optimizing light scattering using this model approach has been made \cite{Moser2019}.

\begin{figure}[]
\includegraphics[width=0.99\columnwidth]{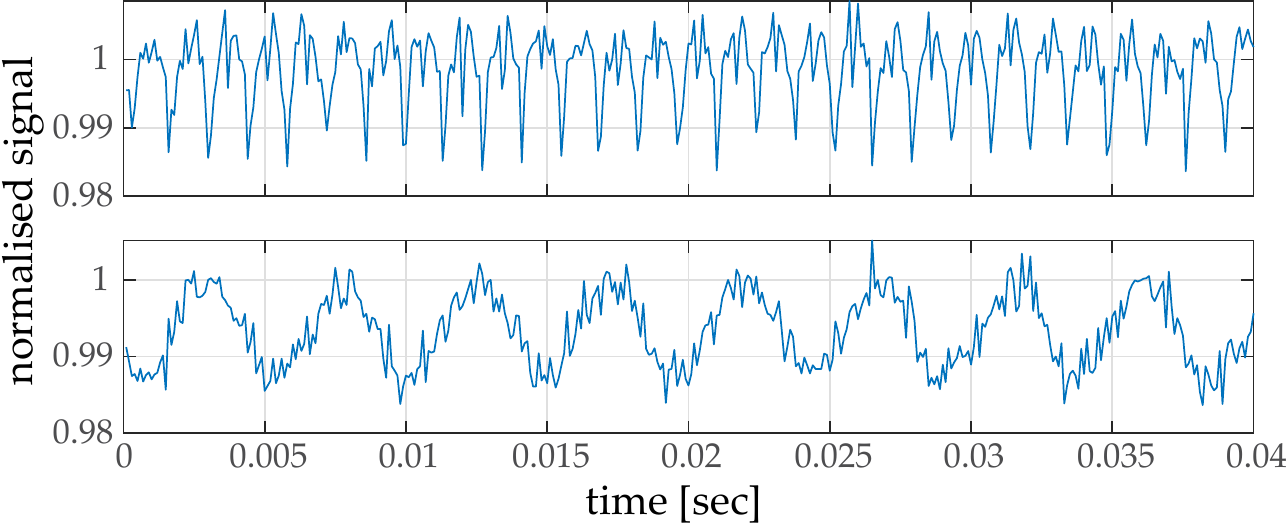}
\caption{Normalised diffracted intensity measured from LCD-SLMs HSP512L and P1920 (Meadowlark), respectively top and bottom traces.}
\label{fig:slmpower}
\end{figure}

Some of the earlier work on LCD-SLMs may give some impression that temporal flicker is a serious problem with using these devices for experiments with BECs. At that time, ca. 2000, this may have been a significant concern as the technology was still very much linked to the consumer display market where high frequency flicker or amplitude/phase coupling had no meaningful impact on the quality of the product. This is generally not the case with many of the LCD-SLMs available today. Many are designed for lab applications where scientists are concerned about the consistency of states output by the devices. There is no doubt that any light projection device requiring element state refresh cannot do so without changing the momentum and amount of light produced. To all practical extent these effects will be on the order of 1\% \cite{Haase2017}. We performed a measurement of intensity of the light diffracted from a customized HSP512L-785-1064 (Meadowlark Optics) and P1920-600-1300 (Meadowlark Optics) spatial light modulators. The normalized intensities sampled at 10kHz over 0.04\,s are shown in \figo{fig:slmpower}. Both SLMs display slightly more than 1\% noise in the diffracted signal as expected. HSP512L has a $\sim 2\times$2~cm area with a measured $\sim$211~Hz back-plane control signal (16-bit 10V output DVI-D controller). P1920 has a $\sim 2\times$1~cm area with a measured $\sim$127~Hz back-plane control signal (standard HDMI 8-bit controller). 
Ultimately, the flicker in LCD-SLMs are the product of the spatial and temporal fluctuations that arise due to the mismatch between the continuous fields driving a device with finite-, discrete-elements, and digital addressing. This means that they are not ideal for every situation, but as can be seen in this review they have successfully been applied to trap, interact, and dynamically manipulate cold and ultracold atomic systems and are worth considering.

\begin{table*}[t!]
\begin{center}
\begin{tabular}{| l  l | m{1.9cm} | m{1.9cm} | m{1.9cm} | m{1.9cm}|}
\hline
& \hspace{-3em} \textbf{Device features} & \textbf{AOD} & \textbf{EOD} & \textbf{DMD-SLM} & \textbf{LCD-SLM}\\  
\hline
I. &High Temporal Bandwidth  & $\checkmark$ & $\checkmark\checkmark$ & $\checkmark$ & $\times$ \\ 
\hline
II. &High Spatial Bandwidth   & $\times$ & $\times$ & $\checkmark$ & $\checkmark\checkmark$ \\
\hline
III. &Optical Phase Control   & $\times$ & $\times$ & $\checkmark$ & $\checkmark\checkmark$\\
\hline
IV. &High Output Light Power  & $\checkmark$ & $\checkmark\checkmark$ & $\times$ & $\checkmark$\\  
\hline
V. &Flicker Considerations  & $\checkmark$ & $\checkmark$ & $\checkmark$ & $\times$\\
\hline
VI. &Direct imaging & $\times$ & $\times$ & $\checkmark\checkmark$ & $\checkmark$\\
\hline
VII. &Fourier plane implementation& $\checkmark$ & $\checkmark$ & $\checkmark$ & $\checkmark\checkmark$\\
\hline
VIII. &Cost & $\checkmark\checkmark$ & $\checkmark$ & $\times$ & $\times$\\
\hline
\end{tabular}
\end{center}
\caption{Relative comparative guide to high-end research-grade technologies for creating configurable planar optical traps for atoms, indicating preferred devices for each of the categories. See text for details regarding each category.
 }
\label{table:Comparisons}
\end{table*}

\subsubsection{Hardware Selection}
The popularity of the device means that a wide variety of products are available. Nematic devices are one of the most common due to their flexibility and potentially high diffraction efficiency. Twisted nematic and ferroelectric devices are less common, but they offer different capabilities. Twisted nematics modulate both phase and amplitude and can be tuned to prioritize one over the other. Ferroelectric devices are faster, having comparatively small switching times, but are limited to two well defined phase states. Typically these liquid crystal cells will, either as a product or with add-ons, have controllers that will connect to graphics ports on standard computers. The speed at which signals can be changed is dictated by the video adapter used, HDMI for example will not exceed 120~Hz refresh rate. To circumvent this, some products have options to be controlled with custom adapters up to 2~kHz. LCD-SLMs designed for scientific use are usually monochrome and thus some manufacturers will have options for either higher bit-depth or increase the rate at which signals can be sent over a video adapter by using all of the color channels, e.g. $3\times8$-bit levels, or $3\times 60$~fps.


\section{Concluding Remarks}
\label{sec:conclusion}
In this review we have described the suite of technologies available to the experimenter for creating configurable optical potentials for ultracold atoms, primarily discussing AOD and EOD optical deflectors, and SLMs including DMD-SLMs and LCD-SLMs. These techniques have established new paradigms for the control and manipulation of ultracold neutral atoms. 

\subsection{Device Comparisons}
The devices described in this Review present the experimenter with a variety of approaches to achieving the desired geometry for a particular application. Some of the key comparative attributes, desirable for configurable trapping applications, are listed in Table~\ref{table:Comparisons}. In this comparison, we refer to  high-grade research devices. Device capabilities may vary significantly within the device families on the basis of the device cost, or the technology options within a single family, i.e. the several different technologies for LCD-SLMs. Table~\ref{table:Comparisons} can nonetheless be used as a general guide for selecting a device based on the features desired for a particular application. Details for each of the device features listed in Table~\ref{table:Comparisons} are now summarized.\\

\noindent \textit{I. High Temporal Bandwidth --} Having a high temporal bandwidth is useful when evolving a potential in time for study of time dynamics of the atomic system, or for phase imprinting; for a given phase gradient the phase imprinting time must be much smaller than the response time of the atom's center of mass. 
A high temporal bandwidth is also desirable for time-averaging, which allows for an increase in the spatial bandwidth of a given device at the expense of temporal bandwidth. This can only be achieved if the temporal bandwidth of the device is again much higher than the response of the atomic system. Temporal resolution is determined by the access time of the device, and modulation frequencies can range from $\sim 1$~kHz for LCD-SLMs, to $\sim 20$~kHz for DMD-SLMs, to $\sim 10$~kHz to $\sim 10$~GHz for AODs and EODs, respectively.\\

\noindent \textit{II. High Spatial Bandwidth --} The higher the spatial bandwidth of a device the more control over the light levels and light distribution it will have in the direct imaging plane, and the more efficient it will be at accessing different modes in the Fourier plane. Spatial bandwidth is the product of the number of spatial resolution elements and level control for each of these elements. Although DMD-SLMs and LCD-SLMs come in similar sizes in terms of number of diffraction elements, LCD-SLMs will have phase resolution $> 8$ bits, versus a single bit `on'/`off' for DMD-SLMs, which will increase efficiency of the produced modes in the Fourier space and light intensity control for direct imaging.\\

\noindent\textit{III. Optical Phase Control --} LCD-SLMs typically operate by spatially controlling the phase of a single polarization of the light. DMD-SLMs can also be used to modulate the  light phase at the cost of spatial bandwidth using amplitude holography as described in Sec.~\ref{sec:BinaryFourierImaging}.\\ 

\noindent\textit{IV. High Output Light Power --} The higher the light output power is for a given optical system, the easier it will be to create deep trapping potentials. The output light power depends on three factors, the device damage threshold, its usable area that can be illuminated, and its efficiency. EODs have the highest damage threshold of the devices considered. AODs have smaller damage threshold than EODs. LCD-SLMs have similar damage threshold to DMD-SLMs, but their diffraction efficiency tends to be higher since the light is directed into the produced mode, while DMD-SLMs will diffract light into many unwanted orders (see Sec.~\ref{sec:DMDs}).\\

\noindent\textit{V. Flickering Considerations --} Flickering will decrease the lifetime of the trapped atoms by transferring energy to the system through cyclical modulation of the output power. Flickering is not present in AODs and EODs. DMD-SLMs mirrors are usually unlatched and re-latched to prevent the mirrors from bonding to the substrate, resulting in a pulses at the refresh rate where the mirrors switch and resettle over $\sim 5~\mu$s~\cite{klaus2017note:}. Electrolytic degeneration and current leakage can occur in LCD-SLMs exposed to strong DC fields. An AC field is applied to prevent this at the cost of a phase fluctuation and a change in output intensity at that frequency. If this control signal is fast compared to the atomic response rate then no appreciable atom loss occurs. Flicker may be mitigated in DMD-SLMs by bypassing the mirror refresh clock~\cite{klaus2017note:}. Some types of smectic liquid crystal have two equilibrium states for particular substrates and can be switched in a similar manner to a DMD-SLM.\\ 

\noindent\textit{VI. Direct Imaging --} Direct imaging is usually easier using DMD-SLMs which have lower computational complexity and better contrast ratio in comparison with LCD-SLMs, which can only achieve high contrast ratios in direct imaging with additional steps, such as an intermediate phase-contrast~\cite{mogensen2000dynamic}.\\ 

\noindent\textit{VII. Fourier Plane Implementation --} AODs and EODs are inherently Fourier plane devices, where the deflection of the beam maps to different positions in the Fourier space. As mentioned in 
criteria~\textit{II.} above, LCD-SLMs have a higher spatial bandwidth which makes them better suited for mode manipulation in the Fourier plane than DMD-SLMs.\\ 

\noindent\textit{VIII. Cost --} For comparable research-grade high-end devices AOMs will be the cheapest followed by EODs. DMD-SLMs with their controllers have a comparable price to LCD-SLMs with their controllers.\\ 

New classes of experiments have been enabled by these techniques, and although the trade-offs in attributes may not select an overall winning technology. In certain applications one may be more suitable than the other. Recent results in the literature suggest a preference for particular approaches. Direct imaging of DMD-SLMs has received significant recent interest, and has provided new avenues for cold atom experiments, producing box traps~\cite{aidelsburger_relaxation_2017,ville_loading_2017, gauthier_direct_2016,gauthier_giant_2019,johnstone_evolution_2019, eckel_rapidly_2018}. DMDs have also enabled high-resolution dynamic control, thanks to large on-board memories and relatively high temporal bandwidth, and have facilitated new studies of two dimensional quantum turbulence~\cite{gauthier_giant_2019,johnstone_evolution_2019,stockdale_universal_2020} and condensate evolution in response to rapidly quenched trapping potentials~\cite{eckel_rapidly_2018,aidelsburger_relaxation_2017}. Furthermore, the ease with which the DMD-SLM trap can be configured has enabled groundbreaking studies in the emerging field of atomtronics, taking advantage of the straightforward tuning of the system parameters~\cite{hausler2017scanning,fritsch_creating_2020,gauthier_quantitative_2019,kwon2020strongly,luick_ideal_2020}.

LCD-SLMs in the Fourier plane (or, to a more limited degree, DMD-SLMs in the Fourier plane) are generally better suited than direct imaging of DMD-SLMs for the implementation of three dimensional trapping, as the resulting potential can confine along the beam propagation without the addition of the confining sheet as described in Sec.~\ref{sec:GaussianTraps}. LCD-SLMS are thus the technology of choice for arbitrary arrays of steerable atom traps, demonstrating deep and configurable trapping~\cite{PhysRevA.102.063107,Barredo2016,Bergamini2004Nov,nogrette_single-atom_2014,labuhn2016tunable}. Box traps have also been produced as a combination of TEM and LG modes created with an LCD-SLM, resulting in a uniform three dimensional BEC~\cite{gaunt_bose_2013}.

AODs have enabled steerable arrays of atom traps, and are better suited for dynamically rearranging trapped atoms than LCD-SLMs~\cite{doi:10.1063/1.5041481,endres2016atom,kaufman2014two}, facilitating quantum simulation experiments. Time-averaging, enabled by the high temporal bandwidth of AOD traps, enables the production of large and smooth trapping potentials~\cite{bell2016bose,1367-2630-17-9-092002,bell2020PhD}. While EODs have not yet been extensively used in atom trapping, their superior optical power handling (see Table~\ref{table:Comparisons}) may be attractive to future applications requiring deep optical traps, and their very large temporal bandwidth will improve time-averaging performance.

It is worth emphasizing these configurable technologies are not the only approaches to realizing boutique optical traps -- several static methods have been implemented with great success~\cite{boiron1998cold,newell2003dense,bakr2009quantum,tempone2017high,eckel2016contact,scherer2007vortex}. In particular, static holograms can provide substantial advantages for the generation of Laguerre-Gaussian and higher order Hermite Gaussian modes~\cite{meyrath2005high,campbell2012generation,tempone2017high}. However, it is in the dynamic manipulation of ultracold atom systems that these programmable technologies have enabled new classes of experiments to be performed. 

\subsection{Future Directions}
As trapping technologies continue to advance, future experiments may take advantage of better optical properties, such as higher efficiency coatings, larger pixel array SLMs and faster optical deflectors, permitting larger time-averaged optical patterns with finer details. The update rates of SLMs are in particular a limiting factor for some implementations such as time-averaging~\cite{gauthier_direct_2016,PhDGauthier2019,bell2020PhD}, which is particularly challenging for lightweight atoms such as $^6$Li~\cite{klaus2017note:}. Future experiments may more substantially develop three-dimensional configurable traps using the technologies, as have been demonstrated for arrays of single atom traps created with an LCD-SLM~\cite{barredo_synthetic_2018}. Substantial progress has been made in other applications using DMD-SLMs to rapidly create three-dimensional patterns, leading to applications such as rapid 3D printing of nanostructures~\cite{gauvin2012microfabrication}. Similarly, combinations of multiple technologies such as AODs coupled with SLMs, also realizing three-dimensional trapping, have been demonstrating in optical tweezers applications~\cite{tanaka20133d,Akselrod2006}. These approaches can combine the best properties of the various devices, such as using the fast deflection properties of AODs with the phase control of LCD-SLMs, or the rapid optical switching properties of DMD-SLMs. However, these approaches have yet to be extensively implemented in cold atoms trapping, with only a few initial demonstrations, see for example~\cite{Fatemi2007Mar}.

\begin{acknowledgments}
\noindent This research was supported by the Australian Research Council Centre of Excellence for Engineered Quantum Systems (EQUS, CE170100009) and the Commonwealth Defence Science and Technology Group NGTF-QT95 program. G. G. acknowledges support of ARC Discovery Project number DP200102239, and A.B.S. acknowledges support of ARC Discovery Project number DP180101002. T.W.N. acknowledges the support of an Australian Research Council Future Fellowship FT190100306.  
\end{acknowledgments}


%

\end{document}